\documentclass[%
 reprint,
 superscriptaddress,
 amsmath,amssymb,
 aps,
]{revtex4-2}

\usepackage{graphicx}
\usepackage{dcolumn}
\usepackage{bm}
\usepackage{hyperref}
\hypersetup{colorlinks=true,breaklinks,urlcolor=blue,linkcolor=blue,citecolor=blue}

\usepackage{soul}
\usepackage{xcolor}

\begin{document}

\preprint{APS/123-QED} 

\title{Ternary oxides of \textit{s}- and \textit{p}-block metals for \\
 photocatalytic solar-to-hydrogen conversion}

\author{Simon Gelin}
\email{gelin@psu.edu;}
\affiliation{Department of Materials Science and Engineering, and Materials Research Institute, The Pennsylvania State University, University Park, PA 16802, USA}
\author{Nicole E. Kirchner-Hall}
\affiliation{Department of Materials Science and Engineering, and Materials Research Institute, The Pennsylvania State University, University Park, PA 16802, USA}
\affiliation{Science \& Technology, Corning Incorporated, Corning, New York, USA}
\author{Rowan R. Katzbaer}
\affiliation{Department of Chemistry, The Pennsylvania State University, University Park, PA 16802, USA}
\author{Monica J. Theibault}
\affiliation{Department of Chemistry and Chemical Biology, Cornell University, 245 Feeney Way, Ithaca, New York 14850, United States}
\author{Yihuang Xiong}
\affiliation{Department of Materials Science and Engineering, and Materials Research Institute, The Pennsylvania State University, University Park, PA 16802, USA}
\author{Wayne Zhao}
\affiliation{Department of Materials Science and Engineering, and Materials Research Institute, The Pennsylvania State University, University Park, PA 16802, USA}
\affiliation{Department of Materials Science and Engineering, University of California, Berkeley, Berkeley, California 94720, USA}
\author{Mohammed M. Khan}
\affiliation{Department of Materials Science and Engineering, and Materials Research Institute, The Pennsylvania State University, University Park, PA 16802, USA}
\affiliation{Physical Sciences and Engineering Division, King Abdullah University of Science and Technology, Thuwal 23955-6900, Kingdom of Saudi Arabia}
\author{Eric Andrewlavage}
\author{Paul Orbe}
\author{Steven M. Baksa}
\affiliation{Department of Materials Science and Engineering, and Materials Research Institute, The Pennsylvania State University, University Park, PA 16802, USA}
\author{Matteo Cococcioni}
\affiliation{Department of Physics, University of Pavia, via Bassi 6, Pavia I-27100, Italy}
\author{Iurii Timrov}
\affiliation{Theory and Simulation of Materials (THEOS), and National Centre for Computational Design and Discovery of
Novel Materials (MARVEL), École Polytechnique Fédérale de Lausanne (EPFL), Lausanne CH-1015, Switzerland}
\author{Quinn Campbell}
\affiliation{Center for Computing Research, Sandia National Laboratories, Albuquerque, NM, USA}
\author{H\'ector Abru\~na}
\affiliation{Department of Chemistry and Chemical Biology, Cornell University, 245 Feeney Way, Ithaca, New York 14850, United States}
\author{Raymond E. Schaak}
\affiliation{Department of Chemistry, Department of Chemical Engineering, and Materials Research Institute, The Pennsylvania State University, University Park, PA 16802, USA}
\author{Ismaila Dabo}
\email{dabo@psu.edu}
\thanks{\\S.G. and N.E.K. contributed equally to this work.}
\affiliation{Department of Materials Science and Engineering, and Materials Research Institute, The Pennsylvania State University, University Park, PA 16802, USA}

\begin{abstract}
Oxides containing metals or metalloids from the {\it p}-block of the periodic table ({\it e.g.}, In, Sn, Sb, Pb, Bi) are of technological interest as transparent conductors and light absorbers for solar energy conversion due to the tunability of their electronic conductivity and optical absorption. Comparatively, these oxides have found limited applications in hydrogen photoelectrolysis primarily due to their high electronegativity, which impedes electron transfer for reducing protons into hydrogen. We have shown recently that inserting {\it s}-block cations into {\it p}-block metal oxides is effective at lowering electronegativities while affording further control of band gaps. Here, we explain the origins of this dual tunability by demonstrating the mediator role of {\it s}-block cations in modulating orbital hybridization while not contributing to frontier electronic states. From this result, we carry out a comprehensive computational study of 109 ternary oxides of {\it s}- and {\it p}-block metal elements as candidate photocatalysts for solar hydrogen generation. We downselect the most desirable materials using band gaps and band edges obtained from Hubbard-corrected density-functional theory with Hubbard parameters computed entirely from first principles, evaluate the stability of these oxides in aqueous conditions, and characterize experimentally four of the remaining materials, synthesized with high phase uniformity, to validate and further develop the computational models. We thus propose nine oxide semiconductors, including CsIn$_3$O$_5$, Sr$_2$In$_2$O$_5$, and KSbO$_2$ which, to the extent of our literature review, have not been previously considered as water-splitting photocatalysts.
\end{abstract}


\maketitle

\section{Introduction}
\label{sec:intro}

Solar hydrogen generation is pivotal to diversifying the global energy supply away from fossil fuels in the transportation sector and across major industries (\textit{e.g.}, ammonia synthesis, process metallurgy, and (bio)hydrocarbon production)~\cite{Pinaud:2013, Pivovar:2018}. Photoelectrolysis holds promise as a sustainable alternative to carbon-emitting hydrogen generation processes if scalable photocatalysts of high solar-to-hydrogen efficiency can be discovered, optimized, and widely deployed~\cite{Takata:2019}. First-principles calculations can expedite the preselection of candidate photocatalysts to be prioritized for synthesis and (photo)electrochemical characterization, which in turn can provide benchmark data to refine the precision of the computational models~\cite{Maeda:2007, Wu:2012}.

\begin{figure*}[t]
\includegraphics[width=0.9\textwidth]{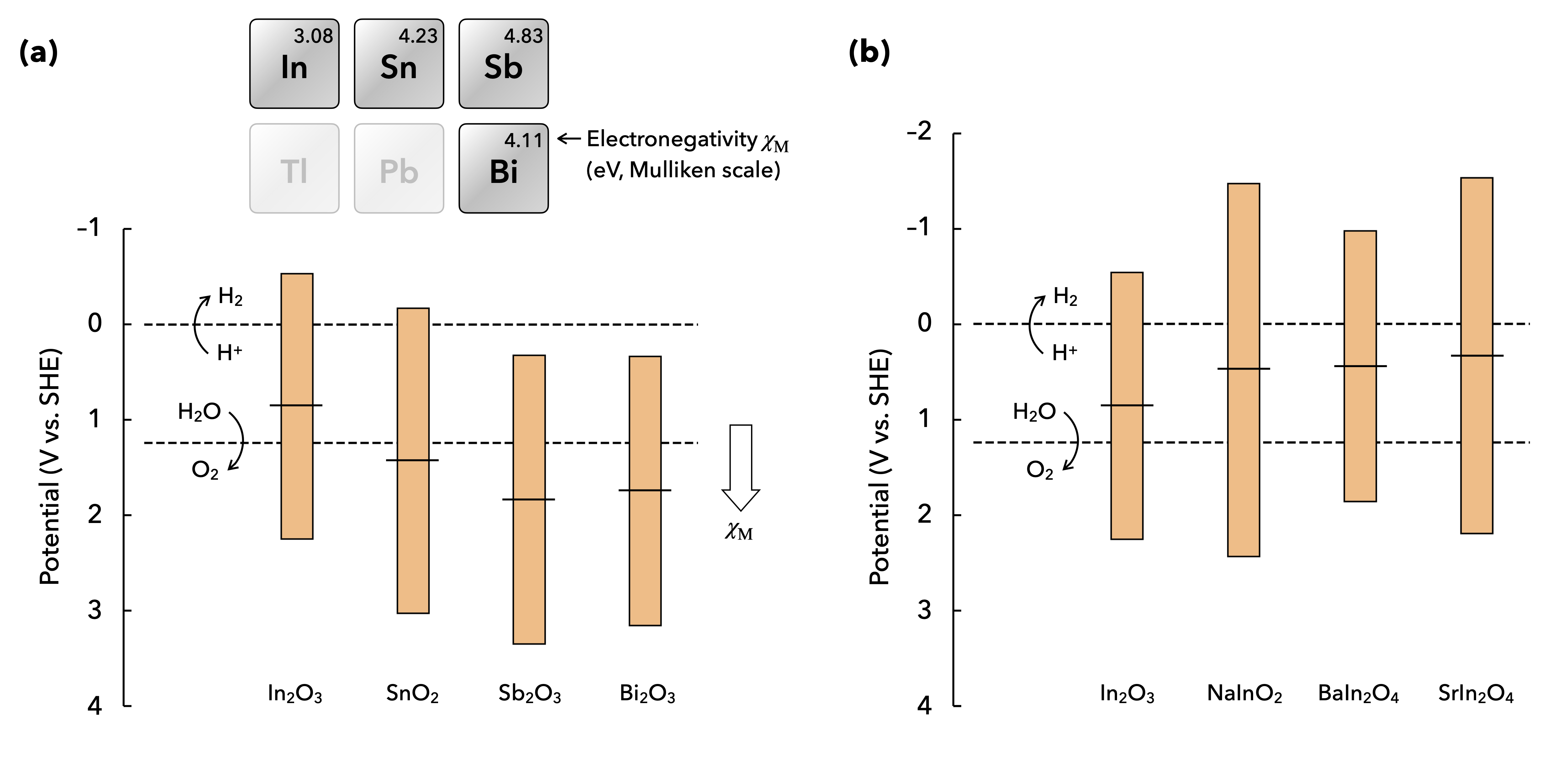}
\caption{Inserting {\it s}-block cations into {\it p}-block metal oxides enables band realignment. (a) Experimental band alignment with respect to the redox potentials of water for representative {\it p}-block metal oxides (In$_2$O$_3$~\cite{Liu:2013}, SnO$_2$~\cite{Liu:2013}, Sb$_2$O$_3$~\cite{He:2013}, Bi$_2$O$_3$~\cite{Jiang:2012}) with the mid-band-gap potential (approximating the flatband potential) shown as a horizontal line. As indicated by the open arrow, the band edges of an oxide tend to shift toward less reducing electrode potentials as the electronegativity of its cations increases. (b) Experimental band alignment for In$_2$O$_3$~\cite{Liu:2013} compared to that of In$_2$O$_3$-based compounds with the addition of {\it s}-block metal cations (NaInO$_2$, BaIn$_2$O$_4$, SrIn$_2$O$_4$~\cite{Xiong:2021}) that cause the band edges to shift toward more reducing potentials.}
\label{fig:binary-ternary-edges}
\end{figure*}

Xiong and coworkers~applied this approach to predict, test, and validate six water-splitting photocatalysts that were downselected from a pool of 70,150 candidates. Most of the downselected candidates were ternary oxides composed of alkali and alkaline earth ({\it s}-block) metals, and {\it p}-block metal cations ({\it e.g.}, In, Sn, Sb, Pb, Bi)~\cite{Xiong:2021}. From this study, it was inferred that the incorporation of {\it s}-block atoms into binary \textit{p}-block metal oxides shifts the band edges up (toward more reducing potentials) by lowering the electronegativities of the oxides, thereby promoting the electrochemical conversion of protons into hydrogen. As an example, Fig.~\ref{fig:binary-ternary-edges} shows the band edges of binary oxides of \textit{p}-block elements relative to the redox potentials of the hydrogen and oxygen evolution reactions [Fig.~\ref{fig:binary-ternary-edges}(a)]. This diagram also illustrates the changes in band edges arising from the addition of {\it s}-block cations into indium oxide [Fig.~\ref{fig:binary-ternary-edges}(b)]. The shift in the redox potentials of In$_2$O$_3$ originates from the low electronegativity of Na, Ba, and Sr compared to that of In [Fig.~\ref{fig:binary-ternary-edges}(b)]. The resulting variation increases the probability that the material can provide a reducing environment at its conduction band minimum~\cite{Butler:1978, Xu:2000, Pinaud:2013, Wu:2012}. This band realignment could be especially beneficial to Sb$_2$O$_3$ and Bi$_2$O$_3$ whose band edges do not straddle the hydrogen redox potential [Fig.~\ref{fig:binary-ternary-edges}(a)].

\begin{figure*}[t]
\includegraphics[scale=1]{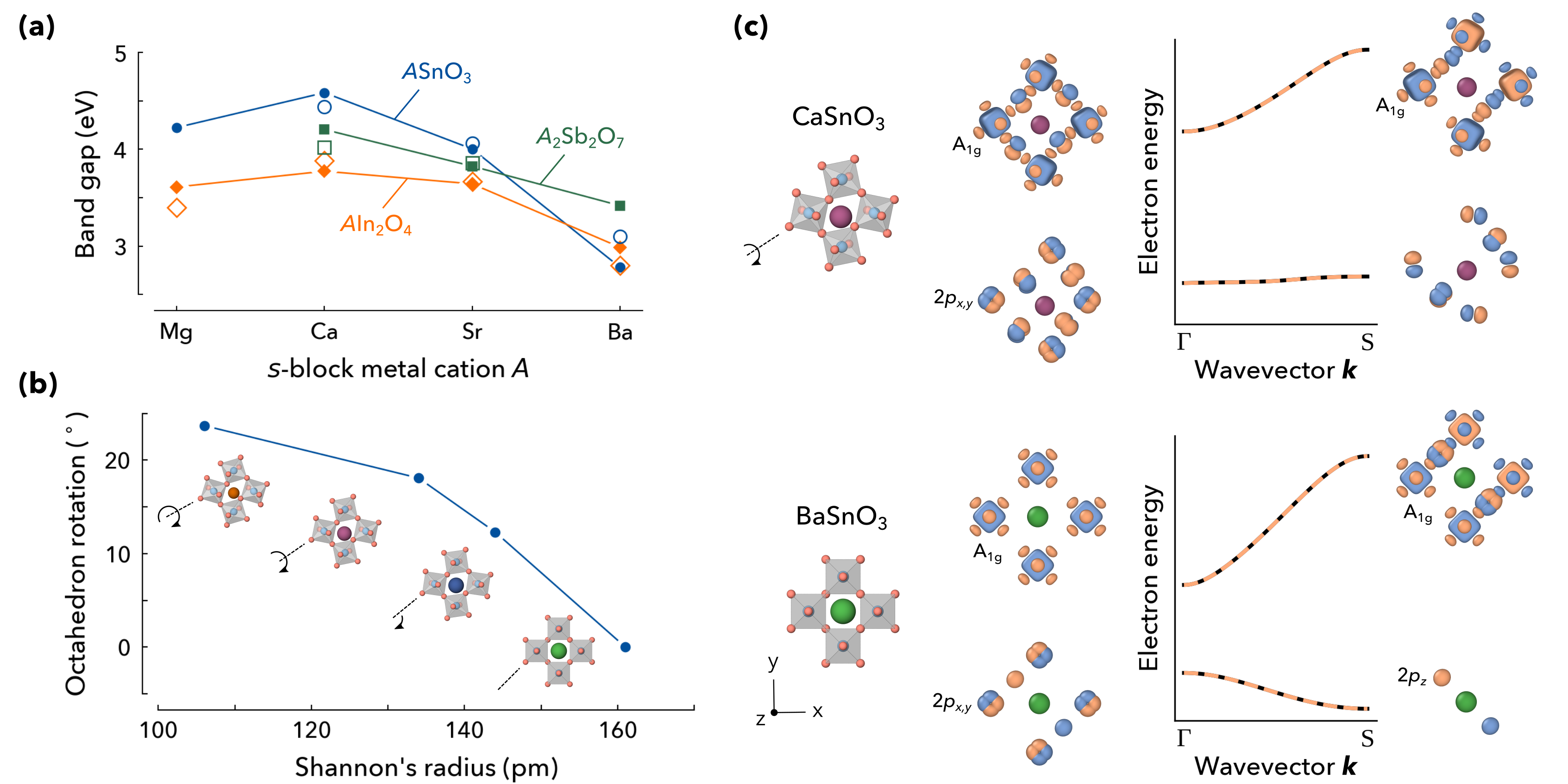}
\caption{Inserting {\it s}-block cations into {\it p}-block metal oxides enables band gap tuning. (a) Experimental and numerical band gaps (open and filled symbols, respectively) of stannate-based ternary oxides with perovskite~\cite{Zhang:2007}, pyrochlore~\cite{Lin:2006}, and spinel~\cite{Ueda:1992,Xiong:2021} structures. Band gaps are calculated using generalized-gradient density functionals, with scissors corrections of 2 eV (\textit{A}SnO$_\mathrm{3}$),  1.7 eV (\textit{A}In$_\mathrm{2}$O$_\mathrm{4}$), and 2.3 eV (\textit{A}$_\mathrm{2}$Sb$_\mathrm{2}$O$_\mathrm{7}$). (b) Octahedra of the perovskite \textit{A}SnO$_\mathrm{3}$ materials rotate away from the vertical axis as the radius of the {\it s}-block metal cation $A$ decreases, creating the distortions visible in the atomic representations. (c) Electronic structures of the highest occupied and lowest unoccupied bands in CaSnO$_\mathrm{3}$ (top) and BaSnO$_\mathrm{3}$ (bottom). The black and orange lines are energy bands calculated from the Kohn--Sham Hamiltonian and by Wannier interpolation~\cite{Marzari:2012,Pizzi:2020}, respectively. The decomposition of Bloch electrons into maximally localized Wannier orbitals at high-symmetry wavevectors --- $\Gamma=(0,0,0)$ and ${\rm S}=(\frac 12,\frac 12,0)$ --- shows that top valence bands are composed of oxygen $2p$ orbitals, whereas the lowest conduction band is a mixture of oxygen $2p$ orbitals and A$_{\text{1g}}$ orbitals centered around stannate cations. The centers of the Wannier orbitals are displaced away from the central {\it s}-block cation for ease of visualization.}
\label{fig:band-gap-trends}
\end{figure*}

While adding {\it s}-block cations into {\it p}-block metal oxides is beneficial to band alignment, it is expected to widen the band gap to the detriment of solar absorption. In fact, {\it s}-block metals have lower electronegativities than {\it p}-block metals, and the band gaps of oxides tend to increase with the difference of electronegativity between oxygen and their cations~\cite{QuartoSunseriPiazzaRomano1997}. If this empirical trend indeed dictates band gap variations in ternary oxides of \textit{s}- and \textit{p}-block metals, how is it that a majority of the newly proposed photocatalysts~\cite{Xiong:2021} belong to that same family of ternary semiconductors? To answer this question, we examine the band gaps of representative ternary oxides containing alkaline earth and {\it p}-block metal cations in Fig.~\ref{fig:band-gap-trends}. In accordance with the electronegativity trend, the band gap increases upon replacing Mg$^{2+}$ with Ca$^{2+}$; the Mulliken electronegativity of calcium is 0.75~eV lower than that of magnesium. However, the band gap systematically decreases when replacing Ca$^{2+}$ with Sr$^{2+}$ and Sr$^{2+}$ with Ba$^{2+}$, even though strontium has an electronegativity 0.20 eV lower than calcium and barium has an electronegativity 0.19 eV lower than strontium [Fig.~\ref{fig:band-gap-trends}(a)]. These unexpected variations can be traced back to distortions of the crystal structures induced by changes in the ionic radii of {\it s}-block metal cations [Fig.~\ref{fig:band-gap-trends}(b)]. Due to their very low electronegativities, alkaline earth metals easily oxidize and only contribute to low-lying valence states or high-energy excited states. As a result, band-edges states tend to be dominated by oxygen and {\it p}-block metal orbitals. This is best evidenced with the perovskite structure $A$SnO$_3$ ($A =$ Mg, Ca, Sr, or Ba), whose band-edge states consist only of oxygen 2$p$ orbitals and A$_{\text{1g}}$ molecular orbitals located at the center of the SnO$_6$ octahedra [Fig.~\ref{fig:band-gap-trends}(c)]. Although {\it s}-block metal cations do not contribute to electronic states around the band gap, they control structural distortions, thereby altering orbital hybridization. Notably, the conduction bands of perfectly cubic BaSnO$_3$ are much wider than those of distorted CaSnO$_3$ [Fig.~\ref{fig:band-gap-trends}(c)], because the Bloch modulation along the direction of two octahedra induces larger tight-binding interactions when the A$_{\text{1g}}$ orbitals are perfectly aligned and have maximal overlap. Since large bandwidths are associated with small band gaps, based on this analysis, we expect that reducing the distortions of a ternary oxide structure by varying the ionic radius of the {\it s}-block cation offers a possibility to counteract the electronegativity trend and fine-tune the band gap. This result provides a strong motivation to further explore the family of {\it s}- and {\it p}-block metal oxides in search of materials combining optical absorption and catalytic activity.

This combinatorial exploration necessitates reliable band-structure predictions at moderate computational cost. Figure~\ref{fig:band-gap-trends}(a) compares the experimental band gaps of representative ternary oxides containing alkaline earth metals with those computed from density-functional theory (DFT) within the generalized-gradient approximation. Applying a constant, positive scissors correction (\textit{i.e.}, a rigid shift of the conduction manifold~\cite{Godby:1988}) to materials having the same crystal structure yields band gaps that accurately match experimental results. In particular, the non-monotonic variations of the band gaps across the spinel structures  (\textit{A}In$_\mathrm{2}$O$_\mathrm{4}$) that contain alkaline earth metals (\textit{A} = Mg, Ca, Sr, Ba) are well captured by these calculations. This observation validates the capability of commonly used density-functional approximations to compute band-gap shifts as a function of composition, but the need for a scissors correction also indicates that the same approximations largely underestimate band gaps. This known limitation, which originates from self-interaction errors~\cite{Cohen:2008,Perdew:1981}, precludes the use of semilocal functionals to assess the solar compatibility and band alignment of the candidate photocatalysts. In principle, band gap predictions could be improved using many-body perturbation theories~\cite{Onida:2002} or hybrid functionals~\cite{Heyd:2003}; however, the computational cost of these methods limits their systematic applications to high-throughput materials screening. An alternative approach consists of restoring the invariance of the energy of an electronic state as a function of its occupation number by mapping the density-functional energy onto a Hubbard Hamiltonian, from which the self-interaction-correction terms can be readily derived (the DFT+$U$ method)~\cite{Anisimov:1991, Anisimov:1997, Dudarev:1998, Cococcioni:2005, Campo:2010}. Nevertheless, a central caveat limiting the predictive ability of the DFT+$U$ method is that the \textit{U} parameters (which control the strength of the self-interaction correction) are typically determined from experimental data~\cite{Yang:2018, Paudel:2008, Lalitha:2007, Gautam:2018, Long:2020}. 

To enable materials discovery, it is instead necessary to determine the Hubbard \textit{U} parameter non-empirically, from first principles using, \textit{e.g.}, constrained DFT (cDFT)~\cite{Dederichs:1984, Mcmahan:1988, Gunnarsson:1989, Hybertsen:1989, Gunnarsson:1990, Pickett:1998, Cococcioni:2005, Solovyev:2005, Nakamura:2006, Shishkin:2016}, constrained random phase approximation (cRPA)~\cite{Springer:1998, Kotani:2000, Aryasetiawan:2004, Aryasetiawan:2006, Sasioglu:2011, Vaugier:2012, Amadon:2014}, and Hartree-Fock-based methods~\cite{Mosey:2007, Mosey:2008, Andriotis:2010, Agapito:2015}. Within cDFT, a linear-response formulation has been developed~\cite{Cococcioni:2005} and recast in the framework of density-functional perturbation theory (DFPT)~\cite{Timrov:2018, Timrov:2021}. In this approach, $U$ parameters can be calculated in primitive unit cells of minimal size using multiple monochromatic perturbations of the on-site potentials, rather than on prohibitively large supercells with localized electronic perturbations~\cite{Timrov:2018, Timrov:2021}.

In the following, we apply this newly implemented linear-response method and validate it by sensitive electrochemical characterization to co-optimize band alignment, optical absorption, and aqueous stability in {\it p}-block metal oxides with {\it s}-block substituents. From this joint analysis, we recommend CsIn$_3$O$_5$, Sr$_2$In$_2$O$_5$, and KSbO$_2$ as oxide materials for solar-to-hydrogen photocatalysis.

\section{Methods and protocols}
\label{sec:methods}

\subsection{Candidate materials}
\label{sec:materials}

We investigated oxide compounds containing {\it s}-block (alkali and alkaline earth metal) cations (Li$^{+}$, Na$^{+}$, K$^{+}$, Rb$^{+}$, Cs$^{+}$, Mg$^{2+}$, Ca$^{2+}$, Sr$^{2+}$, Ba$^{2+}$) and closed-shell ($d^{10}$) \textit{p}-block elements (In$^{+}$, In$^{3+}$, Sn$^{2+}$, Sn$^{4+}$, Sb$^{3+}$, Sb$^{5+}$, Pb$^{2+}$, Pb$^{4+}$, Bi$^{3+}$, Bi$^{5+}$). The \textit{p}-block metal cations were limited to closed-shell ionizations as these electronic configurations correlate with superior photocatalytic performance and electrochemical durability~\cite{Wu:2012}. The resulting list of ternary oxides was further screened to only retain materials that are tabulated in the \textit{Materials Project} database (\textit{i.e.}, materials that have been synthesized or computationally proposed), have computationally predicted structures that match experimentally determined crystal structures, and have less than 50 atoms per unit cell to limit computational cost~\cite{Jain:2013}. The elements in these materials were examined to eliminate those of high toxicity, high radioactivity, or limited abundance following the criteria defined in Ref.~\onlinecite{Xiong:2021}. In explicit terms, all of the elements studied in this work have a median lethal dose value (LD$_{50}$) greater than 250 mg/kg~\cite{CDC:web, Arifin:2015, Asakura:2008, Sano:2005}; none of these elements are classified as radioactive~\cite{radioactivity}; and all of them have an abundance higher than that of gold (0.004 ppm by mass). Table S1 of the Supplemental Material (SM) enumerates the resulting 109 indate, stannate, antimonate, plumbate, and bismuthate materials (the 14 materials with a unit cell containing more than 50 atoms are also listed in the SM).

\subsection{Electronic-structure predictions}
\label{sec:DFTU}

First-principles calculations were initially performed within the generalized-gradient approximation. These predictions were then refined by applying the Hubbard model, which is a flexible and effective framework to correct self-interaction in density-functional approximations. In this model, the manifold of one-electron (Kohn--Sham) eigenstates $\psi_{n\boldsymbol{k}\sigma}$ with occupancies $f_{n\boldsymbol{k}\sigma}$ is mapped onto a set of (fixed) atom-centered orbitals $\varphi^{I\ell}_{m\sigma}$ whose (variable) occupancies can be expressed as
\begin{equation}
n^{I\ell}_{m\sigma} = \frac{\Omega}{(2\pi)^3}\sum_n \int_{\cal B}  f_{n\boldsymbol{k}\sigma} |\langle \varphi^{I\ell}_{m\sigma}| \psi_{n\boldsymbol{k}\sigma}\rangle|^2 d\boldsymbol{k},
\end{equation}
where $n$ and $\boldsymbol k$ are the index and wavevector of the Bloch state, $\ell$ is its angular momentum, $m$ is the orbital index (typically, the magnetic quantum number), $\sigma$ is the spin of the orbital, $I$ identifies the atomic site, $\Omega$ is the unit cell volume, and the integral spans the Brillouin zone $\mathcal B$. The contribution $\tilde E^{I\ell}$ from atomic site $I$ and angular momentum $\ell$ to the total energy is estimated, within the zero-configurational-width approximation~\cite{Hubbard:1963}, by counting the electrons and electron pairs at that site:
\begin{eqnarray}
\tilde E^{I\ell} & = &  \varepsilon^{I\ell} \Big( \sum_{m\sigma} n^{I\ell}_{m\sigma} \Big) + \frac{{u}^{I\ell}}2 \Big( \sum_{m\sigma} n^{I\ell}_{m\sigma} \Big)\Big( \sum_{m\sigma} n^{I\ell}_{m\sigma} - 1 \Big) \nonumber \\
& - & \frac{{j}^{I\ell}}2 \sum_\sigma \Big( \sum_{m} n^{I\ell}_{m\sigma} \Big)\Big( \sum_{m} n^{I\ell}_{m\sigma} - 1 \Big),
\label{eqn:hubbard-energy}
\end{eqnarray}
where $\varepsilon^{I\ell}$ is the energy cost of placing a lone electron at site $I$, and $u^{I\ell}$ ($j^{I\ell}$) is the electrostatic (exchange-correlation) energy of an electron pair, which is assumed to be the same for all of the pairs. This model is not self-interaction-free since the orbital energy $ \tilde \varepsilon^{I\ell}_{m\sigma} = \varepsilon^{I\ell} + u^{I\ell} ( \sum_{m'\sigma'} n^{I\ell}_{m'\sigma'} - \frac 12 ) - j^{I\ell} ( \sum_{m'} n^{I\ell}_{m'\sigma} - \frac 12 )$ depends on the occupancy $n^{I\ell}_{m\sigma}$. In other words, it does not fulfill the generalized  Koopmans theorem~\cite{Dabo:2010, Dabo:2014}. This error can be rectified by carrying out a more detailed (orbital-by-orbital) electron-pair counting:
\begin{eqnarray}
E^{I\ell} & = &  \varepsilon^{I\ell} \Big( \sum_{m\sigma} n^{I\ell}_{m\sigma} \Big) \nonumber \\
&+& \frac{u^{I\ell}}2 \Big( \sum_{m\sigma} n^{I\ell}_{m\sigma} \Big( \sum_{m'\sigma'} n^{I\ell}_{m'\sigma'} - n^{I\ell}_{m\sigma} \Big)\Big) \nonumber \\ & - & \frac{j^{I\ell}}2  \sum_\sigma \Big( \sum_{m} n^{I\ell}_{m\sigma} \Big( \sum_{m'} n^{I\ell}_{m'\sigma} - n^{I\ell}_{m\sigma} \Big)\Big).
\label{eqn:corrected-energy}
\end{eqnarray}
From Janak's theorem~\cite{Janak:1978}, the energy of a given orbital $\varphi^{I\ell}_{m\sigma}$ is thus equal to $\varepsilon^{I\ell}_{m\sigma} = \varepsilon^{I\ell} + u^{I\ell} ( \sum_{m'\sigma'} n^{I\ell}_{m'\sigma'} - n^{I\ell}_{m\sigma} )- j^{I\ell} ( \sum_{m'} n^{I\ell}_{m'\sigma} - n^{I\ell}_{m\sigma} )$ and does not vary with its own occupation number $n^{I\ell}_{m\sigma}$. 

Comparing Eq.~\eqref{eqn:hubbard-energy} to Eq.~\eqref{eqn:corrected-energy}, the self-interaction correction $\Delta E^{I\ell}$ to the Hubbard model can be written as
\begin{equation}
\Delta E^{I\ell} =\frac{U^{I\ell}}2  \sum_{m\sigma} n^{I\ell}_{m\sigma} \Big( 1 - n^{I\ell}_{m\sigma} \Big),
\label{eqn:hubbard-correction}
\end{equation}
where the corrective factor $U^{I\ell} \equiv u^{I\ell}-j^{I\ell}$ is called the Hubbard $U$ parameter. Implementing Eq.~\eqref{eqn:hubbard-correction} to predict band gaps requires one to compute $U^{I\ell}$ non-empirically. From Eq.~\eqref{eqn:hubbard-energy}, $U^{I\ell}$ can be identified as the second-order coefficient ${\partial^2 E}/{\partial (n^{I\ell}_{m\sigma})^2}$ (averaged over all orbitals $\varphi^{I\ell}_{m\sigma}$) of a quadratic expansion of the energy, which can be computed within density-functional theory as
\begin{eqnarray}
U^{I\ell} & = & \frac 1 {2(2\ell+1)}\sum_{m\sigma} \frac{\partial \varepsilon^{I\ell}_{m\sigma}}{\partial n^{I\ell}_{m\sigma}} \nonumber \\
& =& \frac 1 {2(2\ell+1)}\sum_{m\sigma} \langle \varphi^{I\ell}_{m\sigma} | \Delta^{I\ell}_{m\sigma}v_{\rm eff}| \varphi^{I\ell}_{m\sigma} \rangle .
\label{hubbard-u}
\end{eqnarray}
In Eq.~\eqref{hubbard-u}, $\Delta^{I\ell}_{m\sigma}v_{\rm eff}$ is the self-consistent linear response of the effective potential $v_{\rm eff}$ with respect to $n^{I\ell}_{m\sigma}$:
\begin{equation}
\Delta^{I\ell}_{m\sigma}v_{\rm eff}(\boldsymbol{r}) = \int \varepsilon^{-1}(\boldsymbol{r},\boldsymbol{r}') \Delta^{I\ell}_{m\sigma}v_0(\boldsymbol{r}')d\boldsymbol{r}'
\end{equation}
with
\begin{eqnarray}
\varepsilon^{-1} &=& ({\mathbb I}-k \cdot \chi_0)^{-1} \nonumber \\
&=& {\mathbb I}+k \cdot\chi_0+k \cdot \chi_0 \cdot k \cdot \chi_0+\cdots,
\end{eqnarray}
where $\varepsilon$ stands for the dielectric permittivity of the material, $\chi_0 = \delta \rho/\delta v_{\rm tot}$ is its bare susceptibility in response to the total (effective plus external) potential $v_{\rm tot}$, $k= \delta v_{\rm eff}/\delta \rho$ is the kernel of the effective potential $v_{\rm eff}$, $\Delta^{I\ell}_{m\sigma}v_0=k|\varphi^{I\ell}_{m\sigma}|^2$ is the bare response of $v_{\rm eff}$ to a change in $n^{I\ell}_{m\sigma}$, and the integral is carried over the entire space.

To resolve indeterminacies associated to the arbitrary rotations of the orbitals $\varphi^{I\ell}_{m\sigma}$, DFT+$U$ calculations were carried out in the rotationally invariant formulation of Dudarev \textit{et al.}~\cite{Dudarev:1998}, in which the DFT energy is recast as
\begin{equation}
E = \tilde E + \sum_{I\ell}\frac{U^{I\ell}}2  \sum_{mm'\sigma} n^{I\ell}_{mm'\sigma} \Big( \delta_{m'm} - n^{I\ell}_{m'm\sigma} \Big),
\end{equation}
where the doubly indexed occupation numbers
$$
n^{I\ell}_{mm'\sigma} = \frac{\Omega}{(2\pi)^3}\sum_n \int_{\cal B} f_{n\boldsymbol{k}\sigma} \langle  \psi_{n\boldsymbol{k}\sigma}|\varphi^{I\ell}_{m\sigma}\rangle \langle \varphi^{I\ell}_{m'\sigma}| \psi_{n\boldsymbol{k}\sigma}\rangle d\boldsymbol{k} 
$$
are the coefficients of the occupation matrix associated with the angular momentum $\ell$ at atomic site $I$.

As an alternative to Eq.~\eqref{hubbard-u}, the $U$ parameters can be computed (more efficiently) using a converse approach relying on density-functional perturbation theory~\cite{Timrov:2021, Timrov:2018}:
\begin{equation}
U^{I\ell} = \frac 1 {2(2\ell+1)}\sum_{m\sigma} \left( \Delta n^{I\ell}_{m\sigma} \right)^{-1},
\end{equation}
where $\Delta n^{I\ell}_{m\sigma} = {\partial n^{I\ell}_{m\sigma}}/{\partial \varepsilon^{I\ell}_{m\sigma}}$ stands for the self-consistent linear response of $n^{I\ell}_{m\sigma}$ to an on-site, orbital-specific perturbation of the potential $v_{\rm eff}$~\cite{Cococcioni:2005}. This converse approach has the distinct advantage of enabling one to consider monochromatic perturbations of the on-site potential, which removes the need for computationally expensive supercells by recasting the response to an isolated on-site perturbation into the response to a sum of wavevector-dependent perturbations~\cite{Timrov:2018}:
\begin{equation}
\Delta n^{I\ell}_{m\sigma} = \frac{\Omega}{(2\pi)^3}\int_{\cal B} \Delta_{\boldsymbol{q}} n^{I\ell }_{m\sigma} d\boldsymbol{q} ,
\label{eq:dnq}
\end{equation}
where $\Delta_{\boldsymbol{q}} n^{I\ell }_{m\sigma}$ is the linear response of the occupation $n^{I\ell }_{m\sigma}$ to a ${\boldsymbol{q}}$-modulated periodic perturbation of the potential acting on orbital $\varphi^{I\ell }_{m\sigma}$. Within density-functional perturbation theory, these calculations can be completed independently for each ${\boldsymbol{q}}$, allowing for efficient computer parallelization, as further described in Appendix \ref{sec:computational_procedures}~\cite{Timrov:2018}. 

\subsection{Downselection criteria}
\label{sec:pblock_methods_screen}

\begin{figure}[h]
\includegraphics[width=\columnwidth]{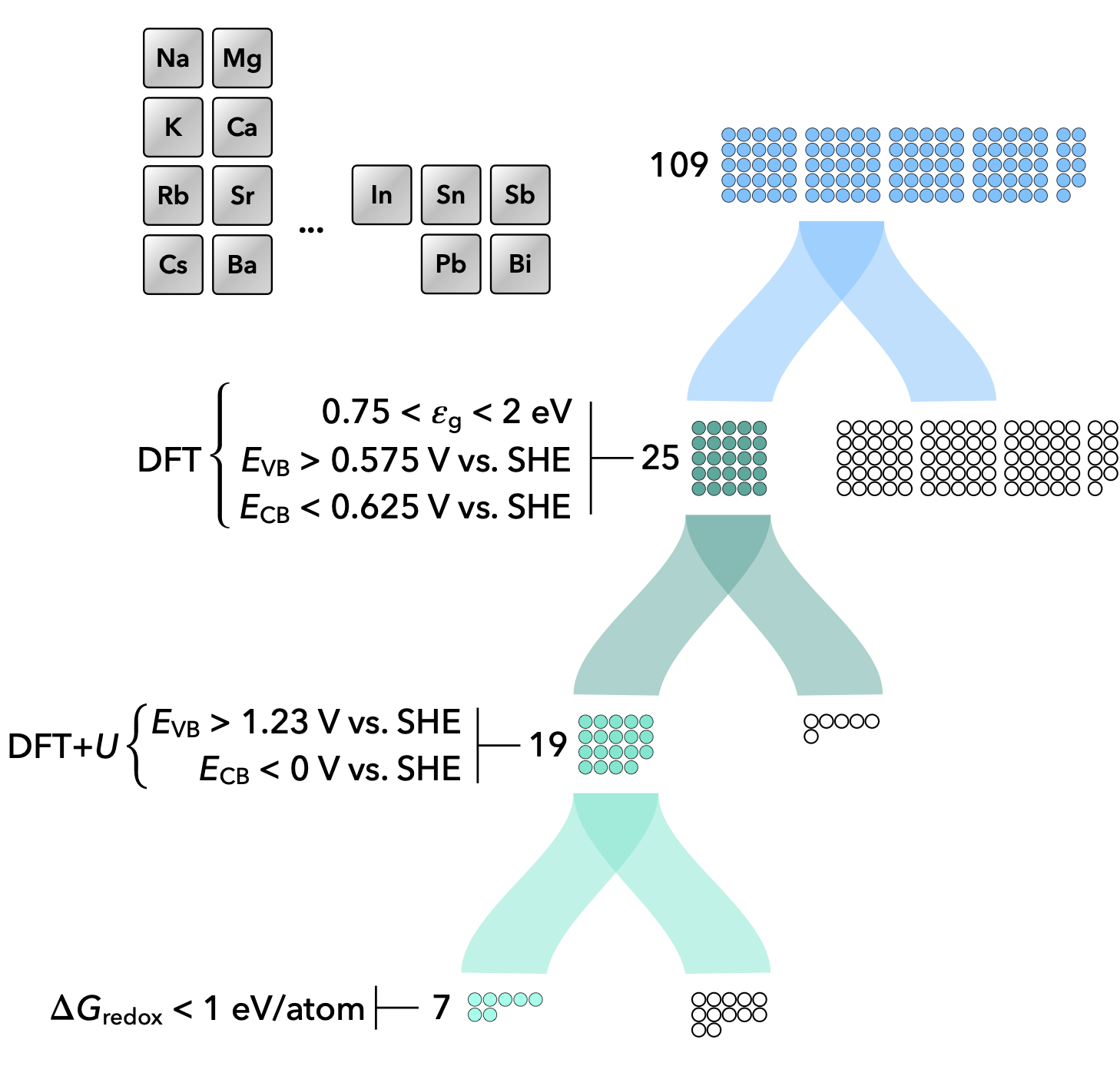}
\caption{Downselection of the 109 ternary oxides containing $s$- and $p$-block metals using band gaps ($\varepsilon_{\rm g}$) and band edges ($E_{\rm VB}$ and $E_{\rm CB}$) predicted at the DFT (step 1) and DFT+$U$ (step 2) levels, and Pourbaix free energies of decomposition under electrochemical conditions in aqueous environments.}
\label{fig:screening}
\end{figure}

The screening procedure is outlined in Fig.~\ref{fig:screening}. Band gaps and band edges of the 109 candidate materials (cf.~Sec.~\ref{sec:materials}) were first calculated at the DFT level. DFT predictions were then corrected using the DFT+$U$ method for materials that passed the first stage of the screening (cf.~Sec.~\ref{sec:DFTU}). In evaluating the valence band maximum ($E_{\text{VB}}$) and conduction band minimum ($E_{\text{CB}}$), the geometric mean of the Mulliken electronegativity of the constituent elements, $\langle \chi\rangle$, was used as an estimate for the flatband potential of the compound, through $E_{\text{FB}}=\langle \chi \rangle/e$, where $e$ is the elementary charge (cf.~Refs.~\onlinecite{Butler:1978,Xu:2000} and Sec.~IV of the SM). The valence and conduction band potentials were then calculated using the band gap, $\varepsilon_{\rm g}$, as $E_{\text{VB}} = E_{\text{FB}} + \varepsilon_\textrm{g} / (2e)$ and $E_{\text{CB}} = E_{\text{FB}} - \varepsilon_\textrm{g} / (2e)$.

Using the calculated band gaps and band edges, the compounds were screened to identify photocatalytic materials. Taking into account the fact that DFT with semilocal functionals underestimates band gaps by a typical margin of 20-50\% in semiconductors and insulators~\cite{Perdew:1983}, the screening criteria for solar absorption and band alignment were initially scaled down by a corrective factor of 0.5-0.8. Based on the derivation in Ref.~\onlinecite{Xiong:2021}, this rescaling yields the following criteria for band gaps and band edges calculated at the DFT level: ({\it i}) 0.75 $< \varepsilon_{\rm g} <$ 2 eV, ({\it ii}) $E_{\text{VB}} >$ 0.575 V vs.~SHE, and ({\it iii}) $E_{\text{CB}} <$ 0.625 V vs.~SHE.

DFT+\textit{U} calculations were then completed for the remaining candidate materials to obtain refined band gap and band alignment predictions. Materials were screened out at this step if they did not meet the following criteria: ({\it iv}) $E_{\text{VB}} >$ 1.23 V vs.~SHE (to provide an oxidizing environment for oxygen evolution), and ({\it v}) $E_{\text{CB}} <$ 0 V vs.~SHE (to provide a reducing environment for hydrogen evolution)~\cite{Maeda:2007, Wu:2012}. Pourbaix diagrams were then extracted from the {\it Materials Project} database for the remaining candidate materials to obtain stable decomposition products and electrochemical decomposition free energies were calculated at their flatband potential under pH 7 (cf.~Sec.~V in the SM). The materials were categorized as having low, moderate, or high predicted stability in water. Materials with a free energy of electrochemical decomposition $\Delta G_{\rm redox}$ of less than 0.5 eV/atom were labeled as having potentially high aqueous stability~\cite{Singh:2017}. Materials with $\Delta G_{\rm redox}$ between 0.5 and 1 eV/atom with stable solid decomposition products were labeled as having moderate aqueous stability~\cite{Singh:2017}. All remaining materials were labeled as having low stability in water and were screened out at this point. A literature review was finally conducted on the remaining high- and moderate-stability materials to assess their practical synthesizability and stability.

\section{Results and discussion}
\label{sec:pblock_results}

Following the methodology described in Sec.~\ref{sec:pblock_methods_screen}, the band gaps of the 109 candidate ternary oxide photocatalysts were computed at the DFT level, and their band edges were estimated from the geometric means of their elemental electronegativities. By considering the first set of screening conditions listed in Fig.~\ref{fig:screening}, we obtained 25 materials showing suitable preliminary band alignment. Before applying the Hubbard \textit{U} correction, we critically examined the contributions from the atomic orbitals to the density of states of the 25 candidate materials to determine the electronic states of highest intensity near the valence and conduction edges, on which the Hubbard $U$ corrections should be applied \cite{KirchnerHall:2021}. The projected density of states, shown in Figs.~S1 and S2 (SM), confirmed that the states at the valence band maximum are predominantly of O-2\textit{p} character. These states exhibit a large bandwidth in accordance with previous computational and experimental observations that {\it p}-block metal oxides tend to have low electron effective masses, which favors electron transport and electron--hole separation~\cite{Hautier:2014, Zhang:2015, Suzuki:2020}. 

Conversely, for most of these 25 materials, the conduction band minimum consists of a mixture of low intensity $s$ and $p$ states from the $p$-block elements, with some materials (\textit{e.g}., CsIn$_3$O$_5$) having high-intensity $d$ states deeper in the conduction band. This analysis showed that these materials tend to not have a high-intensity contribution from the $p$-block elements near the conduction or valence band edges. Hence, Hubbard $U$ corrections were not applied to the orbitals of $p$-block elements and only limited to the O-2\textit{p} orbitals, consistent with previous studies~\cite{KirchnerHall:2021}.

\begin{table*}[t]
\centering
\caption{DFT and DFT+\textit{U} band gaps and band edges for the 25 alkali and alkaline earth \textit{p}-block ternary oxides that passed the DFT screening step, grouped by indates, stannates, antimonates, plumbates, and bismuthates. The nine compounds remaining after the entire screening are highlighted by a star symbol ${(\star)}$. Hubbard parameters are given for the O-2\textit{p} states of each material, and multiple values indicate the presence of symmetrically non-equivalent oxygen atoms. Computed average electron effective mass values ($m^*$) are from the \textit{Electronic Transport Properties} dataset of Ricci \textit{et al.}~\cite{Ricci:2017}. Experimental band gaps from the literature are provided for comparison, if available. Band gaps measured in this study by Tauc analysis are highlighted by a dagger symbol ($\dagger$).}

\label{tab:25result}
\begin{tabular*}{0.9\textwidth}{@{\extracolsep{\fill}} c c c c  c  c c c  c c c  c c }
\\
\hline\\
Chemical & Space & Volume & $m^*$  & $U_{\textrm{O}}$& \multicolumn{3}{c}{DFT (eV)} & \multicolumn{3}{c}{DFT+$\boldsymbol{\textit{U}}$ (eV)} & \multicolumn{1}{c}{HSE06} & \multicolumn{1}{c}{Expt.} \\

formula & group & (\r{A}$^3$) & ($m_{\rm e}$) & (eV) & \multicolumn{1}{c}{$\varepsilon_{\text{g}}$} & $eE_{\text{VB}}$ & $eE_{\text{CB}}$ & $\varepsilon_{\text{g}}$ & $eE_{\text{VB}}$ & $eE_{\text{CB}}$ & $\varepsilon_{\text{g}}$ (eV)& $\varepsilon_{\text{g}}$ (eV) \\ \\

\hline \\

LiInO$_{2}$$^{(\star)}$                    & $I4_1/amd$ &  43.5 & 0.288 & 10.21        & 1.96 & 1.31  & --0.64  & 3.94  & 2.30 & --1.63 & 3.12 & 4.0$^\dagger$, 3.5~\cite{Xu:2017}, 4.3~\cite{Kushida:2006} \\
{CsIn$_{3}$O$_{5}$}$^{(\star)}$             & $Pnma$     & 139.9 & 0.273 & 9.20-9.63    & 1.59 & 1.21  & --0.38  & 3.42  & 2.13 & --1.30 & 2.95 & -- \\
{Sr$_2$In$_{2}$O$_{5}$}$^{(\star)}$         & $Ima2$     & 138.7 & 0.308 & 9.05-9.24    & 1.02 & 1.04  & 0.03  & 2.25  & 1.66 & --0.59 & 2.08 & -- \\
{SrIn$_{2}$O$_{4}$}$^{(\star)}$           & $Pnma$     &  92.3 & 0.228 & 9.31-9.45    & 1.96 & 1.61  & --0.35  & 3.92  & 2.59 & --1.33 & 3.29 & 3.8$^\dagger$, 3.67~\cite{Xiong:2021} \\
{Ba$_{3}$In$_2$O$_{6}$}$^{(\star)}$         & $I4/mmm$   & 188.7 & 0.368 & 10.43, 11.28 & 1.03 & 0.89  & --0.13  & 2.66  & 1.71 & --0.95 & 1.85 & 2.80~\cite{Yin:2002} \\

\\

K$_2$Sn$_2$O$_3$                             & $I2_13$    & 142.6 & 0.956 &  8.88        & 1.35 & 0.86  & --0.50  & 1.65  & 1.00 & --0.64 & --   & -- \\

\\

{KSbO$_{2}$}$^{(\star)}$                        & $C2/c$     & 73.8  & 0.919 & 8.84  & 1.98 & 1.62  & --0.35  & 2.82  & 2.05 & --0.77 & 2.83 & -- \\
{Ca$_{2}$Sb$_{2}$O$_{7}$}$^{(\star)}$  & $Imma$     & 140.2 & 0.333 &  8.59-8.95   & 1.94 & 2.44  & 0.49  & 4.09  & 3.51 & --0.58 & 3.56 & 4.5$^\dagger$, 4.5~\cite{Huang:2012} \\
{Sr$_{2}$Sb$_{2}$O$_{7}$}$^{(\star)}$   & $Imma$     & 150.2 & 0.324 &  8.97-9.19   & 1.56 & 2.17  & 0.61  & 3.66  & 3.22 & --0.44 & 3.15 & 4.3$^\dagger$, 4.2~\cite{Xue:2008}\\

\\

Li$_2$PbO$_3$                                & $C2/c$     & 64.3  & 0.368 & 9.98, 10.05  & 1.13 & 1.09  & --0.04  & 2.08  & 1.57 & --0.51 & --   & -- \\
Li$_4$PbO$_4$                                & $Cmcm$     & 97.9  & 0.527 & 10.00, 10.03 & 1.49 & 0.96  & --0.53  & 3.04  & 1.73 & --1.31 & --   & -- \\
Na$_4$PbO$_4$                                & $P\bar{1}$ & 139.9 & 0.423 & 9.68-9.77    & 1.36 & 0.78  & --0.58  & 3.22  & 1.71 & --1.51 & --   & -- \\
K$_2$Pb$_2$O$_3$                             & $I2_13$    & 147.7 & 0.482 &  8.63        & 1.80 & 0.97  & --0.83  & 2.25  & 1.19 & --1.05 & --   & -- \\
K$_2$PbO$_3$                                 & $P6_3/mcm$ & 96.6  & 0.416 & 9.09         & 1.31 & 0.84  & --0.47  & 2.02  & 1.19 & --0.83 & --   & -- \\
Rb$_2$PbO$_3$                                & $Pnma$     & 120.6 & 0.521 & 9.01, 9.77   & 1.39 & 0.82  & --0.57  & 2.45  & 1.35 & --1.10 & --   & -- \\
Cs$_2$PbO$_3$                                & $Cmc2_1$   & 132.2 & 0.562 & 8.82, 9.13   & 1.60 & 0.82  & --0.77  & 2.39  & 1.22 & --1.17 & --   & -- \\
{Ca$_{2}$PbO$_{4}$}$^{(\star)}$             & $Pbam$     & 95.7  & 0.394 & 8.53, 8.79   & 1.52 & 1.63  &   0.10  & 2.47  & 2.10 & --0.37 & 2.54 & 2.94~\cite{Xiong:2021} \\
CaPbO$_3$                                    & $Pnma$     & 68.4  & 0.542 & 8.63, 8.78   & 1.01 & 1.58  &   0.57  & 1.04  & 1.60 &   0.56 & --   & -- \\
Sr$_2$PbO$_4$                                & $Pbam$     & 108.3 & 0.396 & 9.07         & 1.46 & 1.50  &   0.04  & 2.32  & 1.93 & --0.40 & 2.45 & 1.75~\cite{Zhao:2015} \\
SrPbO$_3$                                    & $Pnma$     &  73.7 & 0.760 & 9.04, 9.10   & 0.91 & 1.46  &   0.55  & 0.64  & 1.32 &   0.68 & 1.14 & 1.80~\cite{Hadjarab:2007} \\
Ba$_2$PbO$_4$                                & $I4/mmm$   & 123.2 & 0.471 & 10.50, 10.60 & 1.40 & 1.36  & --0.03  & 2.36  & 1.84 & --0.52 & 2.09 & 1.45~\cite{Xiong:2021}, 1.7~\cite{Medicherla:2007} \\

\\

Li$_3$BiO$_4$                                & $P4_2/mnm$ & 80.3  & 0.541 & 9.63--10.25  & 1.31 & 1.17  & --0.14  & 2.41  & 1.72 & --0.69 & --   & -- \\
Li$_5$BiO$_5$                                & $Cm$       & 115.6 & 0.598 & 9.81-10.10   & 1.60 & 1.06  & --0.55  & 2.82  & 1.67 & --1.15 & --   & -- \\
Na$_3$BiO$_4$                                & $P2/c$     & 104.4 & 0.408 & 9.00, 9.94   & 1.18 & 1.00  & --0.18  & 2.42  & 1.62 & --0.80 & 2.43 & 2.61~\cite{Katz:2022}\\
K$_3$BiO$_4$                                 & $P\bar{1}$ & 100.0 & 0.710 & 8.77-9.68    & 1.37 & 0.81  & --0.56  & 2.60  & 1.43 & --1.18 & --   & -- \\
\\

\hline
\end{tabular*}
\end{table*}

We thus proceeded to calculate the electronic properties of the 25 candidates using the DFT+\textit{U} approach. The computed band gaps, band edges, and electron effective masses are presented in Table~\ref{tab:25result}. These DFT+$U$ results show a systematic increase (of 80\%, on average) in the band gap in comparison with DFT predictions (except for SrPbO$_3$, whose band gap decreases by 30\% when applying the correction), thereby bringing the calculated band gap in closer agreement with available experimental data. In specific terms, the Hubbard $U$ correction improves the precision of the computed band gaps for seven of the ten materials whose optical properties have been measured experimentally (namely, LiInO$_2$, SrIn$_2$O$_4$, Ba$_3$In$_2$O$_6$, Ca$_2$Sb$_2$O$_7$, Sr$_2$Sb$_2$O$_7$, Ca$_2$PbO$_4$, and Na$_3$BiO$_4$). With a mean absolute error (MAE) of 0.2 eV relative to experimental band gaps, these seven DFT+$U$ results lie within the range of experimental uncertainties. For the other three compounds with experimentally known band gaps, we found that the Hubbard $U$ correction either significantly overestimates the band gap (in Sr$_2$PbO$_4$ and Ba$_2$PbO$_4$) or further underestimates it compared to the DFT result (in SrPbO$_3$), suggesting that mid-gap states may be present or that the Hubbard correction should also be applied to Pb cations. Overall, DFT+$U$ predictions exhibit a mean absolute error of 0.45 eV with respect to experimental measurements, which reduces errors by more than 1 eV compared to DFT and 0.22 eV compared to the HSE06 hybrid functional calculations (cf.~Appendix~\ref{sec:computational_procedures} for details on computational methods, and Fig.~S3 of the SM for a comparison with experimental data). Furthermore, as mentioned above, the 25 screened oxides generally exhibit low electron effective masses, with 15 of them showing an effective mass $m^*$ lower than 0.5 \textit{m$_e$} within DFT+$U$. Consistent with the expected positive shift of the band edges (cf.~Sec.~\ref{sec:intro}), 24 of the 25 materials showed a cathodic realignment of their redox potentials upon alkali and alkaline earth incorporation. Finally, within subsets of ternary oxides belonging to related space groups, made of the same {\it p}-block metal, and sharing the same stoichiometry (namely, Ca$_2$Sb$_2$O$_7$ and Sr$_2$Sb$_2$O$_7$; CaPbO$_3$ and SrPbO$_3$; Ca$_2$PbO$_4$, Sr$_2$PbO$_4$, and Ba$_2$PbO$_4$), the band gap systematically decreases when the ionic radius of the {\it s}-block metal cation increases, as is clearly captured by the increase in unit cell volume. This observation is in line with the trend evidenced in Fig.~\ref{fig:band-gap-trends}: decreasing the ionic radius of alkaline earth metal cations accentuates the distortions of the structure, thereby reducing the overlap between the frontier orbitals. The resulting valence and conduction bands are narrower, causing the band gaps to increase.

Using the second set of criteria in Fig.~\ref{fig:screening}, the DFT+\textit{U} band gaps and band edges enabled us to narrow the search to 19 candidate materials. As a next step, we focused on the aqueous stability of these materials. To this end, Pourbaix diagrams were computed, and these results were used to examine the suitability of the candidate materials based on their electrochemical window and stable decomposition products at pH 7 and under the calculated flatband potential. Table S2 (SM) reports the flatband potentials, decomposition energies, and stable decomposition products for the 19 materials. Using these data, seven materials (LiInO$_2$, CsIn$_3$O$_5$, Sr$_2$In$_{2}$O$_{5}$, SrIn$_{2}$O$_{4}$, Ca$_2$Sb$_2$O$_7$, Sr$_2$Sb$_2$O$_7$, and Ca$_{2}$PbO$_{4}$) were classified as having high stability in water ($\Delta G_{\rm redox}$ $<$ 0.5 eV/atom with solid decomposition products)~\cite{Singh:2017}, whereas two materials (Ba$_{3}$In$_2$O$_{6}$ and KSbO$_{2}$) were classified as having moderate stability (0.5 $<$ $\Delta G_{\rm redox}$ $<$ 1 eV/atom with solid decomposition products).

\begin{figure}[t]
\includegraphics[width=\columnwidth]{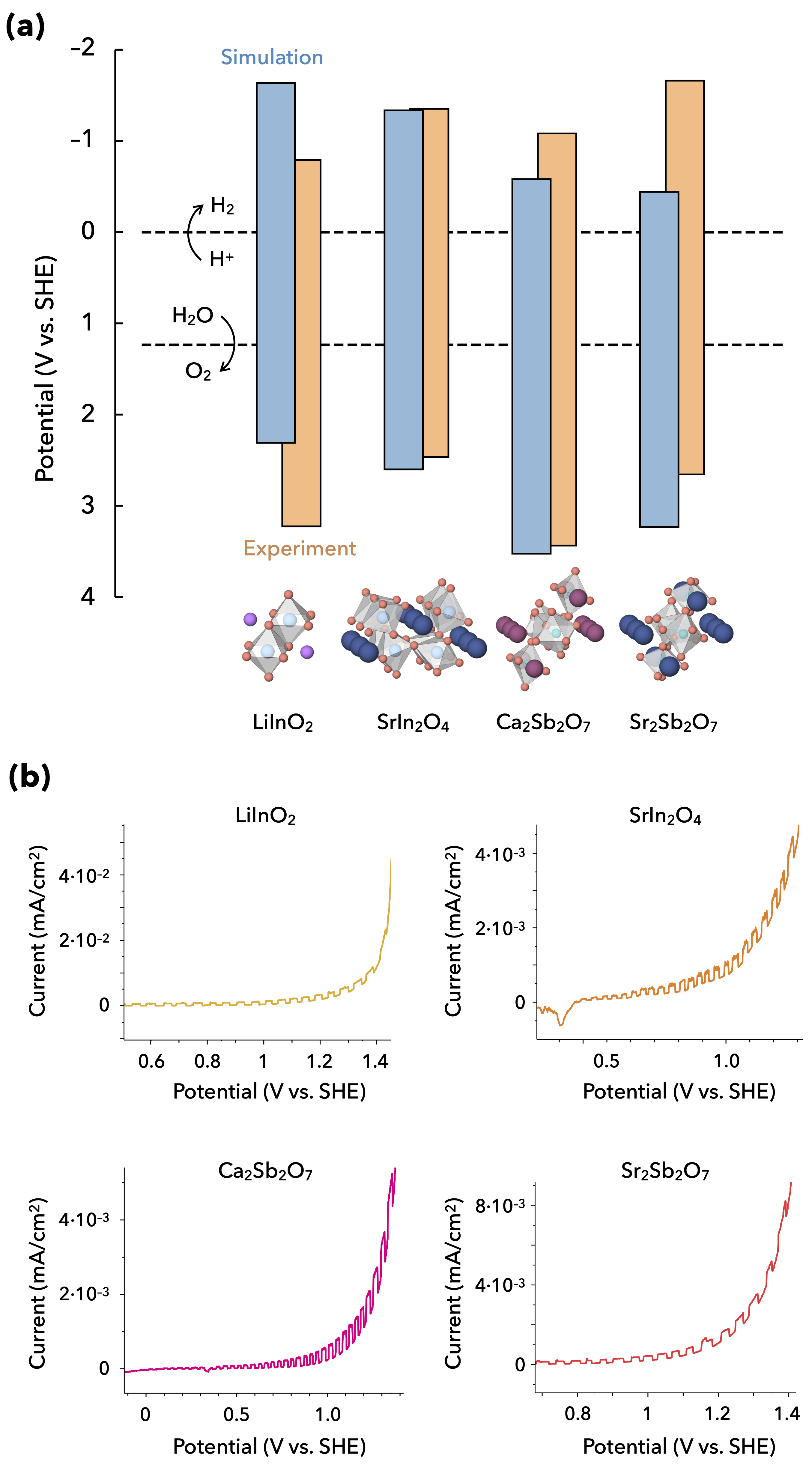}
\caption{(a) Comparison of the computationally determined (within DFT+$U$) and experimentally measured conduction and valence band edges for the $s$- and $p$-block metal oxides that were synthesized and whose phases were confirmed to be uniform. (b) Photoelectrochemical characterization of the synthesized oxides via voltage-dependent photocurrent measurements under chopped illumination.}
\label{fig:validation}
\end{figure}

Of the two materials exhibiting moderate water stability, Ba$_{3}$In$_2$O$_{6}$ is known to be a water-splitting photocatalyst~\cite{Yin:2002}, while, to the extent of our literature search, KSbO$_2$ has yet to be experimentally studied as a water-splitting photocatalyst and warrants further experimental examination. Among the seven materials with expected high aqueous stability, LiInO$_2$, Ca$_{2}$PbO$_{4}$, and SrIn$_2$O$_4$ are known wide-band-gap photocatalysts~\cite{Xiong:2021,Xu:2017}. In the literature, Sr$_2$In$_2$O$_5$ has been synthesized at 2,273~K from a single-crystal melt of SrO and In$_2$O$_3$, making its synthesis challenging with conventional solid-state techniques~\cite{Istomin:2014}. Ca$_2$Sb$_2$O$_7$ and Sr$_2$Sb$_2$O$_7$ have already been extensively studied as water-splitting photocatalysts at the experimental level, using RuO$_2$ as co-catalyst to facilitate oxygen evolution~\cite{Sato:2002}. In contrast, CsIn$_3$O$_5$ has not been studied as a photocatalyst, suggesting further computational and experimental verification of its water-splitting performance.

To directly assess the validity of our computational predictions, powder samples of the compounds were synthesized using solid-state reactions by precursor mixing and calcination at elevated temperatures, as described in Appendix~\ref{sec:experimental_procedure}. The powder X-ray diffraction patterns, shown in Fig.~S7 (SM), confirmed phase uniformity (in the range of 95-99\%) for four of the screened compounds (LiInO$_2$, SrIn$_2$O$_4$, Ca$_2$Sb$_2$O$_7$, and Sr$_2$Sb$_2$O$_7$). The band gaps of these materials were obtained by Tauc analysis of their optical absorption as a function of the incident photon energy. The Tauc plots, which are reported in Fig.~S8 (SM), did not reveal any perceivable contribution from mid-gap states and provided reliable band-gap estimates, enabling us to benchmark computational predictions against experimental measurements (Table \ref{tab:25result}). 

Chopped-illumination voltammetry was used to determine the flatband potential of these four phase-uniform materials via direct measurement of the potential at which the photocurrent shifts from anodic to cathodic (Fig.~\ref{fig:validation}). For LiInO$_2$, the chopped illumination photocurrent is shifted to a slightly higher potential as compared to the theoretical prediction, whereas for SrIn$_2$O$_4$, Ca$_2$Sb$_2$O$_7$, and Sr$_2$Sb$_2$O$_7$, the experimental photocurrent is shifted to a slightly lower potential (Fig.~\ref{fig:validation}a). The discrepancies between theoretical predictions and experimental measurements may stem from variations in the surface chemical composition compared to the bulk (estimations of LiInO$_2$ redox potentials, based on supercell surface calculations, are presented in Sec.~IV of the SM). In all cases, the experimental photocurrent still straddles the redox potentials of the hydrogen and oxygen evolution reactions, indicating the potential use of these materials as water-splitting photocatalysts. Of the four measured materials, LiInO$_2$ had the largest photocurrent density, albeit only on the order of $\sim$0.1~mA/cm$^2$ in these initial experiments (Fig.~\ref{fig:validation}b). Combining these materials with a co-catalyst may potentially improve their photocurrent density and water-splitting catalytic capabilities. Additionally, the increase in photocurrent as the sample is illuminated is smooth, as opposed to exhibiting a sharp spike and rapid decrease to an equilibrium illuminated current, indicating that little charge recombination occurs upon illumination.

\begin{figure}[!b]
\includegraphics[width=\columnwidth]{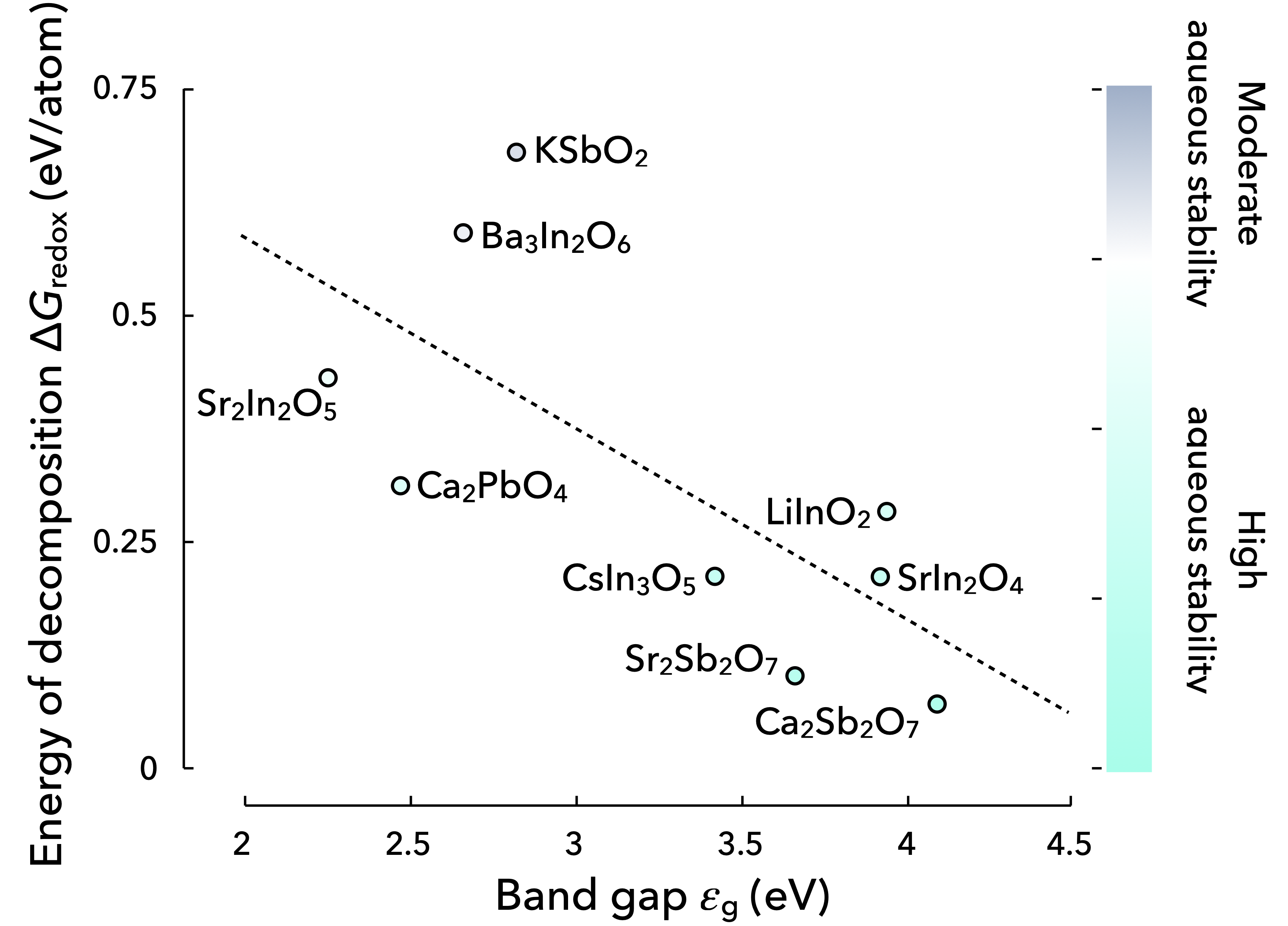}
\caption{ Dependence of the free energy of electrochemical decomposition $\Delta G_{\rm redox}$, calculated at pH 7 and at the flatband potential of the oxide, as a function of its band gap $\varepsilon_{\rm g}$, showing superior aqueous stability for wider energy gaps.}
\label{fig:decomposition}
\end{figure}

In closing, the dependence of the free energies of electrochemical decomposition $\Delta G_{\rm redox}$ of the recommended nine materials with respect to their band gaps $\varepsilon_{\rm g}$ is shown in Fig.~\ref{fig:decomposition}, indicating an increase in aqueous stability as a function of optical transparency. This dependence highlights the inherent tradeoff between electrochemical stability and optical absorption in developing solar-to-hydrogen photocatalysts, which reflects the competition between chemical bonding and electron delocalization in oxide systems~\cite{Burdett:1995}. Furthermore, it helps explain why LiInO$_2$, SrIn$_2$O$_4$, Ca$_2$Sb$_2$O$_7$, and Sr$_2$Sb$_2$O$_7$ could be obtained with high phase uniformity, as those four compounds are predicted to have the highest band gap and thus expected to be most thermodynamically stable. The observed inverse correlation between optical absorption and corrosion resistance may inspire future development strategies to optimize photocatalytic activity and durability, \textit{e.g.}, by designing core--shell structures that consist of an electrochemically stable, protective shell surrounding a light-absorbing core of the same metal-oxide family.

\section{Conclusion}
\label{sec:conclusions}

We have presented a computational investigation of 109 ternary oxides containing \textit{s}- and \textit{p}-block metals for use as water-splitting photocatalysts. Band gaps and band edges calculated from DFT with semilocal functionals were initially applied for a preliminary screening, and  DFT+\textit{U} calculations were conducted on the remaining candidates using Hubbard \textit{U} parameters calculated from first principles for the O-2\textit{p} states of these ternary oxides~\cite{Timrov:2021, Timrov:2018, KirchnerHall:2021}. Refined screening criteria were then considered to select 19 potential water-splitting photocatalysts based on their band alignment. For the 19 screened materials, stability in water was investigated via computed Pourbaix diagrams, enabling us to identify LiInO$_2$, CsIn$_3$O$_5$, Sr$_2$In$_{2}$O$_{5}$, SrIn$_{2}$O$_{4}$, Ca$_2$Sb$_2$O$_7$, Sr$_2$Sb$_2$O$_7$, and Ca$_{2}$PbO$_{4}$ as having potentially high stability, and Ba$_{3}$In$_2$O$_{6}$ and KSbO$_{2}$ as having moderate stability in water. We then experimentally investigated four of these predicted materials by synthesizing them, and measuring their band edges and voltage-dependent photocurrents, providing a critical assessment and direct validation of the computational predictions. This study demonstrates that the addition of alkali and alkaline earth metals to binary oxides of \textit{p}-block metals shifts up the band edges while potentially enabling further tuning of the band gap by distortion of the local coordination environment of the $p$-block metal cations. Moreover, this work highlights nine ternary oxides as potential photocatalysts. A literature search reveals that three of them, namely, CsIn$_3$O$_5$, Sr$_2$In$_2$O$_5$, and KSbO$_2$, have not been studied as water-splitting photocatalysts and suggests further experimental and computational analysis of their photocatalytic performance.

\begin{acknowledgments}
This work was primarily supported by the DMREF and INFEWS programs of the National Science Foundation under Grant No.~DMREF-1729338. I.T.~acknowledges support from the Swiss National Science Foundation (SNSF), through Grant No.~200021-179138, and its National Centre of Competence in Research (NCCR) MARVEL (Grant No.~205602). Contributions included work within the Research Experiences for Teachers program funded by the Center for Nanoscale Science under Grant No.~DMR-1420620 and DMR-2011839 (NSF funded MRSEC). Calculations were performed on the Roar supercomputer of the Institute for Computational and Data Sciences (ICDS) at the Pennsylvania State University. Sandia National Laboratories is a multimission laboratory managed and operated by National Technology and Engineering Solutions of Sandia, LLC, a wholly owned subsidiary of Honeywell International Inc., for the U.S.~Department of Energy’s National Nuclear Security Administration under Contract No.~DE-NA0003525. This paper describes objective technical results and analysis. Any subjective views or opinions that might be expressed in the paper do not necessarily represent the views of the U.S. Department of Energy or the United States Government.
\end{acknowledgments}

\appendix

\section{Computational procedures}
\label{sec:computational_procedures}

{\it First-principles calculations ---} Density-functional theory calculations were completed using the open-source Quantum ESPRESSO software, which implements the plane-wave pseudopotential method~\cite{Giannozzi:2009, Giannozzi:2017, Giannozzi:2020}. Within this framework, we used the generalized-gradient approximation for the exchange-correlation functional, with the Perdew–Burke–Ernzerhof parameterization revised for solids (PBEsol), and ultrasoft pseudopotentials from the SSSP Precision library (version 1.1.2)~\cite{Perdew:2008, Prandini:2018, Lejaeghere:2016}. Kinetic energy and charge density cutoffs of 60 and 480 Ry, respectively, were selected.  Brillouin zone sampling for computing electronic ground states used a Monkhorst-Pack~\cite{Monkhorst:1976} $\Gamma$-centered ${\boldsymbol{k}}$-point mesh with a uniform spacing of 0.04 \AA$^{-1}$. A full geometry optimization was completed for each material at the DFT level of theory~\cite{Katz:2022}. After structural optimization, Hubbard parameter calculations were completed using DFPT as implemented in the HP code~\cite{Timrov:2022}, which is part of Quantum ESPRESSO. The response due to isolated perturbations in supercells was recast into a sum of monochromatic perturbations in the Brillouin zone for a primitive unit cell~\cite{Timrov:2018}.  Thus, the supercell ${\boldsymbol{k}}$-points were converted to a ${\boldsymbol{q}}$-point sampling of the primitive cell using a ${\boldsymbol{k}}$-to-${\boldsymbol{q}}$-point ratio of 2~\cite{KirchnerHall:2021}.  

L\"owdin orthogonalized atomic orbitals were selected as projectors for the Hubbard manifold, and Hubbard corrections were only calculated for the O-2\textit{p} states~\cite{KirchnerHall:2021}. Projected density of states (PDOS) were computed at the DFT and DFT+$U$ levels of theory to ascertain the orbitals that contribute most to the valence and conduction band edges, and subsequently support the application of Hubbard corrections to O-2\textit{p} states but not orbitals of $p$-block metal cations. For these calculations, Gaussian smearing was used with a broadening parameter of $5 \times 10^{-3}$~Ry. The Hubbard parameters calculated for the O-2\textit{p} states of each material are reported in Table~\ref{tab:25result}. After the $U$ parameter calculations, the Hubbard corrections were applied to the O-2\textit{p} states and a self-consistent field calculation was conducted to compute the band gap (with the PBEsol-relaxed structure fixed). The band edges were calculated using the geometric mean of the Mulliken electronegativities as discussed in Sec.~\ref{sec:pblock_methods_screen}. 

Hybrid functional calculations of band gaps were performed using the HSE06 functional (with 75\% of the short-range part of the exchange energy computed using the PBEsol exchange energy and the remaining 25\% computed as the exact-exchange energy)~\cite{Heyd:2003}, and norm-conserving pseudopotentials from the PseudoDojo library~\cite{Setten:2018}. A kinetic energy cutoff of 80 Ry was used for the plane-wave expansion of both Kohn--Sham wavefunctions and the exact-exchange term. The $\Gamma$-centered ${\boldsymbol{k}}$-point mesh was kept constant or densified so that the $\Gamma$-centered ${\boldsymbol{q}}$-point mesh used to sample the Fock operator could be coarsened by a factor two in each directions of the reciprocal lattice. The Gygi--Baldereschi scheme was used to treat the singularity of the Coulomb interaction in Fourier space at small ${\boldsymbol{q}}$-vectors~\cite{Gygi:1986}.

Data and metadata used to produce the results of this paper are available in the Materials Cloud Archive \cite{MaterialsCloudArchive2022}.

{\it Estimation of Shannon ionic radii ---} Shannon effective ionic radii of alkaline earth cations were taken from Ref.~\cite{Shannon:1976}, except for Mg$^{2+}$, whose radius is not listed at the oxygen coordination number of 12. The radius $R$ of Mg$^{2+}$ was instead estimated using an empirical bond-strength–bond-length relation: $s = s_0  \left(R / R_0\right)^{-N}$, where the Mg-O bond-strength $s$ is given by the ratio of the oxidation number of Mg$^{2+}$ over its oxygen coordination number, and $R_0$, $s_0$, and $N$ are fitting parameters given in Ref.~\onlinecite{Brown:1973}.

\section{Experimental procedures}
\label{sec:experimental_procedure}

{\it Synthesis methods ---} Sr$_2$Sb$_2$O$_7$ was synthesized from stoichiometric amounts of Sr(NO$_3$)$_2$ and Sb$_2$O$_3$, calcined at 600 $^\circ$C for 48 hours, then annealed at 1050 $^\circ$C for 24 hours. Ca$_2$Sb$_2$O$_7$ was synthesized from stoichiometric amounts of Ca(NO$_3$)$_2\cdot$4H$_2$O and Sb$_2$O$_3$, calcined at 600 $^\circ$C for 48 hours, then annealed at 1050 $^\circ$C for 24 hours. The procedure for both of these syntheses was adapted from Ref.~\onlinecite{Lin:2006}. LiInO$_2$ was synthesized from In$_2$O$_3$ and Li$_2$CO$_3$, with 20\% molar excess of Li$_2$CO$_3$. The sample was annealed at 1000 $^\circ$C for one hour. The procedure was modified from that reported in Ref.~\onlinecite{Kawakami:2003}. SrIn$_2$O$_4$ was synthesized from stoichiometric amounts of SrCO$_3$ and In$_2$O$_3$, annealed at 1200  $^\circ$C for 3 hours. This procedure follows that reported in Ref.~\onlinecite{Guan:2015}.

{\it Powder X-ray diffraction ---} Powder X-ray diffraction (PXRD) scans were performed on a Malvern Panalytical Empyrean diffractometer with a copper source, using line focus and reflection mode. Powders were pressed to an even height in a well-plate zero-background silicon holder. 

{\it Ultraviolet-visible spectroscopy ---} A Perkin Elmer Lambda 950 UV-Visible spectrometer was used to measure diffuse reflectance spectra with a 150 mm integrating sphere, in diffuse reflection mode. Spectra were collected from 250-2500 nm, with 2 nm steps. The reference spectrum for total reflectance was measured against a Spectralon disc standard. Samples were suspended in ethanol and drop cast on quartz glass slides (Electron Microscopy Sciences). The slides were dried with nitrogen before measurement. Plots of the Kubelka--Munk function, raised to the power of $\frac 12$ or 2 for indirect and direct band gaps, respectively, vs. energy in eV were constructed. Band gaps were calculated by fitting the linear region in the onset of absorption and extrapolating to the  intercept of the horizontal energy scale.

{\it Device fabrication ---} Each material was ground in a mortar and pestle to ensure fineness. Then, inks were made in concentrations of 0.004 mM material in ethanol, and these inks were sonicated for one hour. 100 $\mu$L of each ink was deposited on a 5 $\times$ 8 mm fluorine-doped tin oxide (FTO) slide, in 20 $\mu$L increments, and these inks were annealed at 400 $^\circ$C in a Thermolyne muffle furnace under ambient atmosphere for 2 hours.

{\it Mott--Schottky experiments ---} Mott--Schottky measurements were run on a Biologic SP-150 potentiostat using the Staircase Potentiometric electrochemical impedance spectroscopy (EIS) function, with data taken at 7 frequencies between 30 kHz and 5 kHz. An initial scan was run from 1.5 to --1.5 V (vs.~Ag/AgCl in saturated KCl), with subsequent scans for each material taken in a smaller range. The electrolyte used was a pH 8 sodium phosphate buffer (Hydrion, 1 g/100 mL), the reference electrode was Ag/AgCl in saturated KCl, and the counter electrode was a graphite rod. The working electrode was a device made from each material deposited on FTO, as previously described.

{\it Chopped-illumination experiments ---} Measurements were carried out on a Biologic SP-150 potentiostat using linear sweep voltammetry (LSV) at 5 mv/s, from 1.2 V to --1.0 V (E vs Ag/AgCl in saturated KCl), and in pH 8 aqueous sodium phosphate buffer ($\approx$1 g/100 mL). The reference electrode was Ag/AgCl in saturated KCl, the counter electrode was a graphite rod, and the working electrode was the material on FTO. The samples were illuminated through a Starna cells quartz cuvette with a Newport Xenon 300 W lamp under an illumination of one sun. The chopping frequency was of 0.2–0.3 Hz. 

{\it Light-saturated open-circuit voltammetry ---} Measurements were carried out on a Biologic SP-150 potentiostat using the ``open circuit potential'' function, and were measured over 30 seconds. The samples were illuminated through a Starna cells quartz cuvette with a Newport Xenon 300 W lamp under an illumination of one sun, and the reference, working, and counter electrodes were the same as described previously. Each working electrode was allowed to equilibrate under illumination for 5 minutes, or until the measured open-circuit potential stabilized, before the measurement was taken. 

{\it Discussion on flatband-potential determination ---} Three methods were used to determine the position of the flatband potential of the four synthesized materials: Mott--Schottky, chopped-illumination, and light-saturated open-circuit potential measurement. The Mott--Schottky method, while the most often used, is an indirect measurement of the flatband potential, whereas the chopped-illumination and light-saturated methods are direct measurements. Mott--Schottky relies on several assumptions, such as a negligible capacitance from the Helmholtz layer, which can artificially skew the data toward more negative potentials. In contrast, chopped illumination is a direct measurement of the flatband potential using photocurrent, and light-saturated open-circuit potential (OCP) measurements seemed to confirm the general trends of the chopped-illumination measurements. Thus, chopped illumination was consistently used in this work to determine the flatband potential and the band edges. 

\bibliographystyle{unsrtnat}

\begin{thebibliography}{94}%
\makeatletter
\providecommand \@ifxundefined [1]{%
 \@ifx{#1\undefined}
}%
\providecommand \@ifnum [1]{%
 \ifnum #1\expandafter \@firstoftwo
 \else \expandafter \@secondoftwo
 \fi
}%
\providecommand \@ifx [1]{%
 \ifx #1\expandafter \@firstoftwo
 \else \expandafter \@secondoftwo
 \fi
}%
\providecommand \natexlab [1]{#1}%
\providecommand \enquote  [1]{``#1''}%
\providecommand \bibnamefont  [1]{#1}%
\providecommand \bibfnamefont [1]{#1}%
\providecommand \citenamefont [1]{#1}%
\providecommand \href@noop [0]{\@secondoftwo}%
\providecommand \href [0]{\begingroup \@sanitize@url \@href}%
\providecommand \@href[1]{\@@startlink{#1}\@@href}%
\providecommand \@@href[1]{\endgroup#1\@@endlink}%
\providecommand \@sanitize@url [0]{\catcode `\\12\catcode `\$12\catcode
  `\&12\catcode `\#12\catcode `\^12\catcode `\_12\catcode `\%12\relax}%
\providecommand \@@startlink[1]{}%
\providecommand \@@endlink[0]{}%
\providecommand \url  [0]{\begingroup\@sanitize@url \@url }%
\providecommand \@url [1]{\endgroup\@href {#1}{\urlprefix }}%
\providecommand \urlprefix  [0]{URL }%
\providecommand \Eprint [0]{\href }%
\providecommand \doibase [0]{https://doi.org/}%
\providecommand \selectlanguage [0]{\@gobble}%
\providecommand \bibinfo  [0]{\@secondoftwo}%
\providecommand \bibfield  [0]{\@secondoftwo}%
\providecommand \translation [1]{[#1]}%
\providecommand \BibitemOpen [0]{}%
\providecommand \bibitemStop [0]{}%
\providecommand \bibitemNoStop [0]{.\EOS\space}%
\providecommand \EOS [0]{\spacefactor3000\relax}%
\providecommand \BibitemShut  [1]{\csname bibitem#1\endcsname}%
\let\auto@bib@innerbib\@empty
\bibitem [{\citenamefont {Pinaud}\ \emph {et~al.}(2013)\citenamefont {Pinaud},
  \citenamefont {Benck}, \citenamefont {Seitz}, \citenamefont {Forman},
  \citenamefont {Chen}, \citenamefont {Deutsch}, \citenamefont {James},
  \citenamefont {Baum}, \citenamefont {Baum}, \citenamefont {Ardo},
  \citenamefont {Wang}, \citenamefont {Miller},\ and\ \citenamefont
  {Jaramillo}}]{Pinaud:2013}%
  \BibitemOpen
  \bibfield  {author} {\bibinfo {author} {\bibfnamefont {B.}~\bibnamefont
  {Pinaud}}, \bibinfo {author} {\bibfnamefont {J.}~\bibnamefont {Benck}},
  \bibinfo {author} {\bibfnamefont {L.}~\bibnamefont {Seitz}}, \bibinfo
  {author} {\bibfnamefont {A.}~\bibnamefont {Forman}}, \bibinfo {author}
  {\bibfnamefont {Z.}~\bibnamefont {Chen}}, \bibinfo {author} {\bibfnamefont
  {T.}~\bibnamefont {Deutsch}}, \bibinfo {author} {\bibfnamefont
  {B.}~\bibnamefont {James}}, \bibinfo {author} {\bibfnamefont
  {K.}~\bibnamefont {Baum}}, \bibinfo {author} {\bibfnamefont {G.}~\bibnamefont
  {Baum}}, \bibinfo {author} {\bibfnamefont {S.}~\bibnamefont {Ardo}}, \bibinfo
  {author} {\bibfnamefont {H.}~\bibnamefont {Wang}}, \bibinfo {author}
  {\bibfnamefont {E.}~\bibnamefont {Miller}},\ and\ \bibinfo {author}
  {\bibfnamefont {T.}~\bibnamefont {Jaramillo}},\ }\bibfield  {title} {\bibinfo
  {title} {{Technical and economic feasibility of centralized facilities for
  solar hydrogen production via photocatalysis and photoelectrochemistry}},\
  }\href@noop {} {\bibfield  {journal} {\bibinfo  {journal} {Energy Environ.
  Sci.}\ }\textbf {\bibinfo {volume} {6}},\ \bibinfo {pages} {1983} (\bibinfo
  {year} {2013})}\BibitemShut {NoStop}%
\bibitem [{\citenamefont {Pivovar}\ \emph {et~al.}(2018)\citenamefont
  {Pivovar}, \citenamefont {Rustagi},\ and\ \citenamefont
  {Satyapal}}]{Pivovar:2018}%
  \BibitemOpen
  \bibfield  {author} {\bibinfo {author} {\bibfnamefont {B.}~\bibnamefont
  {Pivovar}}, \bibinfo {author} {\bibfnamefont {N.}~\bibnamefont {Rustagi}},\
  and\ \bibinfo {author} {\bibfnamefont {S.}~\bibnamefont {Satyapal}},\
  }\bibfield  {title} {\bibinfo {title} {{Hydrogen at scale (H2@Scale): Key to
  a clean, economic, and sustainable energy system}},\ }\href@noop {}
  {\bibfield  {journal} {\bibinfo  {journal} {Electrochemical Society
  Interface}\ }\textbf {\bibinfo {volume} {27}},\ \bibinfo {pages} {47}
  (\bibinfo {year} {2018})}\BibitemShut {NoStop}%
\bibitem [{\citenamefont {Takata}\ and\ \citenamefont
  {Domen}(2019)}]{Takata:2019}%
  \BibitemOpen
  \bibfield  {author} {\bibinfo {author} {\bibfnamefont {T.}~\bibnamefont
  {Takata}}\ and\ \bibinfo {author} {\bibfnamefont {K.}~\bibnamefont {Domen}},\
  }\bibfield  {title} {\bibinfo {title} {{Particulate photocatalysts for water
  splitting: Recent advances and future prospects}},\ }\href@noop {} {\bibfield
   {journal} {\bibinfo  {journal} {ACS Energy Lett.}\ }\textbf {\bibinfo
  {volume} {4}},\ \bibinfo {pages} {542} (\bibinfo {year} {2019})}\BibitemShut
  {NoStop}%
\bibitem [{\citenamefont {Maeda}\ and\ \citenamefont
  {Domen}(2007)}]{Maeda:2007}%
  \BibitemOpen
  \bibfield  {author} {\bibinfo {author} {\bibfnamefont {K.}~\bibnamefont
  {Maeda}}\ and\ \bibinfo {author} {\bibfnamefont {K.}~\bibnamefont {Domen}},\
  }\bibfield  {title} {\bibinfo {title} {{New non-oxide photocatalysts designed
  for overall water splitting under visible light}},\ }\href@noop {} {\bibfield
   {journal} {\bibinfo  {journal} {J. Phys. Chem. C}\ }\textbf {\bibinfo
  {volume} {111}},\ \bibinfo {pages} {7851} (\bibinfo {year}
  {2007})}\BibitemShut {NoStop}%
\bibitem [{\citenamefont {Wu}\ \emph {et~al.}(2012)\citenamefont {Wu},
  \citenamefont {Lazic}, \citenamefont {Hautier}, \citenamefont {Persson},\
  and\ \citenamefont {Ceder}}]{Wu:2012}%
  \BibitemOpen
  \bibfield  {author} {\bibinfo {author} {\bibfnamefont {Y.}~\bibnamefont
  {Wu}}, \bibinfo {author} {\bibfnamefont {P.}~\bibnamefont {Lazic}}, \bibinfo
  {author} {\bibfnamefont {G.}~\bibnamefont {Hautier}}, \bibinfo {author}
  {\bibfnamefont {K.}~\bibnamefont {Persson}},\ and\ \bibinfo {author}
  {\bibfnamefont {G.}~\bibnamefont {Ceder}},\ }\bibfield  {title} {\bibinfo
  {title} {{First principles high throughput screening of oxynitrides for
  water-splitting photocatalysts}},\ }\href@noop {} {\bibfield  {journal}
  {\bibinfo  {journal} {Energy Environ. Sci.}\ }\textbf {\bibinfo {volume}
  {6}},\ \bibinfo {pages} {157} (\bibinfo {year} {2012})}\BibitemShut {NoStop}%
\bibitem [{\citenamefont {Liu}\ \emph {et~al.}(2013)\citenamefont {Liu},
  \citenamefont {Zheng}, \citenamefont {Zhang}, \citenamefont {Xiao},\ and\
  \citenamefont {Wang}}]{Liu:2013}%
  \BibitemOpen
  \bibfield  {author} {\bibinfo {author} {\bibfnamefont {X.}~\bibnamefont
  {Liu}}, \bibinfo {author} {\bibfnamefont {H.}~\bibnamefont {Zheng}}, \bibinfo
  {author} {\bibfnamefont {J.}~\bibnamefont {Zhang}}, \bibinfo {author}
  {\bibfnamefont {Y.}~\bibnamefont {Xiao}},\ and\ \bibinfo {author}
  {\bibfnamefont {Z.}~\bibnamefont {Wang}},\ }\bibfield  {title} {\bibinfo
  {title} {{Photoelectric properties and charge dynamics for a set of solid
  state solar cells with Cu$_4$Bi$_4$S$_9$ as the absorber layer}},\
  }\href@noop {} {\bibfield  {journal} {\bibinfo  {journal} {J. Mater. Chem.
  A}\ }\textbf {\bibinfo {volume} {1}},\ \bibinfo {pages} {10703} (\bibinfo
  {year} {2013})}\BibitemShut {NoStop}%
\bibitem [{\citenamefont {He}\ \emph {et~al.}(2013)\citenamefont {He},
  \citenamefont {Liang}, \citenamefont {Ou}, \citenamefont {Liu}, \citenamefont
  {Fang},\ and\ \citenamefont {Xu}}]{He:2013}%
  \BibitemOpen
  \bibfield  {author} {\bibinfo {author} {\bibfnamefont {G.}~\bibnamefont
  {He}}, \bibinfo {author} {\bibfnamefont {C.}~\bibnamefont {Liang}}, \bibinfo
  {author} {\bibfnamefont {Y.}~\bibnamefont {Ou}}, \bibinfo {author}
  {\bibfnamefont {D.}~\bibnamefont {Liu}}, \bibinfo {author} {\bibfnamefont
  {Y.}~\bibnamefont {Fang}},\ and\ \bibinfo {author} {\bibfnamefont
  {Y.}~\bibnamefont {Xu}},\ }\bibfield  {title} {\bibinfo {title} {{Preparation
  of novel Sb$_2$O$_3/$WO$_3$ photocatalysts and their activities under visible
  light irradiation}},\ }\href@noop {} {\bibfield  {journal} {\bibinfo
  {journal} {Materials Research Bulletin}\ }\textbf {\bibinfo {volume} {48}},\
  \bibinfo {pages} {2244} (\bibinfo {year} {2013})}\BibitemShut {NoStop}%
\bibitem [{\citenamefont {Jiang}\ \emph {et~al.}(2012)\citenamefont {Jiang},
  \citenamefont {Cheng},\ and\ \citenamefont {Lin}}]{Jiang:2012}%
  \BibitemOpen
  \bibfield  {author} {\bibinfo {author} {\bibfnamefont {H.}~\bibnamefont
  {Jiang}}, \bibinfo {author} {\bibfnamefont {K.}~\bibnamefont {Cheng}},\ and\
  \bibinfo {author} {\bibfnamefont {J.}~\bibnamefont {Lin}},\ }\bibfield
  {title} {\bibinfo {title} {{Crystalline metallic Au nanoparticle-loaded
  $\alpha$-Bi$_2$O$_3$ microrods for improved photocatalysis}},\ }\href@noop {}
  {\bibfield  {journal} {\bibinfo  {journal} {Phys. Chem. Chem. Phys.}\
  }\textbf {\bibinfo {volume} {14}},\ \bibinfo {pages} {12114} (\bibinfo {year}
  {2012})}\BibitemShut {NoStop}%
\bibitem [{\citenamefont {Xiong}\ \emph {et~al.}(2021)\citenamefont {Xiong},
  \citenamefont {Campbell}, \citenamefont {Fanghanel}, \citenamefont {Badding},
  \citenamefont {Wang}, \citenamefont {Kirchner-Hall}, \citenamefont
  {Theibault}, \citenamefont {Timrov}, \citenamefont {Mondschein},
  \citenamefont {Seth} \emph {et~al.}}]{Xiong:2021}%
  \BibitemOpen
  \bibfield  {author} {\bibinfo {author} {\bibfnamefont {Y.}~\bibnamefont
  {Xiong}}, \bibinfo {author} {\bibfnamefont {Q.~T.}\ \bibnamefont {Campbell}},
  \bibinfo {author} {\bibfnamefont {J.}~\bibnamefont {Fanghanel}}, \bibinfo
  {author} {\bibfnamefont {C.~K.}\ \bibnamefont {Badding}}, \bibinfo {author}
  {\bibfnamefont {H.}~\bibnamefont {Wang}}, \bibinfo {author} {\bibfnamefont
  {N.~E.}\ \bibnamefont {Kirchner-Hall}}, \bibinfo {author} {\bibfnamefont
  {M.~J.}\ \bibnamefont {Theibault}}, \bibinfo {author} {\bibfnamefont
  {I.}~\bibnamefont {Timrov}}, \bibinfo {author} {\bibfnamefont {J.~S.}\
  \bibnamefont {Mondschein}}, \bibinfo {author} {\bibfnamefont
  {K.}~\bibnamefont {Seth}}, \emph {et~al.},\ }\bibfield  {title} {\bibinfo
  {title} {Optimizing accuracy and efficacy in data-driven materials discovery
  for the solar production of hydrogen},\ }\href@noop {} {\bibfield  {journal}
  {\bibinfo  {journal} {Energy \& Environmental Science}\ }\textbf {\bibinfo
  {volume} {14}},\ \bibinfo {pages} {2335} (\bibinfo {year}
  {2021})}\BibitemShut {NoStop}%
\bibitem [{\citenamefont {Butler}\ and\ \citenamefont
  {Ginley}(1978)}]{Butler:1978}%
  \BibitemOpen
  \bibfield  {author} {\bibinfo {author} {\bibfnamefont {M.~A.}\ \bibnamefont
  {Butler}}\ and\ \bibinfo {author} {\bibfnamefont {D.~S.}\ \bibnamefont
  {Ginley}},\ }\bibfield  {title} {\bibinfo {title} {Prediction of flatband
  potentials at semiconductor-electrolyte interfaces from atomic
  electronegativities},\ }\href@noop {} {\bibfield  {journal} {\bibinfo
  {journal} {Journal of The Electrochemical Society}\ }\textbf {\bibinfo
  {volume} {125}},\ \bibinfo {pages} {228} (\bibinfo {year}
  {1978})}\BibitemShut {NoStop}%
\bibitem [{\citenamefont {Xu}\ and\ \citenamefont {Schoonen}(2000)}]{Xu:2000}%
  \BibitemOpen
  \bibfield  {author} {\bibinfo {author} {\bibfnamefont {Y.}~\bibnamefont
  {Xu}}\ and\ \bibinfo {author} {\bibfnamefont {M.~A.}\ \bibnamefont
  {Schoonen}},\ }\bibfield  {title} {\bibinfo {title} {The absolute energy
  positions of conduction and valence bands of selected semiconducting
  minerals},\ }\href@noop {} {\bibfield  {journal} {\bibinfo  {journal}
  {American Mineralogist}\ }\textbf {\bibinfo {volume} {85}},\ \bibinfo {pages}
  {543} (\bibinfo {year} {2000})}\BibitemShut {NoStop}%
\bibitem [{\citenamefont {Quarto}\ \emph {et~al.}(1997)\citenamefont {Quarto},
  \citenamefont {Sunseri}, \citenamefont {Piazza},\ and\ \citenamefont
  {Romano}}]{QuartoSunseriPiazzaRomano1997}%
  \BibitemOpen
  \bibfield  {author} {\bibinfo {author} {\bibfnamefont {F.~D.}\ \bibnamefont
  {Quarto}}, \bibinfo {author} {\bibfnamefont {C.}~\bibnamefont {Sunseri}},
  \bibinfo {author} {\bibfnamefont {S.}~\bibnamefont {Piazza}},\ and\ \bibinfo
  {author} {\bibfnamefont {M.~C.}\ \bibnamefont {Romano}},\ }\bibfield  {title}
  {\bibinfo {title} {Semiempirical correlation between optical band gap values
  of oxides and the difference of electronegativity of the elements. its
  importance for a quantitative use of photocurrent spectroscopy in corrosion
  studies},\ }\href@noop {} {\bibfield  {journal} {\bibinfo  {journal} {The
  Journal of Physical Chemistry B}\ }\textbf {\bibinfo {volume} {101}},\
  \bibinfo {pages} {2519} (\bibinfo {year} {1997})}\BibitemShut {NoStop}%
\bibitem [{\citenamefont {Zhang}\ \emph {et~al.}(2007)\citenamefont {Zhang},
  \citenamefont {Tang},\ and\ \citenamefont {Ye}}]{Zhang:2007}%
  \BibitemOpen
  \bibfield  {author} {\bibinfo {author} {\bibfnamefont {W.}~\bibnamefont
  {Zhang}}, \bibinfo {author} {\bibfnamefont {J.}~\bibnamefont {Tang}},\ and\
  \bibinfo {author} {\bibfnamefont {J.}~\bibnamefont {Ye}},\ } {\bibfield  {journal}
  {\bibinfo  {journal} {Journal of Materials Research}\ }\textbf {\bibinfo
  {volume} {22}},\ \bibinfo {pages} {1859} (\bibinfo {year}
  {2007})}\BibitemShut {NoStop}%
\bibitem [{\citenamefont {Lin}\ \emph {et~al.}(2006)\citenamefont {Lin},
  \citenamefont {Huang}, \citenamefont {Wang}, \citenamefont {Wang},
  \citenamefont {Xia},\ and\ \citenamefont {Shi}}]{Lin:2006}%
  \BibitemOpen
  \bibfield  {author} {\bibinfo {author} {\bibfnamefont {X.}~\bibnamefont
  {Lin}}, \bibinfo {author} {\bibfnamefont {F.}~\bibnamefont {Huang}}, \bibinfo
  {author} {\bibfnamefont {W.}~\bibnamefont {Wang}}, \bibinfo {author}
  {\bibfnamefont {Y.}~\bibnamefont {Wang}}, \bibinfo {author} {\bibfnamefont
  {Y.}~\bibnamefont {Xia}},\ and\ \bibinfo {author} {\bibfnamefont
  {J.}~\bibnamefont {Shi}},\ } {\bibfield  {journal} {\bibinfo  {journal} {Applied Catalysis A: General}\
  }\textbf {\bibinfo {volume} {313}},\ \bibinfo {pages} {218} (\bibinfo {year}
  {2006})}\BibitemShut {NoStop}%
\bibitem [{\citenamefont {Ueda}\ \emph {et~al.}(1992)\citenamefont {Ueda},
  \citenamefont {Omata}, \citenamefont {Hikuma}, \citenamefont {Ueda},
  \citenamefont {Mizoguchi}, \citenamefont {Hashimoto},\ and\ \citenamefont
  {Kawazoe}}]{Ueda:1992}%
  \BibitemOpen
  \bibfield  {author} {\bibinfo {author} {\bibfnamefont {N.}~\bibnamefont
  {Ueda}}, \bibinfo {author} {\bibfnamefont {T.}~\bibnamefont {Omata}},
  \bibinfo {author} {\bibfnamefont {N.}~\bibnamefont {Hikuma}}, \bibinfo
  {author} {\bibfnamefont {K.}~\bibnamefont {Ueda}}, \bibinfo {author}
  {\bibfnamefont {H.}~\bibnamefont {Mizoguchi}}, \bibinfo {author}
  {\bibfnamefont {T.}~\bibnamefont {Hashimoto}},\ and\ \bibinfo {author}
  {\bibfnamefont {H.}~\bibnamefont {Kawazoe}},\ } {\bibfield
  {journal} {\bibinfo  {journal} {Applied Physics Letters}\ }\textbf {\bibinfo
  {volume} {61}},\ \bibinfo {pages} {1954} (\bibinfo {year}
  {1992})}\BibitemShut {NoStop}%
\bibitem [{\citenamefont {Marzari}\ \emph {et~al.}(2012)\citenamefont
  {Marzari}, \citenamefont {Mostofi}, \citenamefont {Yates}, \citenamefont
  {Souza},\ and\ \citenamefont {Vanderbilt}}]{Marzari:2012}%
  \BibitemOpen
  \bibfield  {author} {\bibinfo {author} {\bibfnamefont {N.}~\bibnamefont
  {Marzari}}, \bibinfo {author} {\bibfnamefont {A.~A.}~\bibnamefont {Mostofi}},
  \bibinfo {author} {\bibfnamefont {J.~R.}~\bibnamefont {Yates}}, \bibinfo
  {author} {\bibfnamefont {I.}~\bibnamefont {Souza}},\ and\ \bibinfo {author}
  {\bibfnamefont {D.}~\bibnamefont {Vanderbilt}},\ }\href@noop {} {\bibfield
  {journal} {\bibinfo  {journal} {Rev. Mod. Phys.}\ }\textbf {\bibinfo {volume}
  {84}},\ \bibinfo {pages} {1419} (\bibinfo {year} {2012})}\BibitemShut
  {NoStop}%
\bibitem [{\citenamefont {Pizzi}\ \emph {et~al.}(2020)\citenamefont {Pizzi},
  \citenamefont {Vitale}, \citenamefont {Arita}, \citenamefont {Blügel},
  \citenamefont {Freimuth}, \citenamefont {G{\'{e}}ranton}, \citenamefont
  {Gibertini}, \citenamefont {Gresch}, \citenamefont {Johnson}, \citenamefont
  {Koretsune}, \citenamefont {Iba{\~{n}}ez-Azpiroz}, \citenamefont {Lee},
  \citenamefont {Lihm}, \citenamefont {Marchand}, \citenamefont {Marrazzo},
  \citenamefont {Mokrousov}, \citenamefont {Mustafa}, \citenamefont {Nohara},
  \citenamefont {Nomura}, \citenamefont {Paulatto}, \citenamefont
  {Ponc{\'{e}}}, \citenamefont {Ponweiser}, \citenamefont {Qiao}, \citenamefont
  {Thöle}, \citenamefont {Tsirkin}, \citenamefont {Wierzbowska}, \citenamefont
  {Marzari}, \citenamefont {Vanderbilt}, \citenamefont {Souza}, \citenamefont
  {Mostofi},\ and\ \citenamefont {Yates}}]{Pizzi:2020}%
  \BibitemOpen
  \bibfield  {author} {\bibinfo {author} {\bibfnamefont {G.}~\bibnamefont
  {Pizzi}}, \bibinfo {author} {\bibfnamefont {V.}~\bibnamefont {Vitale}},
  \bibinfo {author} {\bibfnamefont {R.}~\bibnamefont {Arita}}, \bibinfo
  {author} {\bibfnamefont {S.}~\bibnamefont {Blügel}}, \bibinfo {author}
  {\bibfnamefont {F.}~\bibnamefont {Freimuth}}, \bibinfo {author}
  {\bibfnamefont {G.}~\bibnamefont {G{\'{e}}ranton}}, \bibinfo {author}
  {\bibfnamefont {M.}~\bibnamefont {Gibertini}}, \bibinfo {author}
  {\bibfnamefont {D.}~\bibnamefont {Gresch}}, \bibinfo {author} {\bibfnamefont
  {C.}~\bibnamefont {Johnson}}, \bibinfo {author} {\bibfnamefont
  {T.}~\bibnamefont {Koretsune}}, \bibinfo {author} {\bibfnamefont
  {J.}~\bibnamefont {Iba{\~{n}}ez-Azpiroz}}, \bibinfo {author} {\bibfnamefont
  {H.}~\bibnamefont {Lee}}, \bibinfo {author} {\bibfnamefont {J.-M.}\
  \bibnamefont {Lihm}}, \bibinfo {author} {\bibfnamefont {D.}~\bibnamefont
  {Marchand}}, \bibinfo {author} {\bibfnamefont {A.}~\bibnamefont {Marrazzo}},
  \bibinfo {author} {\bibfnamefont {Y.}~\bibnamefont {Mokrousov}}, \bibinfo
  {author} {\bibfnamefont {J.~I.}\ \bibnamefont {Mustafa}}, \bibinfo {author}
  {\bibfnamefont {Y.}~\bibnamefont {Nohara}}, \bibinfo {author} {\bibfnamefont
  {Y.}~\bibnamefont {Nomura}}, \bibinfo {author} {\bibfnamefont
  {L.}~\bibnamefont {Paulatto}}, \bibinfo {author} {\bibfnamefont
  {S.}~\bibnamefont {Ponc{\'{e}}}}, \bibinfo {author} {\bibfnamefont
  {T.}~\bibnamefont {Ponweiser}}, \bibinfo {author} {\bibfnamefont
  {J.}~\bibnamefont {Qiao}}, \bibinfo {author} {\bibfnamefont {F.}~\bibnamefont
  {Thöle}}, \bibinfo {author} {\bibfnamefont {S.~S.}\ \bibnamefont {Tsirkin}},
  \bibinfo {author} {\bibfnamefont {M.}~\bibnamefont {Wierzbowska}}, \bibinfo
  {author} {\bibfnamefont {N.}~\bibnamefont {Marzari}}, \bibinfo {author}
  {\bibfnamefont {D.}~\bibnamefont {Vanderbilt}}, \bibinfo {author}
  {\bibfnamefont {I.}~\bibnamefont {Souza}}, \bibinfo {author} {\bibfnamefont
  {A.~A.}\ \bibnamefont {Mostofi}},\ and\ \bibinfo {author} {\bibfnamefont
  {J.~R.}\ \bibnamefont {Yates}},\ }\bibfield  {title} {\bibinfo {title}
  {Wannier90 as a community code: new features and applications},\ } {\bibfield  {journal} {\bibinfo
  {journal} {Journal of Physics: Condensed Matter}\ }\textbf {\bibinfo {volume}
  {32}},\ \bibinfo {pages} {165902} (\bibinfo {year} {2020})}\BibitemShut
  {NoStop}%
\bibitem [{\citenamefont {Godby}\ \emph {et~al.}(1988)\citenamefont {Godby},
  \citenamefont {Schl\"{u}ter},\ and\ \citenamefont {Sham}}]{Godby:1988}%
  \BibitemOpen
  \bibfield  {author} {\bibinfo {author} {\bibfnamefont {R.~W.}\ \bibnamefont
  {Godby}}, \bibinfo {author} {\bibfnamefont {M.}~\bibnamefont {Schl\"{u}ter}},\
  and\ \bibinfo {author} {\bibfnamefont {L.~J.}\ \bibnamefont {Sham}},\
  }\bibfield  {title} {\bibinfo {title} {Self-energy operators and
  exchange-correlation potentials in semiconductors},\ } {\bibfield  {journal} {\bibinfo
  {journal} {Physical Review B}\ }\textbf {\bibinfo {volume} {37}},\ \bibinfo
  {pages} {10159} (\bibinfo {year} {1988})}\BibitemShut {NoStop}%
\bibitem [{\citenamefont {Cohen}\ \emph {et~al.}(2008)\citenamefont {Cohen},
  \citenamefont {Mori-S\'{a}nchez},\ and\ \citenamefont {Yang}}]{Cohen:2008}%
  \BibitemOpen
  \bibfield  {author} {\bibinfo {author} {\bibfnamefont {A.}~\bibnamefont
  {Cohen}}, \bibinfo {author} {\bibfnamefont {P.}~\bibnamefont
  {Mori-S\'{a}nchez}},\ and\ \bibinfo {author} {\bibfnamefont {W.}~\bibnamefont
  {Yang}},\ }\bibfield  {title} {\bibinfo {title} {Insights into current
  limitations of density functional theory},\ }\href@noop {} {\bibfield
  {journal} {\bibinfo  {journal} {Science}\ }\textbf {\bibinfo {volume}
  {321}},\ \bibinfo {pages} {792} (\bibinfo {year} {2008})}\BibitemShut
  {NoStop}%
\bibitem [{\citenamefont {Perdew}\ and\ \citenamefont
  {Zunger}(1981)}]{Perdew:1981}%
  \BibitemOpen
  \bibfield  {author} {\bibinfo {author} {\bibfnamefont {J.~P.}~\bibnamefont
  {Perdew}}\ and\ \bibinfo {author} {\bibfnamefont {A.}~\bibnamefont
  {Zunger}},\ }\bibfield  {title} {\bibinfo {title} {Self-interaction
  correction to density-functional approximations for many-electron systems},\
  }\href@noop {} {\bibfield  {journal} {\bibinfo  {journal} {Phys. Rev. B}\
  }\textbf {\bibinfo {volume} {23}},\ \bibinfo {pages} {5048} (\bibinfo {year}
  {1981})}\BibitemShut {NoStop}%
\bibitem [{\citenamefont {Onida}\ \emph {et~al.}(2002)\citenamefont {Onida},
  \citenamefont {Reining},\ and\ \citenamefont {Rubio}}]{Onida:2002}%
  \BibitemOpen
  \bibfield  {author} {\bibinfo {author} {\bibfnamefont {G.}~\bibnamefont
  {Onida}}, \bibinfo {author} {\bibfnamefont {L.}~\bibnamefont {Reining}},\
  and\ \bibinfo {author} {\bibfnamefont {A.}~\bibnamefont {Rubio}},\ }\bibfield
   {title} {\bibinfo {title} {Electronic excitations: density-functional versus
  many-body green's-function approaches},\ } {\bibfield  {journal} {\bibinfo
  {journal} {Reviews of Modern Physics}\ }\textbf {\bibinfo {volume} {74}},\
  \bibinfo {pages} {601} (\bibinfo {year} {2002})}\BibitemShut {NoStop}%
\bibitem [{\citenamefont {Heyd}\ \emph {et~al.}(2003)\citenamefont {Heyd},
  \citenamefont {Scuseria},\ and\ \citenamefont {Ernzerhof}}]{Heyd:2003}%
  \BibitemOpen
  \bibfield  {author} {\bibinfo {author} {\bibfnamefont {J.}~\bibnamefont
  {Heyd}}, \bibinfo {author} {\bibfnamefont {G.~E.}\ \bibnamefont {Scuseria}},\
  and\ \bibinfo {author} {\bibfnamefont {M.}~\bibnamefont {Ernzerhof}},\
  }\bibfield  {title} {\bibinfo {title} {Hybrid functionals based on a screened
  coulomb potential},\ } {\bibfield
  {journal} {\bibinfo  {journal} {The Journal of Chemical Physics}\ }\textbf
  {\bibinfo {volume} {118}},\ \bibinfo {pages} {8207} (\bibinfo {year}
  {2003})}\BibitemShut {NoStop}%
\bibitem [{\citenamefont {Anisimov}\ \emph {et~al.}(1991)\citenamefont
  {Anisimov}, \citenamefont {Zaanen},\ and\ \citenamefont
  {Andersen}}]{Anisimov:1991}%
  \BibitemOpen
  \bibfield  {author} {\bibinfo {author} {\bibfnamefont {V.~I.}~\bibnamefont
  {Anisimov}}, \bibinfo {author} {\bibfnamefont {J.}~\bibnamefont {Zaanen}},\
  and\ \bibinfo {author} {\bibfnamefont {O.~K.}~\bibnamefont {Andersen}},\
  }\bibfield  {title} {\bibinfo {title} {{Band theory and Mott insulators:
  Hubbard $U$ instead of Stoner $I$}},\ }\href@noop {} {\bibfield  {journal}
  {\bibinfo  {journal} {Phys. Rev. B}\ }\textbf {\bibinfo {volume} {44}},\
  \bibinfo {pages} {943} (\bibinfo {year} {1991})}\BibitemShut {NoStop}%
\bibitem [{\citenamefont {Anisimov}\ \emph {et~al.}(1997)\citenamefont
  {Anisimov}, \citenamefont {Aryasetiawan},\ and\ \citenamefont
  {Liechtenstein}}]{Anisimov:1997}%
  \BibitemOpen
  \bibfield  {author} {\bibinfo {author} {\bibfnamefont {V.}~\bibnamefont
  {Anisimov}}, \bibinfo {author} {\bibfnamefont {F.}~\bibnamefont
  {Aryasetiawan}},\ and\ \bibinfo {author} {\bibfnamefont {A.}~\bibnamefont
  {Liechtenstein}},\ }\bibfield  {title} {\bibinfo {title} {{First-principles
  calculations of the electronic structure and spectra of strongly correlated
  systems: the LDA+$U$ method}},\ }\href@noop {} {\bibfield  {journal}
  {\bibinfo  {journal} {J. Phys.: Condens. Matter}\ }\textbf {\bibinfo {volume}
  {9}},\ \bibinfo {pages} {767} (\bibinfo {year} {1997})}\BibitemShut {NoStop}%
\bibitem [{\citenamefont {Dudarev}\ \emph {et~al.}(1998)\citenamefont
  {Dudarev}, \citenamefont {Botton}, \citenamefont {Savrasov}, \citenamefont
  {Humphreys},\ and\ \citenamefont {Sutton}}]{Dudarev:1998}%
  \BibitemOpen
  \bibfield  {author} {\bibinfo {author} {\bibfnamefont {S.~L.}~\bibnamefont
  {Dudarev}}, \bibinfo {author} {\bibfnamefont {G.~A.}~\bibnamefont {Botton}},
  \bibinfo {author} {\bibfnamefont {S.~Y.}~\bibnamefont {Savrasov}}, \bibinfo
  {author} {\bibfnamefont {C.~J.}~\bibnamefont {Humphreys}},\ and\ \bibinfo
  {author} {\bibfnamefont {A.~P.}~\bibnamefont {Sutton}},\ }\bibfield  {title}
  {\bibinfo {title} {{Electron-energy-loss spectra and the structural stability
  of nickel oxide: An LSDA+$U$ study}},\ }\href@noop {} {\bibfield  {journal}
  {\bibinfo  {journal} {Phys. Rev. B}\ }\textbf {\bibinfo {volume} {57}},\
  \bibinfo {pages} {1505} (\bibinfo {year} {1998})}\BibitemShut {NoStop}%
\bibitem [{\citenamefont {Cococcioni}\ and\ \citenamefont
  {de~Gironcoli}(2005)}]{Cococcioni:2005}%
  \BibitemOpen
  \bibfield  {author} {\bibinfo {author} {\bibfnamefont {M.}~\bibnamefont
  {Cococcioni}}\ and\ \bibinfo {author} {\bibfnamefont {S.}~\bibnamefont
  {de~Gironcoli}},\ }\bibfield  {title} {\bibinfo {title} {{Linear response
  approach to the calculation of the effective interaction parameters in the
  LDA+$U$ method}},\ }\href@noop {} {\bibfield  {journal} {\bibinfo  {journal}
  {Phys. Rev. B}\ }\textbf {\bibinfo {volume} {71}},\ \bibinfo {pages} {035105}
  (\bibinfo {year} {2005})}\BibitemShut {NoStop}%
\bibitem [{\citenamefont {{V.L.~Campo~Jr}}\ and\ \citenamefont
  {Cococcioni}(2010)}]{Campo:2010}%
  \BibitemOpen
  \bibfield  {author} {\bibinfo {author} {\bibnamefont {{V.L.~Campo~Jr}}}\ and\
  \bibinfo {author} {\bibfnamefont {M.}~\bibnamefont {Cococcioni}},\ }\bibfield
   {title} {\bibinfo {title} {{Extended DFT+\textit{U}+\textit{V} method with
  on-site and inter-site electronic interactions}},\ }\href@noop {} {\bibfield
  {journal} {\bibinfo  {journal} {J. Phys.: Condens Matter}\ }\textbf {\bibinfo
  {volume} {22}},\ \bibinfo {pages} {055602} (\bibinfo {year}
  {2010})}\BibitemShut {NoStop}%
\bibitem [{\citenamefont {Tolba}\ \emph {et~al.}(2018)\citenamefont {Tolba},
  \citenamefont {Gameel}, \citenamefont {Ali}, \citenamefont {Almossalami},\
  and\ \citenamefont {Allam}}]{Yang:2018}%
  \BibitemOpen
  \bibfield  {author} {\bibinfo {author} {\bibfnamefont {S.}~\bibnamefont
  {Tolba}}, \bibinfo {author} {\bibfnamefont {K.}~\bibnamefont {Gameel}},
  \bibinfo {author} {\bibfnamefont {B.}~\bibnamefont {Ali}}, \bibinfo {author}
  {\bibfnamefont {H.}~\bibnamefont {Almossalami}},\ and\ \bibinfo {author}
  {\bibfnamefont {N.}~\bibnamefont {Allam}},\ }\href@noop {} {\emph {\bibinfo
  {title} {Density functional calculations-recent progresses of theory and
  application}}},\ edited by\ \bibinfo {editor} {\bibfnamefont
  {G.}~\bibnamefont {Yang}}\ (\bibinfo  {publisher} {IntechOpen},\ \bibinfo
  {address} {London, UK},\ \bibinfo {year} {2018})\BibitemShut {NoStop}%
\bibitem [{\citenamefont {Paudel}\ and\ \citenamefont
  {Lambrecht}(2008)}]{Paudel:2008}%
  \BibitemOpen
  \bibfield  {author} {\bibinfo {author} {\bibfnamefont {T.~R.}~\bibnamefont
  {Paudel}}\ and\ \bibinfo {author} {\bibfnamefont {W.~R.~L.}~\bibnamefont
  {Lambrecht}},\ }\bibfield  {title} {\bibinfo {title} {{First-principles
  calculation of the O vacancy in ZnO: A self-consistent gap-corrected
  approach}},\ }\href@noop {} {\bibfield  {journal} {\bibinfo  {journal} {Phys.
  Rev. B}\ }\textbf {\bibinfo {volume} {77}},\ \bibinfo {pages} {205202}
  (\bibinfo {year} {2008})}\BibitemShut {NoStop}%
\bibitem [{\citenamefont {Lalitha}\ \emph {et~al.}(2007)\citenamefont
  {Lalitha}, \citenamefont {Karazhanov}, \citenamefont {Ravindran},
  \citenamefont {Senthilarasu}, \citenamefont {Sathyamoorthy},\ and\
  \citenamefont {Janabergenov}}]{Lalitha:2007}%
  \BibitemOpen
  \bibfield  {author} {\bibinfo {author} {\bibfnamefont {S.}~\bibnamefont
  {Lalitha}}, \bibinfo {author} {\bibfnamefont {S.}~\bibnamefont {Karazhanov}},
  \bibinfo {author} {\bibfnamefont {P.}~\bibnamefont {Ravindran}}, \bibinfo
  {author} {\bibfnamefont {S.}~\bibnamefont {Senthilarasu}}, \bibinfo {author}
  {\bibfnamefont {R.}~\bibnamefont {Sathyamoorthy}},\ and\ \bibinfo {author}
  {\bibfnamefont {J.}~\bibnamefont {Janabergenov}},\ }\bibfield  {title}
  {\bibinfo {title} {{Electronic structure, structural and optical properties
  of thermally evaporated CdTe thin films}},\ }\href@noop {} {\bibfield
  {journal} {\bibinfo  {journal} {Physica B}\ }\textbf {\bibinfo {volume}
  {387}},\ \bibinfo {pages} {227} (\bibinfo {year} {2007})}\BibitemShut
  {NoStop}%
\bibitem [{\citenamefont {Gautam}\ and\ \citenamefont
  {Carter}(2018)}]{Gautam:2018}%
  \BibitemOpen
  \bibfield  {author} {\bibinfo {author} {\bibfnamefont {G.}~\bibnamefont
  {Sai Gautam}}\ and\ \bibinfo {author} {\bibfnamefont {E.~A.}~\bibnamefont
  {Carter}},\ }\bibfield  {title} {\bibinfo {title} {{Evaluating transistion
  metal oxides within DFT-SCAN and SCAN+\textit{U} frameworks for solar
  thermochemical applications}},\ }\href@noop {} {\bibfield  {journal}
  {\bibinfo  {journal} {Phys. Rev. Materials}\ }\textbf {\bibinfo {volume}
  {2}},\ \bibinfo {pages} {095401} (\bibinfo {year} {2018})}\BibitemShut
  {NoStop}%
\bibitem [{\citenamefont {Long}\ \emph {et~al.}(2020)\citenamefont {Long},
  \citenamefont {Gautam},\ and\ \citenamefont {Carter}}]{Long:2020}%
  \BibitemOpen
  \bibfield  {author} {\bibinfo {author} {\bibfnamefont {O.~Y.}~\bibnamefont
  {Long}}, \bibinfo {author} {\bibfnamefont {G.}~\bibnamefont {Sai Gautam}},\ and\
  \bibinfo {author} {\bibfnamefont {E.~A.}~\bibnamefont {Carter}},\ }\bibfield
  {title} {\bibinfo {title} {{Evaluating optimal \textit{U} for 3$d$
  transition-metal oxides within the SCAN+\textit{U} framework}},\ }\href@noop
  {} {\bibfield  {journal} {\bibinfo  {journal} {Phys. Rev. Materials}\
  }\textbf {\bibinfo {volume} {4}},\ \bibinfo {pages} {045401} (\bibinfo {year}
  {2020})}\BibitemShut {NoStop}%
\bibitem [{\citenamefont {Dederichs}\ \emph {et~al.}(1984)\citenamefont
  {Dederichs}, \citenamefont {Bl\"ugel}, \citenamefont {Zeller},\ and\
  \citenamefont {Akai}}]{Dederichs:1984}%
  \BibitemOpen
  \bibfield  {author} {\bibinfo {author} {\bibfnamefont {P.~H.}~\bibnamefont
  {Dederichs}}, \bibinfo {author} {\bibfnamefont {S.}~\bibnamefont {Bl\"ugel}},
  \bibinfo {author} {\bibfnamefont {R.}~\bibnamefont {Zeller}},\ and\ \bibinfo
  {author} {\bibfnamefont {H.}~\bibnamefont {Akai}},\ }\bibfield  {title}
  {\bibinfo {title} {Ground states of condensed systems: application to cerium
  impurities},\ }\href@noop {} {\bibfield  {journal} {\bibinfo  {journal}
  {Phys. Rev. Lett.}\ }\textbf {\bibinfo {volume} {53}},\ \bibinfo {pages}
  {2512} (\bibinfo {year} {1984})}\BibitemShut {NoStop}%
\bibitem [{\citenamefont {McMahan}\ \emph {et~al.}(1988)\citenamefont
  {McMahan}, \citenamefont {Martin},\ and\ \citenamefont
  {Satpathy}}]{Mcmahan:1988}%
  \BibitemOpen
  \bibfield  {author} {\bibinfo {author} {\bibfnamefont {A.~K.}~\bibnamefont
  {McMahan}}, \bibinfo {author} {\bibfnamefont {R.~M.}~\bibnamefont {Martin}},\
  and\ \bibinfo {author} {\bibfnamefont {S.}~\bibnamefont {Satpathy}},\
  }\bibfield  {title} {\bibinfo {title} {{Calculated Hamiltonian for
  La$_2$CuO$_4$ and solution in the impurity Anderson approximation}},\
  }\href@noop {} {\bibfield  {journal} {\bibinfo  {journal} {Phys. Rev. B}\
  }\textbf {\bibinfo {volume} {38}},\ \bibinfo {pages} {6650} (\bibinfo {year}
  {1988})}\BibitemShut {NoStop}%
\bibitem [{\citenamefont {Gunnarsson}\ \emph {et~al.}(1989)\citenamefont
  {Gunnarsson}, \citenamefont {Andersen}, \citenamefont {Jepsen},\ and\
  \citenamefont {Zaanen}}]{Gunnarsson:1989}%
  \BibitemOpen
  \bibfield  {author} {\bibinfo {author} {\bibfnamefont {O.}~\bibnamefont
  {Gunnarsson}}, \bibinfo {author} {\bibfnamefont {O.~K.}~\bibnamefont
  {Andersen}}, \bibinfo {author} {\bibfnamefont {O.}~\bibnamefont {Jepsen}},\
  and\ \bibinfo {author} {\bibfnamefont {J.}~\bibnamefont {Zaanen}},\
  }\bibfield  {title} {\bibinfo {title} {{Density-functional calculation of the
  parameters in the Anderson model: Application to Mn in CdTe}},\ }\href@noop
  {} {\bibfield  {journal} {\bibinfo  {journal} {Phys. Rev. B}\ }\textbf
  {\bibinfo {volume} {39}},\ \bibinfo {pages} {1708} (\bibinfo {year}
  {1989})}\BibitemShut {NoStop}%
\bibitem [{\citenamefont {Hybertsen}\ \emph {et~al.}(1989)\citenamefont
  {Hybertsen}, \citenamefont {Schl\"uter},\ and\ \citenamefont
  {Christensen}}]{Hybertsen:1989}%
  \BibitemOpen
  \bibfield  {author} {\bibinfo {author} {\bibfnamefont {M.~S.}~\bibnamefont
  {Hybertsen}}, \bibinfo {author} {\bibfnamefont {M.}~\bibnamefont
  {Schl\"uter}},\ and\ \bibinfo {author} {\bibfnamefont {N.~E.}~\bibnamefont
  {Christensen}},\ }\bibfield  {title} {\bibinfo {title} {{Calculation of
  Coulomb-interaction parameters for La$_2$CuO$_4$ using a
  constrained-density-functional approach}},\ }\href@noop {} {\bibfield
  {journal} {\bibinfo  {journal} {Phys. Rev. B}\ }\textbf {\bibinfo {volume}
  {39}},\ \bibinfo {pages} {9028} (\bibinfo {year} {1989})}\BibitemShut
  {NoStop}%
\bibitem [{\citenamefont {Gunnarsson}(1990)}]{Gunnarsson:1990}%
  \BibitemOpen
  \bibfield  {author} {\bibinfo {author} {\bibfnamefont {O.}~\bibnamefont
  {Gunnarsson}},\ }\bibfield  {title} {\bibinfo {title} {Calculations of
  parameters in model hamiltonians},\ }\href@noop {} {\bibfield  {journal}
  {\bibinfo  {journal} {Phys. Rev. B}\ }\textbf {\bibinfo {volume} {41}},\
  \bibinfo {pages} {514} (\bibinfo {year} {1990})}\BibitemShut {NoStop}%
\bibitem [{\citenamefont {Pickett}\ \emph {et~al.}(1998)\citenamefont
  {Pickett}, \citenamefont {Erwin},\ and\ \citenamefont
  {Ethridge}}]{Pickett:1998}%
  \BibitemOpen
  \bibfield  {author} {\bibinfo {author} {\bibfnamefont {W.~E.}~\bibnamefont
  {Pickett}}, \bibinfo {author} {\bibfnamefont {S.~C.}~\bibnamefont {Erwin}},\
  and\ \bibinfo {author} {\bibfnamefont {E.~C.}~\bibnamefont {Ethridge}},\
  }\bibfield  {title} {\bibinfo {title} {{Reformation of the LDA+$U$ method for
  a local-orbital basis}},\ }\href@noop {} {\bibfield  {journal} {\bibinfo
  {journal} {Phys. Rev. B}\ }\textbf {\bibinfo {volume} {58}},\ \bibinfo
  {pages} {1201} (\bibinfo {year} {1998})}\BibitemShut {NoStop}%
\bibitem [{\citenamefont {Solovyev}\ and\ \citenamefont
  {Imada}(2005)}]{Solovyev:2005}%
  \BibitemOpen
  \bibfield  {author} {\bibinfo {author} {\bibfnamefont {I.~V.}~\bibnamefont
  {Solovyev}}\ and\ \bibinfo {author} {\bibfnamefont {M.}~\bibnamefont
  {Imada}},\ }\bibfield  {title} {\bibinfo {title} {Screening of coulomb
  interactions in transition metals},\ }\href@noop {} {\bibfield  {journal}
  {\bibinfo  {journal} {Phys. Rev. B}\ }\textbf {\bibinfo {volume} {71}},\
  \bibinfo {pages} {045103} (\bibinfo {year} {2005})}\BibitemShut {NoStop}%
\bibitem [{\citenamefont {Nakamura}\ \emph {et~al.}(2006)\citenamefont
  {Nakamura}, \citenamefont {Arita}, \citenamefont {Yoshimoto},\ and\
  \citenamefont {Tsuneyuki}}]{Nakamura:2006}%
  \BibitemOpen
  \bibfield  {author} {\bibinfo {author} {\bibfnamefont {K.}~\bibnamefont
  {Nakamura}}, \bibinfo {author} {\bibfnamefont {R.}~\bibnamefont {Arita}},
  \bibinfo {author} {\bibfnamefont {Y.}~\bibnamefont {Yoshimoto}},\ and\
  \bibinfo {author} {\bibfnamefont {S.}~\bibnamefont {Tsuneyuki}},\ }\bibfield
  {title} {\bibinfo {title} {{First-principles calculation of effective onsite
  Coulomb interactions of $3d$ transition metals: Constrained local density
  functional approach with maximally localized Wannier functions}},\
  }\href@noop {} {\bibfield  {journal} {\bibinfo  {journal} {Phys. Rev. B}\
  }\textbf {\bibinfo {volume} {74}},\ \bibinfo {pages} {235113} (\bibinfo
  {year} {2006})}\BibitemShut {NoStop}%
\bibitem [{\citenamefont {Shishkin}\ and\ \citenamefont
  {Sato}(2016)}]{Shishkin:2016}%
  \BibitemOpen
  \bibfield  {author} {\bibinfo {author} {\bibfnamefont {M.}~\bibnamefont
  {Shishkin}}\ and\ \bibinfo {author} {\bibfnamefont {H.}~\bibnamefont
  {Sato}},\ }\bibfield  {title} {\bibinfo {title} {{Self-consistent
  parametrization of DFT+$U$ framework using linear response approach:
  Application to evaluation of redox potentials of battery cathodes}},\
  }\href@noop {} {\bibfield  {journal} {\bibinfo  {journal} {Phys. Rev. B}\
  }\textbf {\bibinfo {volume} {93}},\ \bibinfo {pages} {085135} (\bibinfo
  {year} {2016})}\BibitemShut {NoStop}%
\bibitem [{\citenamefont {Springer}\ and\ \citenamefont
  {Aryasetiawan}(1998)}]{Springer:1998}%
  \BibitemOpen
  \bibfield  {author} {\bibinfo {author} {\bibfnamefont {M.}~\bibnamefont
  {Springer}}\ and\ \bibinfo {author} {\bibfnamefont {F.}~\bibnamefont
  {Aryasetiawan}},\ }\bibfield  {title} {\bibinfo {title} {{Frequency-dependent
  screened interaction in Ni within the random-phase approximation}},\
  }\href@noop {} {\bibfield  {journal} {\bibinfo  {journal} {Phys. Rev. B}\
  }\textbf {\bibinfo {volume} {57}},\ \bibinfo {pages} {4364} (\bibinfo {year}
  {1998})}\BibitemShut {NoStop}%
\bibitem [{\citenamefont {Kotani}(2000)}]{Kotani:2000}%
  \BibitemOpen
  \bibfield  {author} {\bibinfo {author} {\bibfnamefont {T.}~\bibnamefont
  {Kotani}},\ }\bibfield  {title} {\bibinfo {title} {$ab$ $initio$
  random-phase-approximation calculation of the frequency-dependent effective
  interaction between 3$d$ electrons: {Ni}, {Fe}, and {MnO}},\ }\href@noop {}
  {\bibfield  {journal} {\bibinfo  {journal} {J. Phys.: Condens. Matter}\
  }\textbf {\bibinfo {volume} {12}},\ \bibinfo {pages} {2413} (\bibinfo {year}
  {2000})}\BibitemShut {NoStop}%
\bibitem [{\citenamefont {Aryasetiawan}\ \emph {et~al.}(2004)\citenamefont
  {Aryasetiawan}, \citenamefont {Imada}, \citenamefont {Georges}, \citenamefont
  {Kotliar}, \citenamefont {Biermann},\ and\ \citenamefont
  {Lichtenstein}}]{Aryasetiawan:2004}%
  \BibitemOpen
  \bibfield  {author} {\bibinfo {author} {\bibfnamefont {F.}~\bibnamefont
  {Aryasetiawan}}, \bibinfo {author} {\bibfnamefont {M.}~\bibnamefont {Imada}},
  \bibinfo {author} {\bibfnamefont {A.}~\bibnamefont {Georges}}, \bibinfo
  {author} {\bibfnamefont {G.}~\bibnamefont {Kotliar}}, \bibinfo {author}
  {\bibfnamefont {S.}~\bibnamefont {Biermann}},\ and\ \bibinfo {author}
  {\bibfnamefont {A.~I.}~\bibnamefont {Lichtenstein}},\ }\bibfield  {title}
  {\bibinfo {title} {Frequency-dependent local interactions and low-energy
  effective models from electronic structure calculations},\ }\href@noop {}
  {\bibfield  {journal} {\bibinfo  {journal} {Phys. Rev. B}\ }\textbf {\bibinfo
  {volume} {70}},\ \bibinfo {pages} {195104} (\bibinfo {year}
  {2004})}\BibitemShut {NoStop}%
\bibitem [{\citenamefont {Aryasetiawan}\ \emph {et~al.}(2006)\citenamefont
  {Aryasetiawan}, \citenamefont {Karlsson}, \citenamefont {Jepsen},\ and\
  \citenamefont {Sch\"onberger}}]{Aryasetiawan:2006}%
  \BibitemOpen
  \bibfield  {author} {\bibinfo {author} {\bibfnamefont {F.}~\bibnamefont
  {Aryasetiawan}}, \bibinfo {author} {\bibfnamefont {K.}~\bibnamefont
  {Karlsson}}, \bibinfo {author} {\bibfnamefont {O.}~\bibnamefont {Jepsen}},\
  and\ \bibinfo {author} {\bibfnamefont {U.}~\bibnamefont {Sch\"onberger}},\
  }\bibfield  {title} {\bibinfo {title} {{Calculations of Hubbard $U$ from
  first-principles}},\ }\href@noop {} {\bibfield  {journal} {\bibinfo
  {journal} {Phys. Rev. B}\ }\textbf {\bibinfo {volume} {74}},\ \bibinfo
  {pages} {125106} (\bibinfo {year} {2006})}\BibitemShut {NoStop}%
\bibitem [{\citenamefont {Sasioglu}\ \emph {et~al.}(2011)\citenamefont
  {Sasioglu}, \citenamefont {Friedrich},\ and\ \citenamefont
  {Bl\"ugel}}]{Sasioglu:2011}%
  \BibitemOpen
  \bibfield  {author} {\bibinfo {author} {\bibfnamefont {E.}~\bibnamefont
  {Sasioglu}}, \bibinfo {author} {\bibfnamefont {C.}~\bibnamefont
  {Friedrich}},\ and\ \bibinfo {author} {\bibfnamefont {S.}~\bibnamefont
  {Bl\"ugel}},\ }\bibfield  {title} {\bibinfo {title} {{Effective Coulomb
  interaction in transition metals from constrained random-phase
  approximation}},\ }\href@noop {} {\bibfield  {journal} {\bibinfo  {journal}
  {Phys. Rev. B}\ }\textbf {\bibinfo {volume} {83}},\ \bibinfo {pages}
  {121101(R)} (\bibinfo {year} {2011})}\BibitemShut {NoStop}%
\bibitem [{\citenamefont {Vaugier}\ \emph {et~al.}(2012)\citenamefont
  {Vaugier}, \citenamefont {Jiang},\ and\ \citenamefont
  {Biermann}}]{Vaugier:2012}%
  \BibitemOpen
  \bibfield  {author} {\bibinfo {author} {\bibfnamefont {L.}~\bibnamefont
  {Vaugier}}, \bibinfo {author} {\bibfnamefont {H.}~\bibnamefont {Jiang}},\
  and\ \bibinfo {author} {\bibfnamefont {S.}~\bibnamefont {Biermann}},\
  }\bibfield  {title} {\bibinfo {title} {{Hubbard $U$ and Hund exchange $J$ in
  transition metal oxides: Screening versus localization trends from
  constrained random phase approximation}},\ }\href@noop {} {\bibfield
  {journal} {\bibinfo  {journal} {Phys. Rev. B}\ }\textbf {\bibinfo {volume}
  {86}},\ \bibinfo {pages} {165105} (\bibinfo {year} {2012})}\BibitemShut
  {NoStop}%
\bibitem [{\citenamefont {Amadon}\ \emph {et~al.}(2014)\citenamefont {Amadon},
  \citenamefont {Applencourt},\ and\ \citenamefont {Bruneval}}]{Amadon:2014}%
  \BibitemOpen
  \bibfield  {author} {\bibinfo {author} {\bibfnamefont {B.}~\bibnamefont
  {Amadon}}, \bibinfo {author} {\bibfnamefont {T.}~\bibnamefont
  {Applencourt}},\ and\ \bibinfo {author} {\bibfnamefont {F.}~\bibnamefont
  {Bruneval}},\ }\bibfield  {title} {\bibinfo {title} {{Screened Coulomb
  interaction calculations: cRPA implementation and applications to dynamical
  screening and self-consistency in uranium dioxide and cerium}},\ }\href@noop
  {} {\bibfield  {journal} {\bibinfo  {journal} {Phys. Rev. B}\ }\textbf
  {\bibinfo {volume} {89}},\ \bibinfo {pages} {125110} (\bibinfo {year}
  {2014})}\BibitemShut {NoStop}%
\bibitem [{\citenamefont {Mosey}\ and\ \citenamefont
  {Carter}(2007)}]{Mosey:2007}%
  \BibitemOpen
  \bibfield  {author} {\bibinfo {author} {\bibfnamefont {N.~J.}~\bibnamefont
  {Mosey}}\ and\ \bibinfo {author} {\bibfnamefont {E.~A.}~\bibnamefont {Carter}},\
  }\bibfield  {title} {\bibinfo {title} {{$Ab$ $initio$ evaluation of Coulomb
  and exchange parameters for DFT+$U$ calculations}},\ }\href@noop {}
  {\bibfield  {journal} {\bibinfo  {journal} {Phys. Rev. B}\ }\textbf {\bibinfo
  {volume} {76}},\ \bibinfo {pages} {155123} (\bibinfo {year}
  {2007})}\BibitemShut {NoStop}%
\bibitem [{\citenamefont {Mosey}\ \emph {et~al.}(2008)\citenamefont {Mosey},
  \citenamefont {Liao},\ and\ \citenamefont {Carter}}]{Mosey:2008}%
  \BibitemOpen
  \bibfield  {author} {\bibinfo {author} {\bibfnamefont {N.~J.}~\bibnamefont
  {Mosey}}, \bibinfo {author} {\bibfnamefont {P.}~\bibnamefont {Liao}},\ and\
  \bibinfo {author} {\bibfnamefont {E.~A.}~\bibnamefont {Carter}},\ }\bibfield
  {title} {\bibinfo {title} {{Rotationally invariant $ab$ $initio$ evaluation
  of Coulomb and exchange parameters for DFT+$U$ calculations}},\ }\href@noop
  {} {\bibfield  {journal} {\bibinfo  {journal} {J. Chem. Phys.}\ }\textbf
  {\bibinfo {volume} {129}},\ \bibinfo {pages} {014103} (\bibinfo {year}
  {2008})}\BibitemShut {NoStop}%
\bibitem [{\citenamefont {Andriotis}\ \emph {et~al.}(2010)\citenamefont
  {Andriotis}, \citenamefont {Sheetz},\ and\ \citenamefont
  {Menon}}]{Andriotis:2010}%
  \BibitemOpen
  \bibfield  {author} {\bibinfo {author} {\bibfnamefont {A.~N.}~\bibnamefont
  {Andriotis}}, \bibinfo {author} {\bibfnamefont {R.~M.}~\bibnamefont {Sheetz}},\
  and\ \bibinfo {author} {\bibfnamefont {M.}~\bibnamefont {Menon}},\ }\bibfield
   {title} {\bibinfo {title} {{LSDA+$U$ method: A calculation of the $U$ values
  at the Hartree-Fock level of approximation}},\ }\href@noop {} {\bibfield
  {journal} {\bibinfo  {journal} {Phys. Rev. B}\ }\textbf {\bibinfo {volume}
  {81}},\ \bibinfo {pages} {245103} (\bibinfo {year} {2010})}\BibitemShut
  {NoStop}%
\bibitem [{\citenamefont {Agapito}\ \emph {et~al.}(2015)\citenamefont
  {Agapito}, \citenamefont {Curtarolo},\ and\ \citenamefont
  {Buongiorno~Nardelli}}]{Agapito:2015}%
  \BibitemOpen
  \bibfield  {author} {\bibinfo {author} {\bibfnamefont {L.~A.}~\bibnamefont
  {Agapito}}, \bibinfo {author} {\bibfnamefont {S.}~\bibnamefont {Curtarolo}},\
  and\ \bibinfo {author} {\bibfnamefont {M.}~\bibnamefont
  {Buongiorno~Nardelli}},\ }\bibfield  {title} {\bibinfo {title} {{Reformation
  of DFT+\textit{U} as a pseudohybrid density functional for accelerated
  materials discovery}},\ }\href@noop {} {\bibfield  {journal} {\bibinfo
  {journal} {Phys. Rev. X}\ }\textbf {\bibinfo {volume} {5}},\ \bibinfo {pages}
  {011006} (\bibinfo {year} {2015})}\BibitemShut {NoStop}%
\bibitem [{\citenamefont {Timrov}\ \emph {et~al.}(2018)\citenamefont {Timrov},
  \citenamefont {Marzari},\ and\ \citenamefont {Cococcioni}}]{Timrov:2018}%
  \BibitemOpen
  \bibfield  {author} {\bibinfo {author} {\bibfnamefont {I.}~\bibnamefont
  {Timrov}}, \bibinfo {author} {\bibfnamefont {N.}~\bibnamefont {Marzari}},\
  and\ \bibinfo {author} {\bibfnamefont {M.}~\bibnamefont {Cococcioni}},\
  }\bibfield  {title} {\bibinfo {title} {Hubbard parameters from
  density-functional perturbation theory},\ }\href@noop {} {\bibfield
  {journal} {\bibinfo  {journal} {Phys. Rev. B}\ }\textbf {\bibinfo {volume}
  {98}},\ \bibinfo {pages} {085127} (\bibinfo {year} {2018})}\BibitemShut
  {NoStop}%
\bibitem [{\citenamefont {Timrov}\ \emph {et~al.}(2021)\citenamefont {Timrov},
  \citenamefont {Marzari},\ and\ \citenamefont {Cococcioni}}]{Timrov:2021}%
  \BibitemOpen
  \bibfield  {author} {\bibinfo {author} {\bibfnamefont {I.}~\bibnamefont
  {Timrov}}, \bibinfo {author} {\bibfnamefont {N.}~\bibnamefont {Marzari}},\
  and\ \bibinfo {author} {\bibfnamefont {M.}~\bibnamefont {Cococcioni}},\
  }\bibfield  {title} {\bibinfo {title} {{Self-consistent Hubbard parameters
  from density-functional perturbation theory in the ultrasoft and
  projector-augmented wave formulations}},\ }\href@noop {} {\bibfield
  {journal} {\bibinfo  {journal} {Phys. Rev. B}\ }\textbf {\bibinfo {volume}
  {103}},\ \bibinfo {pages} {045141} (\bibinfo {year} {2021})}\BibitemShut
  {NoStop}%
\bibitem [{\citenamefont {Jain}\ \emph {et~al.}(2013)\citenamefont {Jain},
  \citenamefont {Ong}, \citenamefont {Hautier}, \citenamefont {Chen},
  \citenamefont {Richards}, \citenamefont {Dacek}, \citenamefont {Cholia},
  \citenamefont {Gunter}, \citenamefont {Skinner}, \citenamefont {Ceder},\ and\
  \citenamefont {Persson}}]{Jain:2013}%
  \BibitemOpen
  \bibfield  {author} {\bibinfo {author} {\bibfnamefont {A.}~\bibnamefont
  {Jain}}, \bibinfo {author} {\bibfnamefont {S.}~\bibnamefont {Ong}}, \bibinfo
  {author} {\bibfnamefont {G.}~\bibnamefont {Hautier}}, \bibinfo {author}
  {\bibfnamefont {W.}~\bibnamefont {Chen}}, \bibinfo {author} {\bibfnamefont
  {W.}~\bibnamefont {Richards}}, \bibinfo {author} {\bibfnamefont
  {S.}~\bibnamefont {Dacek}}, \bibinfo {author} {\bibfnamefont
  {S.}~\bibnamefont {Cholia}}, \bibinfo {author} {\bibfnamefont
  {D.}~\bibnamefont {Gunter}}, \bibinfo {author} {\bibfnamefont
  {D.}~\bibnamefont {Skinner}}, \bibinfo {author} {\bibfnamefont
  {G.}~\bibnamefont {Ceder}},\ and\ \bibinfo {author} {\bibfnamefont
  {K.}~\bibnamefont {Persson}},\ }\bibfield  {title} {\bibinfo {title}
  {{Commentary: The Materials Project: A materials genome approach to
  accelerating materials innovation}},\ }\href@noop {} {\bibfield  {journal}
  {\bibinfo  {journal} {APL Materials}\ }\textbf {\bibinfo {volume} {1}},\
  \bibinfo {pages} {011002} (\bibinfo {year} {2013})}\BibitemShut {NoStop}%
\bibitem [{CDC()}]{CDC:web}%
  \BibitemOpen
  \href@noop {} {}\bibinfo {note} {{The National Institute for Occupational
  Safety and Health, Immediately Dangerous To Life or Health Values, Centers
  for Disease Control and Prevention, 2014.}}\BibitemShut {Stop}%
\bibitem [{\citenamefont {Arifin}\ \emph {et~al.}(2015)\citenamefont {Arifin},
  \citenamefont {Nuraini}, \citenamefont {Utomo},\ and\ \citenamefont
  {Wardiyati}}]{Arifin:2015}%
  \BibitemOpen
  \bibfield  {author} {\bibinfo {author} {\bibfnamefont {M.}~\bibnamefont
  {Arifin}}, \bibinfo {author} {\bibfnamefont {Y.}~\bibnamefont {Nuraini}},
  \bibinfo {author} {\bibfnamefont {W.}~\bibnamefont {Utomo}},\ and\ \bibinfo
  {author} {\bibfnamefont {T.}~\bibnamefont {Wardiyati}},\ }\bibfield  {title}
  {\bibinfo {title} {{The potential of \textit{Lumbricus rubellus} as a
  bioaccumulator of excess Pb and Cd in organic media}},\ }\href@noop {}
  {\bibfield  {journal} {\bibinfo  {journal} {J. Degraded and Mining Lands
  Management}\ }\textbf {\bibinfo {volume} {2}},\ \bibinfo {pages} {397}
  (\bibinfo {year} {2015})}\BibitemShut {NoStop}%
\bibitem [{\citenamefont {Asakura}\ \emph {et~al.}(2008)\citenamefont
  {Asakura}, \citenamefont {Satoh}, \citenamefont {Chiba}, \citenamefont
  {Okamoto}, \citenamefont {Serizawa}, \citenamefont {Nakano},\ and\
  \citenamefont {Omae}}]{Asakura:2008}%
  \BibitemOpen
  \bibfield  {author} {\bibinfo {author} {\bibfnamefont {K.}~\bibnamefont
  {Asakura}}, \bibinfo {author} {\bibfnamefont {H.}~\bibnamefont {Satoh}},
  \bibinfo {author} {\bibfnamefont {M.}~\bibnamefont {Chiba}}, \bibinfo
  {author} {\bibfnamefont {M.}~\bibnamefont {Okamoto}}, \bibinfo {author}
  {\bibfnamefont {K.}~\bibnamefont {Serizawa}}, \bibinfo {author}
  {\bibfnamefont {M.}~\bibnamefont {Nakano}},\ and\ \bibinfo {author}
  {\bibfnamefont {K.}~\bibnamefont {Omae}},\ }\bibfield  {title} {\bibinfo
  {title} {{Oral toxicity of indium in rats: single and 28-day repeated
  administration studies}},\ }\href@noop {} {\bibfield  {journal} {\bibinfo
  {journal} {J. Occup. Health}\ }\textbf {\bibinfo {volume} {50}},\ \bibinfo
  {pages} {471} (\bibinfo {year} {2008})}\BibitemShut {NoStop}%
\bibitem [{\citenamefont {Sano}\ \emph {et~al.}(2005)\citenamefont {Sano},
  \citenamefont {Satoh}, \citenamefont {Chiba}, \citenamefont {Okamoto},
  \citenamefont {Serizawa}, \citenamefont {Nakashima},\ and\ \citenamefont
  {Omae}}]{Sano:2005}%
  \BibitemOpen
  \bibfield  {author} {\bibinfo {author} {\bibfnamefont {Y.}~\bibnamefont
  {Sano}}, \bibinfo {author} {\bibfnamefont {H.}~\bibnamefont {Satoh}},
  \bibinfo {author} {\bibfnamefont {M.}~\bibnamefont {Chiba}}, \bibinfo
  {author} {\bibfnamefont {M.}~\bibnamefont {Okamoto}}, \bibinfo {author}
  {\bibfnamefont {K.}~\bibnamefont {Serizawa}}, \bibinfo {author}
  {\bibfnamefont {H.}~\bibnamefont {Nakashima}},\ and\ \bibinfo {author}
  {\bibfnamefont {K.}~\bibnamefont {Omae}},\ }\bibfield  {title} {\bibinfo
  {title} {{Oral toxicity of bismuth in rats: single and 28-day repeated
  administration studies}},\ }\href@noop {} {\bibfield  {journal} {\bibinfo
  {journal} {J. Occup. Health}\ }\textbf {\bibinfo {volume} {47}},\ \bibinfo
  {pages} {293} (\bibinfo {year} {2005})}\BibitemShut {NoStop}%
\bibitem [{rad()}]{radioactivity}%
  \BibitemOpen
  \href@noop {} {}\bibinfo {note} {{M. R. Bhat, \textit{Evaluated Nuclear
  Structure Data File, Nuclear Data for Science and Technology}, International
  Atomic Energy Agency, 2017.}}\BibitemShut {Stop}%
\bibitem [{\citenamefont {Hubbard}(1964)}]{Hubbard:1963}%
  \BibitemOpen
  \bibfield  {author} {\bibinfo {author} {\bibfnamefont {J.}~\bibnamefont
  {Hubbard}},\ }\bibfield  {title} {\bibinfo {title} {{Electron correlations in
  narrow energy bands II. The degenerate band case}},\ }\href@noop {}
  {\bibfield  {journal} {\bibinfo  {journal} {Proc. Roy. Soc. A}\ }\textbf
  {\bibinfo {volume} {277}},\ \bibinfo {pages} {237} (\bibinfo {year}
  {1964})}\BibitemShut {NoStop}%
\bibitem [{\citenamefont {Dabo}\ \emph {et~al.}(2010)\citenamefont {Dabo},
  \citenamefont {Ferretti}, \citenamefont {Poilvert}, \citenamefont {Li},
  \citenamefont {Marzari},\ and\ \citenamefont {Cococcioni}}]{Dabo:2010}%
  \BibitemOpen
  \bibfield  {author} {\bibinfo {author} {\bibfnamefont {I.}~\bibnamefont
  {Dabo}}, \bibinfo {author} {\bibfnamefont {A.}~\bibnamefont {Ferretti}},
  \bibinfo {author} {\bibfnamefont {N.}~\bibnamefont {Poilvert}}, \bibinfo
  {author} {\bibfnamefont {Y.}~\bibnamefont {Li}}, \bibinfo {author}
  {\bibfnamefont {N.}~\bibnamefont {Marzari}},\ and\ \bibinfo {author}
  {\bibfnamefont {M.}~\bibnamefont {Cococcioni}},\ }\bibfield  {title}
  {\bibinfo {title} {{Koopmans’ condition for density-functional theory}},\
  }\href@noop {} {\bibfield  {journal} {\bibinfo  {journal} {Phys. Rev. B}\
  }\textbf {\bibinfo {volume} {82}},\ \bibinfo {pages} {115121} (\bibinfo
  {year} {2010})}\BibitemShut {NoStop}%
\bibitem [{\citenamefont {Dabo}\ \emph {et~al.}(2014)\citenamefont {Dabo},
  \citenamefont {Ferretti},\ and\ \citenamefont {Marzari}}]{Dabo:2014}%
  \BibitemOpen
  \bibfield  {author} {\bibinfo {author} {\bibfnamefont {I.}~\bibnamefont
  {Dabo}}, \bibinfo {author} {\bibfnamefont {A.}~\bibnamefont {Ferretti}},\
  and\ \bibinfo {author} {\bibfnamefont {N.}~\bibnamefont {Marzari}},\
  }\bibfield  {title} {\bibinfo {title} {{Piecewise linearity and spectroscopic
  properties from Koopmans-compliant functionals}},\ }in\ \href@noop {} {\emph
  {\bibinfo {booktitle} {Topics in Current Chemistry: First-Principles
  Approaches to Spectroscopic Properties of Complex Materials 347}}},\ \bibinfo
  {editor} {edited by\ \bibinfo {editor} {\bibfnamefont {C.~D.}\ \bibnamefont
  {Valentin}}, \bibinfo {editor} {\bibfnamefont {S.}~\bibnamefont {Botti}},\
  and\ \bibinfo {editor} {\bibfnamefont {M.}~\bibnamefont {Cococcioni}}}\
  (\bibinfo  {publisher} {Springer},\ \bibinfo {address} {Berlin},\ \bibinfo
  {year} {2014})\ p.\ \bibinfo {pages} {193–233}\BibitemShut {NoStop}%
\bibitem [{\citenamefont {Janak}(1978)}]{Janak:1978}%
  \BibitemOpen
  \bibfield  {author} {\bibinfo {author} {\bibfnamefont {J.}~\bibnamefont
  {Janak}},\ }\bibfield  {title} {\bibinfo {title} {{Proof that ${\partial
  E}/{\partial n_i} = \varepsilon$}},\ }\href@noop {} {\bibfield  {journal}
  {\bibinfo  {journal} {Phys. Rev. B}\ }\textbf {\bibinfo {volume} {18}},\
  \bibinfo {pages} {7165} (\bibinfo {year} {1978})}\BibitemShut {NoStop}%
\bibitem [{\citenamefont {Perdew}\ and\ \citenamefont
  {Levy}(1983)}]{Perdew:1983}%
  \BibitemOpen
  \bibfield  {author} {\bibinfo {author} {\bibfnamefont {J.~P.}~\bibnamefont
  {Perdew}}\ and\ \bibinfo {author} {\bibfnamefont {M.}~\bibnamefont {Levy}},\
  }\bibfield  {title} {\bibinfo {title} {{Physical content of the exact
  Kohn-Sham orbital energies: Band gaps and derivative discontinuities}},\
  }\href@noop {} {\bibfield  {journal} {\bibinfo  {journal} {Phys. Rev. Lett.}\
  }\textbf {\bibinfo {volume} {51}},\ \bibinfo {pages} {1884} (\bibinfo {year}
  {1983})}\BibitemShut {NoStop}%
\bibitem [{\citenamefont {Singh}\ \emph {et~al.}(2017)\citenamefont {Singh},
  \citenamefont {Zhou}, \citenamefont {Shinde}, \citenamefont {Suram},
  \citenamefont {Montoya}, \citenamefont {Winston}, \citenamefont {Gregoire},\
  and\ \citenamefont {Persson}}]{Singh:2017}%
  \BibitemOpen
  \bibfield  {author} {\bibinfo {author} {\bibfnamefont {A.}~\bibnamefont
  {Singh}}, \bibinfo {author} {\bibfnamefont {L.}~\bibnamefont {Zhou}},
  \bibinfo {author} {\bibfnamefont {A.}~\bibnamefont {Shinde}}, \bibinfo
  {author} {\bibfnamefont {S.}~\bibnamefont {Suram}}, \bibinfo {author}
  {\bibfnamefont {J.}~\bibnamefont {Montoya}}, \bibinfo {author} {\bibfnamefont
  {D.}~\bibnamefont {Winston}}, \bibinfo {author} {\bibfnamefont
  {J.}~\bibnamefont {Gregoire}},\ and\ \bibinfo {author} {\bibfnamefont
  {K.}~\bibnamefont {Persson}},\ }\bibfield  {title} {\bibinfo {title}
  {{Electrochemcial stability of metastable materials}},\ }\href@noop {}
  {\bibfield  {journal} {\bibinfo  {journal} {Chem. Mater.}\ }\textbf {\bibinfo
  {volume} {29}},\ \bibinfo {pages} {10159} (\bibinfo {year}
  {2017})}\BibitemShut {NoStop}%
\bibitem [{\citenamefont {Kirchner-Hall}\ \emph {et~al.}(2021)\citenamefont
  {Kirchner-Hall}, \citenamefont {Zhao}, \citenamefont {Xiong}, \citenamefont
  {Timrov},\ and\ \citenamefont {Dabo}}]{KirchnerHall:2021}%
  \BibitemOpen
  \bibfield  {author} {\bibinfo {author} {\bibfnamefont {N.}~\bibnamefont
  {Kirchner-Hall}}, \bibinfo {author} {\bibfnamefont {W.}~\bibnamefont {Zhao}},
  \bibinfo {author} {\bibfnamefont {Y.}~\bibnamefont {Xiong}}, \bibinfo
  {author} {\bibfnamefont {I.}~\bibnamefont {Timrov}},\ and\ \bibinfo {author}
  {\bibfnamefont {I.}~\bibnamefont {Dabo}},\ }\bibfield  {title} {\bibinfo
  {title} {{Extensive benchmarking of DFT+\textit{U} calculations for
  predicting band gaps}},\ }\href@noop {} {\bibfield  {journal} {\bibinfo
  {journal} {Appl. Sci.}\ }\textbf {\bibinfo {volume} {11}},\ \bibinfo {pages}
  {2395} (\bibinfo {year} {2021})}\BibitemShut {NoStop}%
\bibitem [{\citenamefont {Hautier}\ \emph {et~al.}(2014)\citenamefont
  {Hautier}, \citenamefont {Miglio}, \citenamefont {Waroquiers}, \citenamefont
  {Rignanese},\ and\ \citenamefont {Gonze}}]{Hautier:2014}%
  \BibitemOpen
  \bibfield  {author} {\bibinfo {author} {\bibfnamefont {G.}~\bibnamefont
  {Hautier}}, \bibinfo {author} {\bibfnamefont {A.}~\bibnamefont {Miglio}},
  \bibinfo {author} {\bibfnamefont {D.}~\bibnamefont {Waroquiers}}, \bibinfo
  {author} {\bibfnamefont {G.}~\bibnamefont {Rignanese}},\ and\ \bibinfo
  {author} {\bibfnamefont {X.}~\bibnamefont {Gonze}},\ }\bibfield  {title}
  {\bibinfo {title} {{How does chemistry influence electron effective mass in
  oxides? A high-throughput computational analysis}},\ }\href@noop {}
  {\bibfield  {journal} {\bibinfo  {journal} {Chem. Mater.}\ }\textbf {\bibinfo
  {volume} {26}},\ \bibinfo {pages} {5447} (\bibinfo {year}
  {2014})}\BibitemShut {NoStop}%
\bibitem [{\citenamefont {Zhang}\ \emph {et~al.}(2015)\citenamefont {Zhang},
  \citenamefont {Yu}, \citenamefont {Liu},\ and\ \citenamefont
  {Liu}}]{Zhang:2015}%
  \BibitemOpen
  \bibfield  {author} {\bibinfo {author} {\bibfnamefont {J.}~\bibnamefont
  {Zhang}}, \bibinfo {author} {\bibfnamefont {W.}~\bibnamefont {Yu}}, \bibinfo
  {author} {\bibfnamefont {J.}~\bibnamefont {Liu}},\ and\ \bibinfo {author}
  {\bibfnamefont {B.}~\bibnamefont {Liu}},\ }\bibfield  {title} {\bibinfo
  {title} {{Illustration of high-active Ag$_2$CrO$_4$ photocatalyst from the
  first-principle calculation of electronic structures and carrier effective
  mass}},\ }\href@noop {} {\bibfield  {journal} {\bibinfo  {journal} {Applied
  Surface Science}\ }\textbf {\bibinfo {volume} {358}},\ \bibinfo {pages} {457}
  (\bibinfo {year} {2015})}\BibitemShut {NoStop}%
\bibitem [{\citenamefont {Suzuki}\ \emph {et~al.}(2020)\citenamefont {Suzuki},
  \citenamefont {Kanno}, \citenamefont {Hada}, \citenamefont {Abe},\ and\
  \citenamefont {Saeki}}]{Suzuki:2020}%
  \BibitemOpen
  \bibfield  {author} {\bibinfo {author} {\bibfnamefont {H.}~\bibnamefont
  {Suzuki}}, \bibinfo {author} {\bibfnamefont {S.}~\bibnamefont {Kanno}},
  \bibinfo {author} {\bibfnamefont {M.}~\bibnamefont {Hada}}, \bibinfo {author}
  {\bibfnamefont {R.}~\bibnamefont {Abe}},\ and\ \bibinfo {author}
  {\bibfnamefont {A.}~\bibnamefont {Saeki}},\ }\bibfield  {title} {\bibinfo
  {title} {{Exploring the relationship between effective mass, transient
  photoconductivity, and photocatalytic activity of Sr$_x$Pb$_{1-x}$BiO$_2$Cl
  (\textit{x}=0-1) oxyhalides}},\ }\href@noop {} {\bibfield  {journal}
  {\bibinfo  {journal} {Chem. Mater.}\ }\textbf {\bibinfo {volume} {32}},\
  \bibinfo {pages} {4166} (\bibinfo {year} {2020})}\BibitemShut {NoStop}%
\bibitem [{\citenamefont {Ricci}\ \emph {et~al.}(2017)\citenamefont {Ricci},
  \citenamefont {Chen}, \citenamefont {Aydemir}, \citenamefont {Snyder},
  \citenamefont {Rignanese}, \citenamefont {Jain},\ and\ \citenamefont
  {Hautier}}]{Ricci:2017}%
  \BibitemOpen
  \bibfield  {author} {\bibinfo {author} {\bibfnamefont {F.}~\bibnamefont
  {Ricci}}, \bibinfo {author} {\bibfnamefont {W.}~\bibnamefont {Chen}},
  \bibinfo {author} {\bibfnamefont {U.}~\bibnamefont {Aydemir}}, \bibinfo
  {author} {\bibfnamefont {G.}~\bibnamefont {Snyder}}, \bibinfo {author}
  {\bibfnamefont {G.}~\bibnamefont {Rignanese}}, \bibinfo {author}
  {\bibfnamefont {A.}~\bibnamefont {Jain}},\ and\ \bibinfo {author}
  {\bibfnamefont {G.}~\bibnamefont {Hautier}},\ }\bibfield  {title} {\bibinfo
  {title} {{Data Descriptor: An \textit{ab initio} electronic transport
  database for inorganic materials}},\ }\href@noop {} {\bibfield  {journal}
  {\bibinfo  {journal} {Sci. Data}\ }\textbf {\bibinfo {volume} {4}},\ \bibinfo
  {pages} {170085} (\bibinfo {year} {2017})}\BibitemShut {NoStop}%
\bibitem [{\citenamefont {Xu}\ \emph {et~al.}(2017)\citenamefont {Xu},
  \citenamefont {Xu}, \citenamefont {Zhang}, \citenamefont {Luo}, \citenamefont
  {Wang},\ and\ \citenamefont {Zhang}}]{Xu:2017}%
  \BibitemOpen
  \bibfield  {author} {\bibinfo {author} {\bibfnamefont {K.}~\bibnamefont
  {Xu}}, \bibinfo {author} {\bibfnamefont {D.}~\bibnamefont {Xu}}, \bibinfo
  {author} {\bibfnamefont {X.}~\bibnamefont {Zhang}}, \bibinfo {author}
  {\bibfnamefont {Z.}~\bibnamefont {Luo}}, \bibinfo {author} {\bibfnamefont
  {Y.}~\bibnamefont {Wang}},\ and\ \bibinfo {author} {\bibfnamefont
  {S.}~\bibnamefont {Zhang}},\ }\bibfield  {title} {\bibinfo {title}
  {{Visible-light activity of N-LiInO$_2$: Band structure modifications through
  interstitial nitrogen doping}},\ }\href@noop {} {\bibfield  {journal}
  {\bibinfo  {journal} {Appl. Surface Sci.}\ }\textbf {\bibinfo {volume}
  {391}},\ \bibinfo {pages} {645} (\bibinfo {year} {2017})}\BibitemShut
  {NoStop}%
\bibitem [{\citenamefont {Kushida}\ and\ \citenamefont
  {Kuriyama}(2006)}]{Kushida:2006}%
  \BibitemOpen
  \bibfield  {author} {\bibinfo {author} {\bibfnamefont {K.}~\bibnamefont
  {Kushida}}\ and\ \bibinfo {author} {\bibfnamefont {K.}~\bibnamefont
  {Kuriyama}},\ }{\bibfield  {journal} {\bibinfo
  {journal} {physica status solidi c}\ }\textbf {\bibinfo {volume} {3}},\
  \bibinfo {pages} {2800} (\bibinfo {year} {2006})}\BibitemShut {NoStop}%
\bibitem [{\citenamefont {Yin}\ \emph {et~al.}(2002)\citenamefont {Yin},
  \citenamefont {Zou},\ and\ \citenamefont {Ye}}]{Yin:2002}%
  \BibitemOpen
  \bibfield  {author} {\bibinfo {author} {\bibfnamefont {J.}~\bibnamefont
  {Yin}}, \bibinfo {author} {\bibfnamefont {Z.}~\bibnamefont {Zou}},\ and\
  \bibinfo {author} {\bibfnamefont {J.}~\bibnamefont {Ye}},\ }\bibfield
  {title} {\bibinfo {title} {{Synthesis and photophysical properties of barium
  indium oxides}},\ }\href@noop {} {\bibfield  {journal} {\bibinfo  {journal}
  {Journal of Materials Research}\ }\textbf {\bibinfo {volume} {17}},\ \bibinfo
  {pages} {2201} (\bibinfo {year} {2002})}\BibitemShut {NoStop}%
\bibitem [{\citenamefont {Huang}\ \emph {et~al.}(2012)\citenamefont {Huang},
  \citenamefont {Xu}, \citenamefont {Zhu}, \citenamefont {Liu}, \citenamefont
  {Yuan}, \citenamefont {Fu}, \citenamefont {Zhang},\ and\ \citenamefont
  {Li}}]{Huang:2012}%
  \BibitemOpen
  \bibfield  {author} {\bibinfo {author} {\bibfnamefont {R.}~\bibnamefont
  {Huang}}, \bibinfo {author} {\bibfnamefont {X.}~\bibnamefont {Xu}}, \bibinfo
  {author} {\bibfnamefont {J.}~\bibnamefont {Zhu}}, \bibinfo {author}
  {\bibfnamefont {W.}~\bibnamefont {Liu}}, \bibinfo {author} {\bibfnamefont
  {R.}~\bibnamefont {Yuan}}, \bibinfo {author} {\bibfnamefont {X.}~\bibnamefont
  {Fu}}, \bibinfo {author} {\bibfnamefont {Y.}~\bibnamefont {Zhang}},\ and\
  \bibinfo {author} {\bibfnamefont {Z.}~\bibnamefont {Li}},\ }\bibfield
  {title} {\bibinfo {title} {{Nanocrystalline CaSb$_2$O$_5$(OH)$_2$ and
  Ca$_2$Sb$_2$O$_7$: Controlled syntheses, electronic structures and
  photocatalytic activity}},\ }\href@noop {} {\bibfield  {journal} {\bibinfo
  {journal} {Appl. Catalysis B: Environmental}\ }\textbf {\bibinfo {volume}
  {127}},\ \bibinfo {pages} {205} (\bibinfo {year} {2012})}\BibitemShut
  {NoStop}%
\bibitem [{\citenamefont {Xue}\ \emph {et~al.}(2008)\citenamefont {Xue},
  \citenamefont {Li}, \citenamefont {Wu}, \citenamefont {Ding}, \citenamefont
  {Wang},\ and\ \citenamefont {Fu}}]{Xue:2008}%
  \BibitemOpen
  \bibfield  {author} {\bibinfo {author} {\bibfnamefont {H.}~\bibnamefont
  {Xue}}, \bibinfo {author} {\bibfnamefont {Z.}~\bibnamefont {Li}}, \bibinfo
  {author} {\bibfnamefont {L.}~\bibnamefont {Wu}}, \bibinfo {author}
  {\bibfnamefont {Z.}~\bibnamefont {Ding}}, \bibinfo {author} {\bibfnamefont
  {X.}~\bibnamefont {Wang}},\ and\ \bibinfo {author} {\bibfnamefont
  {X.}~\bibnamefont {Fu}},\ }\bibfield  {title} {\bibinfo {title}
  {{Nanocrystalline ternary wide band gap p-block metal semiconductor
  Sr$_2$Sb$_2$O$_7$: Hydrothermal syntheses and photocatalytic benzene
  degradation}},\ }\href@noop {} {\bibfield  {journal} {\bibinfo  {journal} {J.
  Phys. Chem. C}\ }\textbf {\bibinfo {volume} {112}},\ \bibinfo {pages} {5850}
  (\bibinfo {year} {2008})}\BibitemShut {NoStop}%
\bibitem [{\citenamefont {Zhao}\ \emph {et~al.}(2015)\citenamefont {Zhao},
  \citenamefont {Han}, \citenamefont {Cui}, \citenamefont {Zong},\ and\
  \citenamefont {Li}}]{Zhao:2015}%
  \BibitemOpen
  \bibfield  {author} {\bibinfo {author} {\bibfnamefont {D.}~\bibnamefont
  {Zhao}}, \bibinfo {author} {\bibfnamefont {J.}~\bibnamefont {Han}}, \bibinfo
  {author} {\bibfnamefont {J.}~\bibnamefont {Cui}}, \bibinfo {author}
  {\bibfnamefont {X.}~\bibnamefont {Zong}},\ and\ \bibinfo {author}
  {\bibfnamefont {C.}~\bibnamefont {Li}},\ }\bibfield  {title} {\bibinfo
  {title} {{A new Pb(IV)-based photocathode material Sr$_2$PbO$_4$ with good
  light harvesting ability}},\ }\href@noop {} {\bibfield  {journal} {\bibinfo
  {journal} {J. Mat. Chem. A}\ }\textbf {\bibinfo {volume} {3}},\ \bibinfo
  {pages} {12051} (\bibinfo {year} {2015})}\BibitemShut {NoStop}%
\bibitem [{\citenamefont {Hadjarab}\ \emph {et~al.}(2007)\citenamefont
  {Hadjarab}, \citenamefont {Saadi}, \citenamefont {Bouguelia},\ and\
  \citenamefont {Trari}}]{Hadjarab:2007}%
  \BibitemOpen
  \bibfield  {author} {\bibinfo {author} {\bibfnamefont {B.}~\bibnamefont
  {Hadjarab}}, \bibinfo {author} {\bibfnamefont {S.}~\bibnamefont {Saadi}},
  \bibinfo {author} {\bibfnamefont {A.}~\bibnamefont {Bouguelia}},\ and\
  \bibinfo {author} {\bibfnamefont {M.}~\bibnamefont {Trari}},\ }\bibfield
  {title} {\bibinfo {title} {{Physical properties and photoelectrochemical
  characterization of SrPbO$_3$}},\ }\href@noop {} {\bibfield  {journal}
  {\bibinfo  {journal} {Phys. Stat. Sol.}\ }\textbf {\bibinfo {volume} {204}},\
  \bibinfo {pages} {2369} (\bibinfo {year} {2007})}\BibitemShut {NoStop}%
\bibitem [{\citenamefont {Medicherla}\ \emph {et~al.}(2007)\citenamefont
  {Medicherla}, \citenamefont {Shripathi},\ and\ \citenamefont
  {Lalla}}]{Medicherla:2007}%
  \BibitemOpen
  \bibfield  {author} {\bibinfo {author} {\bibfnamefont {V.}~\bibnamefont
  {Medicherla}}, \bibinfo {author} {\bibfnamefont {T.}~\bibnamefont
  {Shripathi}},\ and\ \bibinfo {author} {\bibfnamefont {N.}~\bibnamefont
  {Lalla}},\ }\bibfield  {title} {\bibinfo {title} {{Electronic structure of
  BaPbO$_3$ and Ba$_2$PbO$_4$}},\ }\href@noop {} {\bibfield  {journal}
  {\bibinfo  {journal} {J. Phys.: Condens. Matter}\ }\textbf {\bibinfo {volume}
  {20}},\ \bibinfo {pages} {035219} (\bibinfo {year} {2007})}\BibitemShut
  {NoStop}%
\bibitem [{\citenamefont {Katz}\ \emph {et~al.}(2022)\citenamefont {Katz},
  \citenamefont {Theibault}, \citenamefont {Kirchner-Hall}, \citenamefont
  {Mao}, \citenamefont {Dabo}, \citenamefont {Abru{\~{n}}a},\ and\
  \citenamefont {Schaak}}]{Katz:2022}%
  \BibitemOpen
  \bibfield  {author} {\bibinfo {author} {\bibfnamefont {R.~J.}\ \bibnamefont
  {Katz}}, \bibinfo {author} {\bibfnamefont {M.~J.}\ \bibnamefont {Theibault}},
  \bibinfo {author} {\bibfnamefont {N.~E.}\ \bibnamefont {Kirchner-Hall}},
  \bibinfo {author} {\bibfnamefont {Z.}~\bibnamefont {Mao}}, \bibinfo {author}
  {\bibfnamefont {I.}~\bibnamefont {Dabo}}, \bibinfo {author} {\bibfnamefont
  {H.~D.}\ \bibnamefont {Abru{\~{n}}a}},\ and\ \bibinfo {author} {\bibfnamefont
  {R.~E.}\ \bibnamefont {Schaak}},\ }\bibfield  {title} {\bibinfo {title}
  {Understanding the photoelectrochemical properties of theoretically predicted
  water-splitting catalysts for effective materials discovery},\ }\href@noop {}
  {\bibfield  {journal} {\bibinfo  {journal} {Advanced Energy Materials}\ ,\
  \bibinfo {pages} {2201869}} (\bibinfo {year} {2022})}\BibitemShut {NoStop}%
\bibitem [{\citenamefont {Istomin}\ \emph {et~al.}(2014)\citenamefont
  {Istomin}, \citenamefont {Antipov}, \citenamefont {Fedotov}, \citenamefont
  {Bredikhin}, \citenamefont {Lyskov}, \citenamefont {Shafeie}, \citenamefont
  {Svensson}, \citenamefont {Liu},\ and\ \citenamefont {Shen}}]{Istomin:2014}%
  \BibitemOpen
  \bibfield  {author} {\bibinfo {author} {\bibfnamefont {S.}~\bibnamefont
  {Istomin}}, \bibinfo {author} {\bibfnamefont {E.}~\bibnamefont {Antipov}},
  \bibinfo {author} {\bibfnamefont {Y.}~\bibnamefont {Fedotov}}, \bibinfo
  {author} {\bibfnamefont {S.}~\bibnamefont {Bredikhin}}, \bibinfo {author}
  {\bibfnamefont {N.}~\bibnamefont {Lyskov}}, \bibinfo {author} {\bibfnamefont
  {S.}~\bibnamefont {Shafeie}}, \bibinfo {author} {\bibfnamefont
  {G.}~\bibnamefont {Svensson}}, \bibinfo {author} {\bibfnamefont
  {Y.}~\bibnamefont {Liu}},\ and\ \bibinfo {author} {\bibfnamefont
  {Z.}~\bibnamefont {Shen}},\ }\bibfield  {title} {\bibinfo {title} {{Crystal
  structure and high-temperature electrical conductivity of novel
  perovskite-related gallium and indium oxides}},\ }\href@noop {} {\bibfield
  {journal} {\bibinfo  {journal} {J. Solid State Electrochem.}\ }\textbf
  {\bibinfo {volume} {18}},\ \bibinfo {pages} {1415} (\bibinfo {year}
  {2014})}\BibitemShut {NoStop}%
\bibitem [{\citenamefont {Sato}\ \emph {et~al.}(2002)\citenamefont {Sato},
  \citenamefont {Saito}, \citenamefont {Nishiyama},\ and\ \citenamefont
  {Inoue}}]{Sato:2002}%
  \BibitemOpen
  \bibfield  {author} {\bibinfo {author} {\bibfnamefont {J.}~\bibnamefont
  {Sato}}, \bibinfo {author} {\bibfnamefont {N.}~\bibnamefont {Saito}},
  \bibinfo {author} {\bibfnamefont {H.}~\bibnamefont {Nishiyama}},\ and\
  \bibinfo {author} {\bibfnamefont {Y.}~\bibnamefont {Inoue}},\ }\bibfield
  {title} {\bibinfo {title} {{Photocatalytic water decomposition by
  RuO$_2$-loaded antimonates, M$_2$Sb$_2$O$_7$ (M=Ca,Sr), CaSb$_2$O$_6$ and
  NaSbO$_3$, with d$^{10}$ configuration}},\ }\href@noop {} {\bibfield
  {journal} {\bibinfo  {journal} {J. Photochemistry and Photobiology A:
  Chemistry}\ }\textbf {\bibinfo {volume} {148}},\ \bibinfo {pages} {85}
  (\bibinfo {year} {2002})}\BibitemShut {NoStop}%
\bibitem{Burdett:1995}
J.~K. Burdett.
\newblock {\em Chemical bonding in solids}.
\newblock Oxford University Press, New York, 1995.
\bibitem [{\citenamefont {Giannozzi}\ \emph {et~al.}(2009)\citenamefont
  {Giannozzi}, \citenamefont {Baroni}, \citenamefont {Bonini}, \citenamefont
  {Calandra}, \citenamefont {Car}, \citenamefont {Cavazzoni}, \citenamefont
  {Ceresoli}, \citenamefont {Chiarotti}, \citenamefont {Cococcioni},
  \citenamefont {Dabo}, \citenamefont {Dal~Corso}, \citenamefont
  {De~Gironcoli}, \citenamefont {Fabris}, \citenamefont {Fratesi},
  \citenamefont {Gebauer}, \citenamefont {Gerstmann}, \citenamefont
  {Gougoussis}, \citenamefont {Kokalj}, \citenamefont {Lazzeri}, \citenamefont
  {Martin-Samos}, \citenamefont {Marzari}, \citenamefont {Mauri}, \citenamefont
  {Mazzarello}, \citenamefont {Paolini}, \citenamefont {Pasquarello},
  \citenamefont {Paulatto}, \citenamefont {Sbraccia}, \citenamefont {Scandolo},
  \citenamefont {Sclauzero}, \citenamefont {Seitsonen}, \citenamefont
  {Smogunov}, \citenamefont {Umari},\ and\ \citenamefont
  {Wentzcovitch}}]{Giannozzi:2009}%
  \BibitemOpen
  \bibfield  {author} {\bibinfo {author} {\bibfnamefont {P.}~\bibnamefont
  {Giannozzi}}, \bibinfo {author} {\bibfnamefont {S.}~\bibnamefont {Baroni}},
  \bibinfo {author} {\bibfnamefont {N.}~\bibnamefont {Bonini}}, \bibinfo
  {author} {\bibfnamefont {M.}~\bibnamefont {Calandra}}, \bibinfo {author}
  {\bibfnamefont {R.}~\bibnamefont {Car}}, \bibinfo {author} {\bibfnamefont
  {C.}~\bibnamefont {Cavazzoni}}, \bibinfo {author} {\bibfnamefont
  {D.}~\bibnamefont {Ceresoli}}, \bibinfo {author} {\bibfnamefont
  {G.}~\bibnamefont {Chiarotti}}, \bibinfo {author} {\bibfnamefont
  {M.}~\bibnamefont {Cococcioni}}, \bibinfo {author} {\bibfnamefont
  {I.}~\bibnamefont {Dabo}}, \bibinfo {author} {\bibfnamefont {A.}~\bibnamefont
  {Dal~Corso}}, \bibinfo {author} {\bibfnamefont {S.}~\bibnamefont
  {De~Gironcoli}}, \bibinfo {author} {\bibfnamefont {S.}~\bibnamefont
  {Fabris}}, \bibinfo {author} {\bibfnamefont {G.}~\bibnamefont {Fratesi}},
  \bibinfo {author} {\bibfnamefont {R.}~\bibnamefont {Gebauer}}, \bibinfo
  {author} {\bibfnamefont {U.}~\bibnamefont {Gerstmann}}, \bibinfo {author}
  {\bibfnamefont {C.}~\bibnamefont {Gougoussis}}, \bibinfo {author}
  {\bibfnamefont {A.}~\bibnamefont {Kokalj}}, \bibinfo {author} {\bibfnamefont
  {M.}~\bibnamefont {Lazzeri}}, \bibinfo {author} {\bibfnamefont
  {L.}~\bibnamefont {Martin-Samos}}, \bibinfo {author} {\bibfnamefont
  {N.}~\bibnamefont {Marzari}}, \bibinfo {author} {\bibfnamefont
  {F.}~\bibnamefont {Mauri}}, \bibinfo {author} {\bibfnamefont
  {R.}~\bibnamefont {Mazzarello}}, \bibinfo {author} {\bibfnamefont
  {S.}~\bibnamefont {Paolini}}, \bibinfo {author} {\bibfnamefont
  {A.}~\bibnamefont {Pasquarello}}, \bibinfo {author} {\bibfnamefont
  {L.}~\bibnamefont {Paulatto}}, \bibinfo {author} {\bibfnamefont
  {C.}~\bibnamefont {Sbraccia}}, \bibinfo {author} {\bibfnamefont
  {S.}~\bibnamefont {Scandolo}}, \bibinfo {author} {\bibfnamefont
  {G.}~\bibnamefont {Sclauzero}}, \bibinfo {author} {\bibfnamefont
  {A.}~\bibnamefont {Seitsonen}}, \bibinfo {author} {\bibfnamefont
  {A.}~\bibnamefont {Smogunov}}, \bibinfo {author} {\bibfnamefont
  {P.}~\bibnamefont {Umari}},\ and\ \bibinfo {author} {\bibfnamefont
  {R.}~\bibnamefont {Wentzcovitch}},\ }\bibfield  {title} {\bibinfo {title}
  {{Q}uantum {ESPRESSO}: {A} modular and open-source software project for
  quantum simulations of materials},\ }
  {\bibfield  {journal} {\bibinfo  {journal} {J. Phys.: Condens. Matter}\
  }\textbf {\bibinfo {volume} {21}},\ \bibinfo {pages} {395502} (\bibinfo
  {year} {2009})}\BibitemShut {NoStop}%
\bibitem [{\citenamefont {Giannozzi}\ \emph {et~al.}(2017)\citenamefont
  {Giannozzi}, \citenamefont {Andreussi}, \citenamefont {Brumme}, \citenamefont
  {Bunau}, \citenamefont {Buongiorno~Nardelli}, \citenamefont {Calandra},
  \citenamefont {Car}, \citenamefont {Cavazzoni}, \citenamefont {Ceresoli},
  \citenamefont {Cococcioni}, \citenamefont {Colonna}, \citenamefont
  {Carnimeo}, \citenamefont {Dal~Corso}, \citenamefont {de~Gironcoli},
  \citenamefont {Delugas}, \citenamefont {Di{S}tasio~{J}r.}, \citenamefont
  {Ferretti}, \citenamefont {Floris}, \citenamefont {Fratesi}, \citenamefont
  {Fugallo}, \citenamefont {Gebauer}, \citenamefont {Gerstmann}, \citenamefont
  {Giustino}, \citenamefont {Gorni}, \citenamefont {Jia}, \citenamefont
  {Kawamura}, \citenamefont {Ko}, \citenamefont {Kokalj}, \citenamefont
  {K\"{u}\c{c}\"{u}kbenli}, \citenamefont {Lazzeri}, \citenamefont {Marsili},
  \citenamefont {Marzari}, \citenamefont {Mauri}, \citenamefont {Nguyen},
  \citenamefont {Nguyen}, \citenamefont {Otero-de-la {R}osa}, \citenamefont
  {Paulatto}, \citenamefont {Ponc\'e}, \citenamefont {Rocca}, \citenamefont
  {Sabatini}, \citenamefont {Santra}, \citenamefont {Schlipf}, \citenamefont
  {Seitsonen}, \citenamefont {Smogunov}, \citenamefont {Timrov}, \citenamefont
  {Thonhauser}, \citenamefont {Umari}, \citenamefont {Vast},\ and\
  \citenamefont {Baroni}}]{Giannozzi:2017}%
  \BibitemOpen
  \bibfield  {author} {\bibinfo {author} {\bibfnamefont {P.}~\bibnamefont
  {Giannozzi}}, \bibinfo {author} {\bibfnamefont {O.}~\bibnamefont
  {Andreussi}}, \bibinfo {author} {\bibfnamefont {T.}~\bibnamefont {Brumme}},
  \bibinfo {author} {\bibfnamefont {O.}~\bibnamefont {Bunau}}, \bibinfo
  {author} {\bibfnamefont {M.}~\bibnamefont {Buongiorno~Nardelli}}, \bibinfo
  {author} {\bibfnamefont {M.}~\bibnamefont {Calandra}}, \bibinfo {author}
  {\bibfnamefont {R.}~\bibnamefont {Car}}, \bibinfo {author} {\bibfnamefont
  {C.}~\bibnamefont {Cavazzoni}}, \bibinfo {author} {\bibfnamefont
  {D.}~\bibnamefont {Ceresoli}}, \bibinfo {author} {\bibfnamefont
  {M.}~\bibnamefont {Cococcioni}}, \bibinfo {author} {\bibfnamefont
  {N.}~\bibnamefont {Colonna}}, \bibinfo {author} {\bibfnamefont
  {I.}~\bibnamefont {Carnimeo}}, \bibinfo {author} {\bibfnamefont
  {A.}~\bibnamefont {Dal~Corso}}, \bibinfo {author} {\bibfnamefont
  {S.}~\bibnamefont {de~Gironcoli}}, \bibinfo {author} {\bibfnamefont
  {P.}~\bibnamefont {Delugas}}, \bibinfo {author} {\bibfnamefont {R.~A.}\
  \bibnamefont {Di{S}tasio~{J}r.}}, \bibinfo {author} {\bibfnamefont
  {A.}~\bibnamefont {Ferretti}}, \bibinfo {author} {\bibfnamefont
  {A.}~\bibnamefont {Floris}}, \bibinfo {author} {\bibfnamefont
  {G.}~\bibnamefont {Fratesi}}, \bibinfo {author} {\bibfnamefont
  {G.}~\bibnamefont {Fugallo}}, \bibinfo {author} {\bibfnamefont
  {R.}~\bibnamefont {Gebauer}}, \bibinfo {author} {\bibfnamefont
  {U.}~\bibnamefont {Gerstmann}}, \bibinfo {author} {\bibfnamefont
  {F.}~\bibnamefont {Giustino}}, \bibinfo {author} {\bibfnamefont
  {T.}~\bibnamefont {Gorni}}, \bibinfo {author} {\bibfnamefont
  {J.}~\bibnamefont {Jia}}, \bibinfo {author} {\bibfnamefont {M.}~\bibnamefont
  {Kawamura}}, \bibinfo {author} {\bibfnamefont {H.-Y.}\ \bibnamefont {Ko}},
  \bibinfo {author} {\bibfnamefont {A.}~\bibnamefont {Kokalj}}, \bibinfo
  {author} {\bibfnamefont {E.}~\bibnamefont {K\"{u}\c{c}\"{u}kbenli}}, \bibinfo
  {author} {\bibfnamefont {M.}~\bibnamefont {Lazzeri}}, \bibinfo {author}
  {\bibfnamefont {M.}~\bibnamefont {Marsili}}, \bibinfo {author} {\bibfnamefont
  {N.}~\bibnamefont {Marzari}}, \bibinfo {author} {\bibfnamefont
  {F.}~\bibnamefont {Mauri}}, \bibinfo {author} {\bibfnamefont {N.~L.}\
  \bibnamefont {Nguyen}}, \bibinfo {author} {\bibfnamefont {H.-V.}\
  \bibnamefont {Nguyen}}, \bibinfo {author} {\bibfnamefont {A.}~\bibnamefont
  {Otero-de-la {R}osa}}, \bibinfo {author} {\bibfnamefont {L.}~\bibnamefont
  {Paulatto}}, \bibinfo {author} {\bibfnamefont {S.}~\bibnamefont {Ponc\'e}},
  \bibinfo {author} {\bibfnamefont {D.}~\bibnamefont {Rocca}}, \bibinfo
  {author} {\bibfnamefont {R.}~\bibnamefont {Sabatini}}, \bibinfo {author}
  {\bibfnamefont {B.}~\bibnamefont {Santra}}, \bibinfo {author} {\bibfnamefont
  {M.}~\bibnamefont {Schlipf}}, \bibinfo {author} {\bibfnamefont
  {A.}~\bibnamefont {Seitsonen}}, \bibinfo {author} {\bibfnamefont
  {A.}~\bibnamefont {Smogunov}}, \bibinfo {author} {\bibfnamefont
  {I.}~\bibnamefont {Timrov}}, \bibinfo {author} {\bibfnamefont
  {T.}~\bibnamefont {Thonhauser}}, \bibinfo {author} {\bibfnamefont
  {P.}~\bibnamefont {Umari}}, \bibinfo {author} {\bibfnamefont
  {N.}~\bibnamefont {Vast}},\ and\ \bibinfo {author} {\bibfnamefont
  {S.}~\bibnamefont {Baroni}},\ }\bibfield  {title} {\bibinfo {title}
  {{A}dvanced capabilities for materials modelling with {Q}uantum {ESPRESSO}},\
  }\href@noop {} {\bibfield  {journal} {\bibinfo  {journal} {J. Phys.: Condens.
  Matter}\ }\textbf {\bibinfo {volume} {29}},\ \bibinfo {pages} {465901}
  (\bibinfo {year} {2017})}\BibitemShut {NoStop}%
\bibitem [{\citenamefont {Giannozzi}\ \emph {et~al.}(2020)\citenamefont
  {Giannozzi}, \citenamefont {Baseggio}, \citenamefont {Bonf\`a}, \citenamefont
  {Brunato}, \citenamefont {Car}, \citenamefont {Carnimeo}, \citenamefont
  {Cavazzoni}, \citenamefont {de~Gironcoli}, \citenamefont {Delugas},
  \citenamefont {Ferrari~Ruffino}, \citenamefont {Ferretti}, \citenamefont
  {Marzari}, \citenamefont {Timrov}, \citenamefont {Urru},\ and\ \citenamefont
  {Baroni}}]{Giannozzi:2020}%
  \BibitemOpen
  \bibfield  {author} {\bibinfo {author} {\bibfnamefont {P.}~\bibnamefont
  {Giannozzi}}, \bibinfo {author} {\bibfnamefont {O.}~\bibnamefont {Baseggio}},
  \bibinfo {author} {\bibfnamefont {P.}~\bibnamefont {Bonf\`a}}, \bibinfo
  {author} {\bibfnamefont {D.}~\bibnamefont {Brunato}}, \bibinfo {author}
  {\bibfnamefont {R.}~\bibnamefont {Car}}, \bibinfo {author} {\bibfnamefont
  {I.}~\bibnamefont {Carnimeo}}, \bibinfo {author} {\bibfnamefont
  {C.}~\bibnamefont {Cavazzoni}}, \bibinfo {author} {\bibfnamefont
  {S.}~\bibnamefont {de~Gironcoli}}, \bibinfo {author} {\bibfnamefont
  {P.}~\bibnamefont {Delugas}}, \bibinfo {author} {\bibfnamefont
  {F.}~\bibnamefont {Ferrari~Ruffino}}, \bibinfo {author} {\bibfnamefont
  {A.}~\bibnamefont {Ferretti}}, \bibinfo {author} {\bibfnamefont
  {N.}~\bibnamefont {Marzari}}, \bibinfo {author} {\bibfnamefont
  {I.}~\bibnamefont {Timrov}}, \bibinfo {author} {\bibfnamefont
  {A.}~\bibnamefont {Urru}},\ and\ \bibinfo {author} {\bibfnamefont
  {S.}~\bibnamefont {Baroni}},\ }\bibfield  {title} {\bibinfo {title} {{Quantum
  ESPRESSO toward the exascale}},\ }\href@noop {} {\bibfield  {journal}
  {\bibinfo  {journal} {J. Chem. Phys.}\ }\textbf {\bibinfo {volume} {152}},\
  \bibinfo {pages} {154105} (\bibinfo {year} {2020})}\BibitemShut {NoStop}%
\bibitem [{\citenamefont {Perdew}\ \emph {et~al.}(2008)\citenamefont {Perdew},
  \citenamefont {Ruzsinszky}, \citenamefont {Csonka}, \citenamefont {Vydrov},
  \citenamefont {Scuseria}, \citenamefont {Constantin}, \citenamefont {Zhou},\
  and\ \citenamefont {Burke}}]{Perdew:2008}%
  \BibitemOpen
  \bibfield  {author} {\bibinfo {author} {\bibfnamefont {J.~P.}~\bibnamefont
  {Perdew}}, \bibinfo {author} {\bibfnamefont {A.}~\bibnamefont {Ruzsinszky}},
  \bibinfo {author} {\bibfnamefont {G.~I.}~\bibnamefont {Csonka}}, \bibinfo
  {author} {\bibfnamefont {O.~A.}~\bibnamefont {Vydrov}}, \bibinfo {author}
  {\bibfnamefont {G.~E.}~\bibnamefont {Scuseria}}, \bibinfo {author}
  {\bibfnamefont {L.~A.}~\bibnamefont {Constantin}}, \bibinfo {author}
  {\bibfnamefont {X.}~\bibnamefont {Zhou}},\ and\ \bibinfo {author}
  {\bibfnamefont {K.}~\bibnamefont {Burke}},\ }\href@noop {} {\bibfield
  {journal} {\bibinfo  {journal} {Phys. Rev. Lett.}\ }\textbf {\bibinfo
  {volume} {100}},\ \bibinfo {pages} {136406} (\bibinfo {year}
  {2008})}\BibitemShut {NoStop}%
\bibitem [{\citenamefont {Prandini}\ \emph {et~al.}(2018)\citenamefont
  {Prandini}, \citenamefont {Marrazzo}, \citenamefont {Castelli}, \citenamefont
  {Mounet},\ and\ \citenamefont {Marzari}}]{Prandini:2018}%
  \BibitemOpen
  \bibfield  {author} {\bibinfo {author} {\bibfnamefont {G.}~\bibnamefont
  {Prandini}}, \bibinfo {author} {\bibfnamefont {A.}~\bibnamefont {Marrazzo}},
  \bibinfo {author} {\bibfnamefont {I.}~\bibnamefont {Castelli}}, \bibinfo
  {author} {\bibfnamefont {N.}~\bibnamefont {Mounet}},\ and\ \bibinfo {author}
  {\bibfnamefont {N.}~\bibnamefont {Marzari}},\ }\href@noop {} {\bibfield
  {journal} {\bibinfo  {journal} {npj Computational Materials}\ }\textbf
  {\bibinfo {volume} {4}},\ \bibinfo {pages} {72} (\bibinfo {year}
  {2018})}\BibitemShut {NoStop}%
\bibitem [{\citenamefont {Lejaeghere}\ \emph {et~al.}(2016)\citenamefont
  {Lejaeghere}, \citenamefont {Bihlmayer}, \citenamefont {Björkman},
  \citenamefont {Blaha}, \citenamefont {Blügel}, \citenamefont {Blum},
  \citenamefont {Caliste}, \citenamefont {Castelli}, \citenamefont {Clark},
  \citenamefont {Corso}, \citenamefont {de~Gironcoli}, \citenamefont {Deutsch},
  \citenamefont {Dewhurst}, \citenamefont {Marco}, \citenamefont {Draxl},
  \citenamefont {Du{\l}ak}, \citenamefont {Eriksson}, \citenamefont
  {Flores-Livas}, \citenamefont {Garrity}, \citenamefont {Genovese},
  \citenamefont {Giannozzi}, \citenamefont {Giantomassi}, \citenamefont
  {Goedecker}, \citenamefont {Gonze}, \citenamefont {Gr{\aa}näs},
  \citenamefont {Gross}, \citenamefont {Gulans}, \citenamefont {Gygi},
  \citenamefont {Hamann}, \citenamefont {Hasnip}, \citenamefont {Holzwarth},
  \citenamefont {Iu{\c{s}}an}, \citenamefont {Jochym}, \citenamefont {Jollet},
  \citenamefont {Jones}, \citenamefont {Kresse}, \citenamefont {Koepernik},
  \citenamefont {Kü{\c{c}}ükbenli}, \citenamefont {Kvashnin}, \citenamefont
  {Locht}, \citenamefont {Lubeck}, \citenamefont {Marsman}, \citenamefont
  {Marzari}, \citenamefont {Nitzsche}, \citenamefont {Nordström},
  \citenamefont {Ozaki}, \citenamefont {Paulatto}, \citenamefont {Pickard},
  \citenamefont {Poelmans}, \citenamefont {Probert}, \citenamefont {Refson},
  \citenamefont {Richter}, \citenamefont {Rignanese}, \citenamefont {Saha},
  \citenamefont {Scheffler}, \citenamefont {Schlipf}, \citenamefont {Schwarz},
  \citenamefont {Sharma}, \citenamefont {Tavazza}, \citenamefont {Thunström},
  \citenamefont {Tkatchenko}, \citenamefont {Torrent}, \citenamefont
  {Vanderbilt}, \citenamefont {van Setten}, \citenamefont {Speybroeck},
  \citenamefont {Wills}, \citenamefont {Yates}, \citenamefont {Zhang},\ and\
  \citenamefont {Cottenier}}]{Lejaeghere:2016}%
  \BibitemOpen
  \bibfield  {author} {\bibinfo {author} {\bibfnamefont {K.}~\bibnamefont
  {Lejaeghere}}, \bibinfo {author} {\bibfnamefont {G.}~\bibnamefont
  {Bihlmayer}}, \bibinfo {author} {\bibfnamefont {T.}~\bibnamefont
  {Björkman}}, \bibinfo {author} {\bibfnamefont {P.}~\bibnamefont {Blaha}},
  \bibinfo {author} {\bibfnamefont {S.}~\bibnamefont {Blügel}}, \bibinfo
  {author} {\bibfnamefont {V.}~\bibnamefont {Blum}}, \bibinfo {author}
  {\bibfnamefont {D.}~\bibnamefont {Caliste}}, \bibinfo {author} {\bibfnamefont
  {I.~E.}\ \bibnamefont {Castelli}}, \bibinfo {author} {\bibfnamefont {S.~J.}\
  \bibnamefont {Clark}}, \bibinfo {author} {\bibfnamefont {A.~D.}\ \bibnamefont
  {Corso}}, \bibinfo {author} {\bibfnamefont {S.}~\bibnamefont {de~Gironcoli}},
  \bibinfo {author} {\bibfnamefont {T.}~\bibnamefont {Deutsch}}, \bibinfo
  {author} {\bibfnamefont {J.~K.}\ \bibnamefont {Dewhurst}}, \bibinfo {author}
  {\bibfnamefont {I.~D.}\ \bibnamefont {Marco}}, \bibinfo {author}
  {\bibfnamefont {C.}~\bibnamefont {Draxl}}, \bibinfo {author} {\bibfnamefont
  {M.}~\bibnamefont {Du{\l}ak}}, \bibinfo {author} {\bibfnamefont
  {O.}~\bibnamefont {Eriksson}}, \bibinfo {author} {\bibfnamefont {J.~A.}\
  \bibnamefont {Flores-Livas}}, \bibinfo {author} {\bibfnamefont {K.~F.}\
  \bibnamefont {Garrity}}, \bibinfo {author} {\bibfnamefont {L.}~\bibnamefont
  {Genovese}}, \bibinfo {author} {\bibfnamefont {P.}~\bibnamefont {Giannozzi}},
  \bibinfo {author} {\bibfnamefont {M.}~\bibnamefont {Giantomassi}}, \bibinfo
  {author} {\bibfnamefont {S.}~\bibnamefont {Goedecker}}, \bibinfo {author}
  {\bibfnamefont {X.}~\bibnamefont {Gonze}}, \bibinfo {author} {\bibfnamefont
  {O.}~\bibnamefont {Gr{\aa}näs}}, \bibinfo {author} {\bibfnamefont
  {E.~K.~U.}\ \bibnamefont {Gross}}, \bibinfo {author} {\bibfnamefont
  {A.}~\bibnamefont {Gulans}}, \bibinfo {author} {\bibfnamefont
  {F.}~\bibnamefont {Gygi}}, \bibinfo {author} {\bibfnamefont {D.~R.}\
  \bibnamefont {Hamann}}, \bibinfo {author} {\bibfnamefont {P.~J.}\
  \bibnamefont {Hasnip}}, \bibinfo {author} {\bibfnamefont {N.~A.~W.}\
  \bibnamefont {Holzwarth}}, \bibinfo {author} {\bibfnamefont {D.}~\bibnamefont
  {Iu{\c{s}}an}}, \bibinfo {author} {\bibfnamefont {D.~B.}\ \bibnamefont
  {Jochym}}, \bibinfo {author} {\bibfnamefont {F.}~\bibnamefont {Jollet}},
  \bibinfo {author} {\bibfnamefont {D.}~\bibnamefont {Jones}}, \bibinfo
  {author} {\bibfnamefont {G.}~\bibnamefont {Kresse}}, \bibinfo {author}
  {\bibfnamefont {K.}~\bibnamefont {Koepernik}}, \bibinfo {author}
  {\bibfnamefont {E.}~\bibnamefont {Kü{\c{c}}ükbenli}}, \bibinfo {author}
  {\bibfnamefont {Y.~O.}\ \bibnamefont {Kvashnin}}, \bibinfo {author}
  {\bibfnamefont {I.~L.~M.}\ \bibnamefont {Locht}}, \bibinfo {author}
  {\bibfnamefont {S.}~\bibnamefont {Lubeck}}, \bibinfo {author} {\bibfnamefont
  {M.}~\bibnamefont {Marsman}}, \bibinfo {author} {\bibfnamefont
  {N.}~\bibnamefont {Marzari}}, \bibinfo {author} {\bibfnamefont
  {U.}~\bibnamefont {Nitzsche}}, \bibinfo {author} {\bibfnamefont
  {L.}~\bibnamefont {Nordström}}, \bibinfo {author} {\bibfnamefont
  {T.}~\bibnamefont {Ozaki}}, \bibinfo {author} {\bibfnamefont
  {L.}~\bibnamefont {Paulatto}}, \bibinfo {author} {\bibfnamefont {C.~J.}\
  \bibnamefont {Pickard}}, \bibinfo {author} {\bibfnamefont {W.}~\bibnamefont
  {Poelmans}}, \bibinfo {author} {\bibfnamefont {M.~I.~J.}\ \bibnamefont
  {Probert}}, \bibinfo {author} {\bibfnamefont {K.}~\bibnamefont {Refson}},
  \bibinfo {author} {\bibfnamefont {M.}~\bibnamefont {Richter}}, \bibinfo
  {author} {\bibfnamefont {G.-M.}\ \bibnamefont {Rignanese}}, \bibinfo {author}
  {\bibfnamefont {S.}~\bibnamefont {Saha}}, \bibinfo {author} {\bibfnamefont
  {M.}~\bibnamefont {Scheffler}}, \bibinfo {author} {\bibfnamefont
  {M.}~\bibnamefont {Schlipf}}, \bibinfo {author} {\bibfnamefont
  {K.}~\bibnamefont {Schwarz}}, \bibinfo {author} {\bibfnamefont
  {S.}~\bibnamefont {Sharma}}, \bibinfo {author} {\bibfnamefont
  {F.}~\bibnamefont {Tavazza}}, \bibinfo {author} {\bibfnamefont
  {P.}~\bibnamefont {Thunström}}, \bibinfo {author} {\bibfnamefont
  {A.}~\bibnamefont {Tkatchenko}}, \bibinfo {author} {\bibfnamefont
  {M.}~\bibnamefont {Torrent}}, \bibinfo {author} {\bibfnamefont
  {D.}~\bibnamefont {Vanderbilt}}, \bibinfo {author} {\bibfnamefont {M.~J.}\
  \bibnamefont {van Setten}}, \bibinfo {author} {\bibfnamefont {V.~V.}\
  \bibnamefont {Speybroeck}}, \bibinfo {author} {\bibfnamefont {J.~M.}\
  \bibnamefont {Wills}}, \bibinfo {author} {\bibfnamefont {J.~R.}\ \bibnamefont
  {Yates}}, \bibinfo {author} {\bibfnamefont {G.-X.}\ \bibnamefont {Zhang}},\
  and\ \bibinfo {author} {\bibfnamefont {S.}~\bibnamefont {Cottenier}},\
  }\bibfield  {title} {\bibinfo {title} {Reproducibility in density functional
  theory calculations of solids},\ }\href@noop {} {\bibfield  {journal}
  {\bibinfo  {journal} {Science}\ }\textbf {\bibinfo {volume} {351}} (\bibinfo
  {year} {2016})}\BibitemShut {NoStop}%
\bibitem [{\citenamefont {Monkhorst}\ and\ \citenamefont
  {Pack}(1976)}]{Monkhorst:1976}%
  \BibitemOpen
  \bibfield  {author} {\bibinfo {author} {\bibfnamefont {H.}~\bibnamefont
  {Monkhorst}}\ and\ \bibinfo {author} {\bibfnamefont {J.}~\bibnamefont
  {Pack}},\ }\bibfield  {title} {\bibinfo {title} {{Special points for
  Brillouin-zone integration Monkhorst and Pack}},\ }\href@noop {} {\bibfield
  {journal} {\bibinfo  {journal} {Phys. Rev. B}\ }\textbf {\bibinfo {volume}
  {13}},\ \bibinfo {pages} {5188} (\bibinfo {year} {1976})}\BibitemShut
  {NoStop}%
\bibitem [{\citenamefont {Timrov}\ \emph {et~al.}(2022)\citenamefont {Timrov},
  \citenamefont {Marzari},\ and\ \citenamefont {Cococcioni}}]{Timrov:2022}%
  \BibitemOpen
  \bibfield  {author} {\bibinfo {author} {\bibfnamefont {I.}~\bibnamefont
  {Timrov}}, \bibinfo {author} {\bibfnamefont {N.}~\bibnamefont {Marzari}},\
  and\ \bibinfo {author} {\bibfnamefont {M.}~\bibnamefont {Cococcioni}},\
  }\bibfield  {title} {\bibinfo {title} {{HP} {\textendash} a code for the
  calculation of hubbard parameters using density-functional perturbation
  theory},\ } {\bibfield
  {journal} {\bibinfo  {journal} {Computer Physics Communications}\ }\textbf
  {\bibinfo {volume} {279}},\ \bibinfo {pages} {108455} (\bibinfo {year}
  {2022})}\BibitemShut {NoStop}%
\bibitem [{\citenamefont {van Setten}\ \emph {et~al.}(2018)\citenamefont {van
  Setten}, \citenamefont {Giantomassi}, \citenamefont {Bousquet}, \citenamefont
  {Verstraete}, \citenamefont {Hamann}, \citenamefont {Gonze},\ and\
  \citenamefont {Rignanese}}]{Setten:2018}%
  \BibitemOpen
  \bibfield  {author} {\bibinfo {author} {\bibfnamefont {M.}~\bibnamefont {van
  Setten}}, \bibinfo {author} {\bibfnamefont {M.}~\bibnamefont {Giantomassi}},
  \bibinfo {author} {\bibfnamefont {E.}~\bibnamefont {Bousquet}}, \bibinfo
  {author} {\bibfnamefont {M.}~\bibnamefont {Verstraete}}, \bibinfo {author}
  {\bibfnamefont {D.}~\bibnamefont {Hamann}}, \bibinfo {author} {\bibfnamefont
  {X.}~\bibnamefont {Gonze}},\ and\ \bibinfo {author} {\bibfnamefont {G.-M.}\
  \bibnamefont {Rignanese}},\ }
  {\bibfield  {journal} {\bibinfo  {journal} {Computer Physics Communications}\
  }\textbf {\bibinfo {volume} {226}},\ \bibinfo {pages} {39} (\bibinfo {year}
  {2018})}\BibitemShut {NoStop}%
\bibitem [{\citenamefont {Gygi}\ and\ \citenamefont
  {Baldereschi}(1986)}]{Gygi:1986}%
  \BibitemOpen
  \bibfield  {author} {\bibinfo {author} {\bibfnamefont {F.}~\bibnamefont
  {Gygi}}\ and\ \bibinfo {author} {\bibfnamefont {A.}~\bibnamefont
  {Baldereschi}},\ }
  {\bibfield {journal} {\bibinfo  {journal} {Physical Review B}\ }\textbf {\bibinfo
  {volume} {34}},\ \bibinfo {pages} {4405} (\bibinfo {year}
  {1986})}\BibitemShut {NoStop}%
\bibitem [{Mat()}]{MaterialsCloudArchive2022}%
  \BibitemOpen
  {\bibinfo {title} {{Materials Cloud Archive:}}}\ \bibinfo
  {howpublished} {\url{https://doi.org/10.24435/materialscloud:zh-14}}\BibitemShut {NoStop}%
\bibitem [{\citenamefont {Shannon}(1976)}]{Shannon:1976}%
  \BibitemOpen
  \bibfield  {author} {\bibinfo {author} {\bibfnamefont {R.~D.}\ \bibnamefont
  {Shannon}},\ }\bibfield  {title} {\bibinfo {title} {Revised effective ionic
  radii and systematic studies of interatomic distances in halides and
  chalcogenides},\ }
  {\bibfield  {journal} {\bibinfo  {journal} {Acta Crystallographica Section
  A}\ }\textbf {\bibinfo {volume} {32}},\ \bibinfo {pages} {751} (\bibinfo
  {year} {1976})}\BibitemShut {NoStop}%
\bibitem [{\citenamefont {Brown}\ and\ \citenamefont
  {Shannon}(1973)}]{Brown:1973}%
  \BibitemOpen
  \bibfield  {author} {\bibinfo {author} {\bibfnamefont {I.~D.}\ \bibnamefont
  {Brown}}\ and\ \bibinfo {author} {\bibfnamefont {R.~D.}\ \bibnamefont
  {Shannon}},\ }\bibfield  {title} {\bibinfo {title} {Empirical
  bond-strength--bond-length curves for oxides},\ } {\bibfield  {journal} {\bibinfo
  {journal} {Acta Crystallographica Section A}\ }\textbf {\bibinfo {volume}
  {29}},\ \bibinfo {pages} {266} (\bibinfo {year} {1973})}\BibitemShut
  {NoStop}%
\bibitem [{\citenamefont {Kawakami}\ \emph {et~al.}(2003)\citenamefont
  {Kawakami}, \citenamefont {Sasaki}, \citenamefont {Tabata}, \citenamefont
  {Shimooka}, \citenamefont {Kohiki}, \citenamefont {Matsushima}, \citenamefont
  {Oku},\ and\ \citenamefont {Shishido}}]{Kawakami:2003}%
  \BibitemOpen
  \bibfield  {author} {\bibinfo {author} {\bibfnamefont {S.}~\bibnamefont
  {Kawakami}}, \bibinfo {author} {\bibfnamefont {M.}~\bibnamefont {Sasaki}},
  \bibinfo {author} {\bibfnamefont {H.}~\bibnamefont {Tabata}}, \bibinfo
  {author} {\bibfnamefont {H.}~\bibnamefont {Shimooka}}, \bibinfo {author}
  {\bibfnamefont {S.}~\bibnamefont {Kohiki}}, \bibinfo {author} {\bibfnamefont
  {S.}~\bibnamefont {Matsushima}}, \bibinfo {author} {\bibfnamefont
  {M.}~\bibnamefont {Oku}},\ and\ \bibinfo {author} {\bibfnamefont
  {T.}~\bibnamefont {Shishido}},\ } {\bibfield
  {journal} {\bibinfo  {journal} {J. Alloys Compd.}\ }\textbf {\bibinfo
  {volume} {359}},\ \bibinfo {pages} {278} (\bibinfo {year}
  {2003})}\BibitemShut {NoStop}%
\bibitem [{\citenamefont {Guan}\ \emph {et~al.}(2015)\citenamefont {Guan},
  \citenamefont {Zheng}, \citenamefont {Mei}, \citenamefont {Molokeev},
  \citenamefont {Xie}, \citenamefont {Yang}, \citenamefont {Wu}, \citenamefont
  {Huang},\ and\ \citenamefont {Huang}}]{Guan:2015}%
  \BibitemOpen
  \bibfield  {author} {\bibinfo {author} {\bibfnamefont {M.}~\bibnamefont
  {Guan}}, \bibinfo {author} {\bibfnamefont {H.}~\bibnamefont {Zheng}},
  \bibinfo {author} {\bibfnamefont {L.}~\bibnamefont {Mei}}, \bibinfo {author}
  {\bibfnamefont {M.~S.}\ \bibnamefont {Molokeev}}, \bibinfo {author}
  {\bibfnamefont {J.}~\bibnamefont {Xie}}, \bibinfo {author} {\bibfnamefont
  {T.}~\bibnamefont {Yang}}, \bibinfo {author} {\bibfnamefont {X.}~\bibnamefont
  {Wu}}, \bibinfo {author} {\bibfnamefont {S.}~\bibnamefont {Huang}},\ and\
  \bibinfo {author} {\bibfnamefont {Z.}~\bibnamefont {Huang}},\ }\href@noop {}
  {\bibfield  {journal} {\bibinfo  {journal} {J. Am. Ceram. Soc.}\ }\textbf
  {\bibinfo {volume} {98}},\ \bibinfo {pages} {1182} (\bibinfo {year}
  {2015})}\BibitemShut {NoStop}%
\end{thebibliography}

\end{document}